\input hyperbasics
\catcode`\@=11
\def\unredoffs{\voffset=13mm \hoffset=6.5truemm} 
\def\redoffs{\voffset=-12.truemm\hoffset=-3truemm} 
\def\speclscape{}
%
\newbox\leftpage \newdimen\fullhsize \newdimen\hstitle \newdimen\hsbody
\newdimen\hdim
\hfuzz=1pt
\ifx\hyperdef\UNd@FiNeD\def\hyperdef#1#2#3#4{#4}\def\hyperref#1#2#3#4{#4}\fi
\def\newans{y }
\def\answb{y }
\ifx\answb\newans\message{(This uses normal fonts.)}%
%
\def\bigans{b }
\def\answ{b }
\ifx\answ\bigans\message{(Format simple colonne 12pts.}
\magnification=1200 \unredoffs\hsize=147truemm\vsize=219truemm
\hsbody=\hsize \hstitle=\hsize 
\else\message{(Format double colonne, 10pts.} \let\l@r=L
\magnification=1000 \vsize=182.5truemm
\redoffs%
\hstitle=122.5truemm\hsbody=122.5truemm\fullhsize=258truemm\hsize=\hsbody 
\output={
  \almostshipout{\leftline{\vbox{\makeheadline\pagebody\makefootline}}}
\advancepageno%
}
\def\almostshipout#1{\if L\l@r \count1=1 \message{[\the\count0.\the\count1]}
      \global\setbox\leftpage=#1 \global\let\l@r=R
 \else \count1=2
  \shipout\vbox{\speclscape{\hsize\fullhsize}
      \hbox to\fullhsize{\box\leftpage\hfil#1}}  \global\let\l@r=L\fi}
\fi

\def\sla#1{\mkern-1.5mu\raise0.4pt\hbox{$\not$}\mkern1.2mu #1\mkern 0.7mu}
\def\Dbar{\mkern-1.5mu\raise0.4pt\hbox{$\not$}\mkern-.1mu {\rm D}\mkern.1mu}
\def\Abar{\mkern1.mu\raise0.4pt\hbox{$\not$}\mkern-1.3mu A\mkern.1mu}
\def\Bbar{\mkern-0.mu\raise0.4pt\hbox{$\not$}\mkern-.3mu B\mkern 0.6mu}
\newskip\tableskipamount \tableskipamount=8pt plus 3pt minus 3pt


\newdimen\chapskip

\font\ssbx=cmssbx10  

\font\caprm=cmr9
\font\capit=cmti9
\font\capbf=cmbx9
\font\capsl=cmsl9
\font\capmi=cmmi9
\font\capex=cmex9
\font\capsy=cmsy9
\chapskip=17.5mm
\def\makeheadline{\vbox to 0pt{\vskip-22.5pt
\line{\vbox to8.5pt{}\the\headline}\vss}\nointerlineskip}
\font\tenbi=cmmib10 
\font\ninebi=cmmib9
\font\sevenbi=cmmib7 
\font\fivebi=cmmib5
\textfont4=\tenbi
\scriptfont4=\sevenbi
\scriptscriptfont4=\fivebi
\font\headrm=cmr10

\font\sixrm=cmr6
\font\fiverm=cmr5
\font\sixmi=cmmi6
\font\fivemi=cmmi5
\font\sixsy=cmsy6
\font\fivesy=cmsy5
\font\sixbf=cmbx6
\font\fivebf=cmbx5
\skewchar\capmi='177 \skewchar\sixmi='177 \skewchar\fivemi='177
\skewchar\capsy='60 \skewchar\sixsy='60 \skewchar\fivesy='60

\def\elevenpoint{
\textfont0=\caprm \scriptfont0=\sixrm \scriptscriptfont0=\fiverm
\def\rm{\fam0\caprm}
\textfont1=\capmi \scriptfont1=\sixmi \scriptscriptfont1=\fivemi
\textfont2=\capsy \scriptfont2=\sixsy \scriptscriptfont2=\fivesy
\textfont3=\capex \scriptfont3=\capex \scriptscriptfont3=\capex
\textfont\itfam=\capit \def\it{\fam\itfam\capit} 
\textfont\slfam=\capsl  \def\sl{\fam\slfam\capsl} 
\textfont\bffam=\capbf \scriptfont\bffam=\sixbf
\scriptscriptfont\bffam=\fivebf
\def\bf{\fam\bffam\capbf} 
\textfont4=\ninebi \scriptfont4=\sevenbi \scriptscriptfont4=\fivebi
\abovedisplayskip=11pt plus 3pt minus 8pt
\belowdisplayskip=\abovedisplayskip
\smallskipamount=2.7pt plus 1pt minus 1pt
\medskipamount=5.4pt plus 2pt minus 2pt
\bigskipamount=10.8pt plus 3.6pt minus 3.6pt
\normalbaselineskip=11pt
\setbox\strutbox=\hbox{\vrule height7.8pt depth3.2pt width0pt}
\normalbaselines \rm}

%
%

\catcode`\@=11

\font\tenmsa=msam10
\font\sevenmsa=msam7
\font\fivemsa=msam5
\font\tenmsb=msbm10
\font\sevenmsb=msbm7
\font\fivemsb=msbm5
\newfam\msafam
\newfam\msbfam
\textfont\msafam=\tenmsa  \scriptfont\msafam=\sevenmsa
  \scriptscriptfont\msafam=\fivemsa
\textfont\msbfam=\tenmsb  \scriptfont\msbfam=\sevenmsb
  \scriptscriptfont\msbfam=\fivemsb

\def\hexnumber@#1{\ifcase#1 0\or1\or2\or3\or4\or5\or6\or7\or8\or9\or
	A\or B\or C\or D\or E\or F\fi }

\font\teneuf=eufm10
\font\seveneuf=eufm7
\font\fiveeuf=eufm5
\newfam\euffam
\textfont\euffam=\teneuf
\scriptfont\euffam=\seveneuf
\scriptscriptfont\euffam=\fiveeuf
\def\frak{\ifmmode\let\next\frak@\else
 \def\next{\Err@{Use \string\frak\space only in math mode}}\fi\next}
\def\goth{\ifmmode\let\next\frak@\else
 \def\next{\Err@{Use \string\goth\space only in math mode}}\fi\next}
\def\frak@#1{{\frak@@{#1}}}
\def\frak@@#1{\fam\euffam#1}

\edef\msa@{\hexnumber@\msafam}
\edef\msb@{\hexnumber@\msbfam}

\mathchardef\boxdot="2\msa@00
\mathchardef\boxplus="2\msa@01
\mathchardef\boxtimes="2\msa@02
\mathchardef\square="0\msa@03
\mathchardef\blacksquare="0\msa@04
\mathchardef\centerdot="2\msa@05
\mathchardef\lozenge="0\msa@06
\mathchardef\blacklozenge="0\msa@07
\mathchardef\circlearrowright="3\msa@08
\mathchardef\circlearrowleft="3\msa@09
\mathchardef\rightleftharpoons="3\msa@0A
\mathchardef\leftrightharpoons="3\msa@0B
\mathchardef\boxminus="2\msa@0C
\mathchardef\Vdash="3\msa@0D
\mathchardef\Vvdash="3\msa@0E
\mathchardef\vDash="3\msa@0F
\mathchardef\twoheadrightarrow="3\msa@10
\mathchardef\twoheadleftarrow="3\msa@11
\mathchardef\leftleftarrows="3\msa@12
\mathchardef\rightrightarrows="3\msa@13
\mathchardef\upuparrows="3\msa@14
\mathchardef\downdownarrows="3\msa@15
\mathchardef\upharpoonright="3\msa@16

\mathchardef\downharpoonright="3\msa@17
\mathchardef\upharpoonleft="3\msa@18
\mathchardef\downharpoonleft="3\msa@19
\mathchardef\rightarrowtail="3\msa@1A
\mathchardef\leftarrowtail="3\msa@1B
\mathchardef\leftrightarrows="3\msa@1C
\mathchardef\rightleftarrows="3\msa@1D
\mathchardef\Lsh="3\msa@1E
\mathchardef\Rsh="3\msa@1F
\mathchardef\rightsquigarrow="3\msa@20
\mathchardef\leftrightsquigarrow="3\msa@21
\mathchardef\looparrowleft="3\msa@22
\mathchardef\looparrowright="3\msa@23
\mathchardef\circeq="3\msa@24
\mathchardef\succsim="3\msa@25
\mathchardef\gtrsim="3\msa@26
\mathchardef\gtrapprox="3\msa@27
\mathchardef\multimap="3\msa@28
\mathchardef\therefore="3\msa@29
\mathchardef\because="3\msa@2A
\mathchardef\doteqdot="3\msa@2B

\mathchardef\triangleq="3\msa@2C
\mathchardef\precsim="3\msa@2D
\mathchardef\lesssim="3\msa@2E
\mathchardef\lessapprox="3\msa@2F
\mathchardef\eqslantless="3\msa@30
\mathchardef\eqslantgtr="3\msa@31
\mathchardef\curlyeqprec="3\msa@32
\mathchardef\curlyeqsucc="3\msa@33
\mathchardef\preccurlyeq="3\msa@34
\mathchardef\leqq="3\msa@35
\mathchardef\leqslant="3\msa@36
\mathchardef\lessgtr="3\msa@37
\mathchardef\backprime="0\msa@38
\mathchardef\risingdotseq="3\msa@3A
\mathchardef\fallingdotseq="3\msa@3B
\mathchardef\succcurlyeq="3\msa@3C
\mathchardef\geqq="3\msa@3D
\mathchardef\geqslant="3\msa@3E
\mathchardef\gtrless="3\msa@3F
\mathchardef\sqsubset="3\msa@40
\mathchardef\sqsupset="3\msa@41
\mathchardef\vartriangleright="3\msa@42
\mathchardef\vartriangleleft="3\msa@43
\mathchardef\trianglerighteq="3\msa@44
\mathchardef\trianglelefteq="3\msa@45
\mathchardef\bigstar="0\msa@46
\mathchardef\between="3\msa@47
\mathchardef\blacktriangledown="0\msa@48
\mathchardef\blacktriangleright="3\msa@49
\mathchardef\blacktriangleleft="3\msa@4A
\mathchardef\vartriangle="0\msa@4D
\mathchardef\blacktriangle="0\msa@4E
\mathchardef\triangledown="0\msa@4F
\mathchardef\eqcirc="3\msa@50
\mathchardef\lesseqgtr="3\msa@51
\mathchardef\gtreqless="3\msa@52
\mathchardef\lesseqqgtr="3\msa@53
\mathchardef\gtreqqless="3\msa@54
\mathchardef\Rrightarrow="3\msa@56
\mathchardef\Lleftarrow="3\msa@57
\mathchardef\veebar="2\msa@59
\mathchardef\barwedge="2\msa@5A
\mathchardef\doublebarwedge="2\msa@5B
\mathchardef\angle="0\msa@5C
\mathchardef\measuredangle="0\msa@5D
\mathchardef\sphericalangle="0\msa@5E
\mathchardef\varpropto="3\msa@5F
\mathchardef\smallsmile="3\msa@60
\mathchardef\smallfrown="3\msa@61
\mathchardef\Subset="3\msa@62
\mathchardef\Supset="3\msa@63
\mathchardef\Cup="2\msa@64

\mathchardef\Cap="2\msa@65

\mathchardef\curlywedge="2\msa@66
\mathchardef\curlyvee="2\msa@67
\mathchardef\leftthreetimes="2\msa@68
\mathchardef\rightthreetimes="2\msa@69
\mathchardef\subseteqq="3\msa@6A
\mathchardef\supseteqq="3\msa@6B
\mathchardef\bumpeq="3\msa@6C
\mathchardef\Bumpeq="3\msa@6D
\mathchardef\lll="3\msa@6E

\mathchardef\ggg="3\msa@6F

\mathchardef\circledS="0\msa@73
\mathchardef\pitchfork="3\msa@74
\mathchardef\dotplus="2\msa@75
\mathchardef\backsim="3\msa@76
\mathchardef\backsimeq="3\msa@77
\mathchardef\complement="0\msa@7B
\mathchardef\intercal="2\msa@7C
\mathchardef\circledcirc="2\msa@7D
\mathchardef\circledast="2\msa@7E
\mathchardef\circleddash="2\msa@7F
\def\ulcorner{\delimiter"4\msa@70\msa@70 }
\def\urcorner{\delimiter"5\msa@71\msa@71 }
\def\llcorner{\delimiter"4\msa@78\msa@78 }
\def\lrcorner{\delimiter"5\msa@79\msa@79 }
\def\yen{\mathhexbox\msa@55 }
\def\checkmark{\mathhexbox\msa@58 }
\def\circledR{\mathhexbox\msa@72 }
\def\maltese{\mathhexbox\msa@7A }
\mathchardef\lvertneqq="3\msb@00
\mathchardef\gvertneqq="3\msb@01
\mathchardef\nleq="3\msb@02
\mathchardef\ngeq="3\msb@03
\mathchardef\nless="3\msb@04
\mathchardef\ngtr="3\msb@05
\mathchardef\nprec="3\msb@06
\mathchardef\nsucc="3\msb@07
\mathchardef\lneqq="3\msb@08
\mathchardef\gneqq="3\msb@09
\mathchardef\nleqslant="3\msb@0A
\mathchardef\ngeqslant="3\msb@0B
\mathchardef\lneq="3\msb@0C
\mathchardef\gneq="3\msb@0D
\mathchardef\npreceq="3\msb@0E
\mathchardef\nsucceq="3\msb@0F
\mathchardef\precnsim="3\msb@10
\mathchardef\succnsim="3\msb@11
\mathchardef\lnsim="3\msb@12
\mathchardef\gnsim="3\msb@13
\mathchardef\nleqq="3\msb@14
\mathchardef\ngeqq="3\msb@15
\mathchardef\precneqq="3\msb@16
\mathchardef\succneqq="3\msb@17
\mathchardef\precnapprox="3\msb@18
\mathchardef\succnapprox="3\msb@19
\mathchardef\lnapprox="3\msb@1A
\mathchardef\gnapprox="3\msb@1B
\mathchardef\nsim="3\msb@1C
\mathchardef\ncong="3\msb@1D

\mathchardef\varsubsetneq="3\msb@20
\mathchardef\varsupsetneq="3\msb@21
\mathchardef\nsubseteqq="3\msb@22
\mathchardef\nsupseteqq="3\msb@23
\mathchardef\subsetneqq="3\msb@24
\mathchardef\supsetneqq="3\msb@25
\mathchardef\varsubsetneqq="3\msb@26
\mathchardef\varsupsetneqq="3\msb@27
\mathchardef\subsetneq="3\msb@28
\mathchardef\supsetneq="3\msb@29
\mathchardef\nsubseteq="3\msb@2A
\mathchardef\nsupseteq="3\msb@2B
\mathchardef\nparallel="3\msb@2C
\mathchardef\nmid="3\msb@2D
\mathchardef\nshortmid="3\msb@2E
\mathchardef\nshortparallel="3\msb@2F
\mathchardef\nvdash="3\msb@30
\mathchardef\nVdash="3\msb@31
\mathchardef\nvDash="3\msb@32
\mathchardef\nVDash="3\msb@33
\mathchardef\ntrianglerighteq="3\msb@34
\mathchardef\ntrianglelefteq="3\msb@35
\mathchardef\ntriangleleft="3\msb@36
\mathchardef\ntriangleright="3\msb@37
\mathchardef\nleftarrow="3\msb@38
\mathchardef\nrightarrow="3\msb@39
\mathchardef\nLeftarrow="3\msb@3A
\mathchardef\nRightarrow="3\msb@3B
\mathchardef\nLeftrightarrow="3\msb@3C
\mathchardef\nleftrightarrow="3\msb@3D
\mathchardef\divideontimes="2\msb@3E
\mathchardef\varnothing="0\msb@3F
\mathchardef\nexists="0\msb@40
\mathchardef\mho="0\msb@66
\mathchardef\eth="0\msb@67
\mathchardef\eqsim="3\msb@68
\mathchardef\beth="0\msb@69
\mathchardef\gimel="0\msb@6A
\mathchardef\daleth="0\msb@6B
\mathchardef\lessdot="3\msb@6C
\mathchardef\gtrdot="3\msb@6D
\mathchardef\ltimes="2\msb@6E
\mathchardef\rtimes="2\msb@6F
\mathchardef\shortmid="3\msb@70
\mathchardef\shortparallel="3\msb@71
\mathchardef\smallsetminus="2\msb@72
\mathchardef\thicksim="3\msb@73
\mathchardef\thickapprox="3\msb@74
\mathchardef\approxeq="3\msb@75
\mathchardef\succapprox="3\msb@76
\mathchardef\precapprox="3\msb@77
\mathchardef\curvearrowleft="3\msb@78
\mathchardef\curvearrowright="3\msb@79
\mathchardef\digamma="0\msb@7A
\mathchardef\varkappa="0\msb@7B
\mathchardef\hslash="0\msb@7D
\mathchardef\hbar="0\msb@7E
\mathchardef\backepsilon="3\msb@7F
\def\Bbb{\ifmmode\let\next\Bbb@\else
 \def\next{\errmessage{Use \string\Bbb\space only in math mode}}\fi\next}
\def\Bbb@#1{{\Bbb@@{#1}}}
\def\Bbb@@#1{\fam\msbfam#1}

\catcode`\@=12

\def\sla#1{\mkern-1.5mu\raise0.4pt\hbox{$\not$}\mkern1.2mu #1\mkern 0.7mu}
\def\Dbar{\mkern-1.5mu\raise0.4pt\hbox{$\not$}\mkern-.1mu {\rm D}\mkern.1mu}
\def\Abar{\mkern1.mu\raise0.4pt\hbox{$\not$}\mkern-1.3mu A\mkern.1mu}
\nopagenumbers
\headline={\ifnum\pageno=1\hfill\else\draftdate\hfil{\headrm\folio}%
\hfil\fi}	 
\else\message{(This uses pseudo 12pts fonts.}
\hoffset=8mm
\voffset=16mm
\input lfont12 

\def\sla#1{\mkern-1.5mu\raise0.5pt\hbox{$\not$}\mkern1.2mu #1\mkern 0.7mu}
\def\Dbar{\mkern-1.5mu\raise0.5pt\hbox{$\not$}\mkern-.1mu {\rm D}\mkern.1mu}
\def\Abar{\mkern1.mu\raise0.5pt\hbox{$\not$}\mkern-1.3mu A\mkern.1mu}
\fi

\newcount\yearltd\yearltd=\year\advance\yearltd by -1900
\newif\ifdraftmode
\draftmodefalse
\def\draft{\draftmodetrue{\count255=\time\divide\count255 by 60
\xdef\hourmin{\number\count255} 
  \multiply\count255 by-60\advance\count255 by\time
  \xdef\hourmin{\hourmin:\ifnum\count255<10 0\fi\the\count255}}}
\def\draftdate{\ifdraftmode{\headrm\quad (\jobname,\ le
\number\day/\number\month/\number\yearltd\ \ \hourmin)}\else{}\fi} 
\newif\iffrancmode
\francmodefalse
\def\e{\mathop{\rm e}\nolimits}

\def\d{{\rm d}}
\def\ud{{\textstyle{1\over 2}}}

\def\tr{\mathop{\rm tr}\nolimits}
\def\det{\mathop{\rm det}\nolimits}

\chardef\sigmat=27

\def\frac#1#2{{\textstyle{#1\over#2}}}

\def\leaderfill{\leaders\hbox to 1em{\hss.\hss}\hfill}
\catcode`\@=11
\def\deqalignno#1{\displ@y\tabskip\centering \halign to
\displaywidth{\hfil$\displaystyle{##}$\tabskip0pt&$\displaystyle{{}##}$
\hfil\tabskip0pt &\quad
\hfil$\displaystyle{##}$\tabskip0pt&$\displaystyle{{}##}$ 
\hfil\tabskip\centering& \llap{$##$}\tabskip0pt \crcr #1 \crcr}}
\def\deqalign#1{\null\,\vcenter{\openup\jot\m@th\ialign{
\strut\hfil$\displaystyle{##}$&$\displaystyle{{}##}$\hfil
&&\quad\strut\hfil$\displaystyle{##}$&$\displaystyle{{}##}$
\hfil\crcr#1\crcr}}\,}
\def\xlabel#1{\expandafter\xl@bel#1}\def\xl@bel#1{#1}
\def\label#1{\l@bel #1{\hbox{}}}
\def\l@bel#1{\ifx\UNd@FiNeD#1\message{label \string#1 is undefined.}%
\xdef#1{?.? }\fi{\let\hyperref=\relax\xdef\next{#1}}%
\ifx\next\em@rk\def\next{}%
\else\def\next{#1}\fi\next}
\def\DefWarn#1{\ifx\UNd@FiNeD#1\else
\immediate\write16{*** WARNING: the label \string#1 is already defined%
***}\fi}%
\newread\ch@ckfile
\def\cinput#1{\def\filen@me{#1 }
\immediate\openin\ch@ckfile=\filen@me
\ifeof\ch@ckfile\message{<< (\filen@me\ DOES NOT EXIST in this pass)>>}\else%
\closein \ch@ckfile\input\filen@me\fi}
\ifx\UNd@FiNeD\prefix\def\prefix{}\fi 
\newread\ch@ckfile
\immediate\openin\ch@ckfile=\jobname.def
\ifeof\ch@ckfile\message{<< (\jobname.def DOES NOT EXIST in this pass) >>}
\else
\def\DefWarn#1{}%
\closein \ch@ckfile%
\input\jobname.def\fi
\def\listcontent{
\immediate\openin\ch@ckfile=\jobname.tab 
\ifeof\ch@ckfile\message{no file \jobname.tab, no table of contents this
pass}%
\else\closein\ch@ckfile\centerline{\bf\iffrancmode Table des
mati\`eres \else Contents\fi}\nobreak\medskip%
{\baselineskip=12pt\parskip=0pt\catcode`\@=11\input\jobname.tab
\catcode`\@=12\bigbreak\bigskip}\fi}
\newcount\nosection
\newcount\nosubsection
\newcount\neqno
\newcount\notenumber
\newcount\nofigure
\newcount\notable
\newif\ifappmode
\def\equation{\jobname.equ}
\newwrite\equa

\newdimen\hulp
\def\maketitle#1{
\edef\oneliner##1{\centerline{##1}}
\edef\twoliner##1{\vbox{\parindent=0pt\leftskip=0pt plus 1fill\rightskip=0pt
plus 1fill 
                     \parfillskip=0pt\relax##1}} 
\setbox0=\vbox{#1}\hulp=0.5\hsize
                 \ifdim\wd0<\hulp\oneliner{#1}\else
                 \twoliner{#1}\fi}
\def\preprint#1{\ifdraftmode\gdef\prepname{\jobname/#1}\else%
\gdef\prepname{#1}\fi\hfill{
\expandafter{\prepname}}\vskip20mm} 
\def\title#1\par{\gdef\titlename{#1}
\maketitle{\ssbx\uppercase\expandafter{\titlename}}
\vskip20truemm
\nosection=0
\neqno=0
\notenumber=0
\nofigure=0
\notable=0
\def\prefix{}
\appmodefalse
\mark{\the\nosection}
\message{#1}
\immediate\openout\equa=\equation
}
\def\abstract{\vskip8mm\iffrancmode\centerline{R\'ESUM\'E}\else%
\centerline{ABSTRACT}\fi \smallskip \begingroup\narrower
\elevenpoint\baselineskip10pt} 
\def\endabstract{\par\endgroup \bigskip}
\def\section#1\par{\vskip0pt plus.1\vsize\penalty-100\vskip0pt plus-.1
\vsize\bigskip\vskip\parskip
\ifnum\nosection=0\ifappmode\relax\else\writetoc
\fi\fi
\advance\nosection by 1\global\nosubsection=0\global\neqno=0
\vbox{\noindent\bf{\hyperdef\hypernoname{section}{\prefix\the\nosection}%
{\prefix\the\nosection}\ #1}}
\writetoca{{\string\hyperref{}{section}{\prefix\the\nosection}%
{\prefix\the\nosection}} {#1}}
\message{\the\nosection\ #1}
\mark{\the\nosection}\bigskip\noindent
}

\def\appendix#1#2\par{\bigbreak\nosection=0
\appmodetrue
\notenumber=0
\neqno=0
\def\prefix{A}
\mark{\the\nosection}
\message{APPENDICES}
{\leftline{APPENDICES} \hyperdef\hypernoname{appendix}{\prefix}{ 
\leftline{\uppercase\expandafter{#1}}
\leftline{\uppercase\expandafter{#2}}}}
\bigskip\noindent\nonfrenchspacing
\writetoca{\string\hyperref{}{appendix}{\prefix}{Appendices}.\ #1 \ #2}%
}
\def\subsection#1\par {\vskip0pt plus.05\vsize\penalty-100\vskip0pt
plus-.05\vsize\bigskip\vskip\parskip\advance\nosubsection by 1
\vbox{\noindent\it{\hyperdef\hypernoname{subsection}{\prefix\the\nosection.%
\the\nosubsection}{\prefix\the\nosection.\the\nosubsection\ #1}}}%
\smallskip\noindent 
\writetoca{{\string\hyperref{}{subsection}{\prefix\the\nosection.%
\the\nosubsection}{\prefix\the\nosection.\the\nosubsection}} {#1}}
\message{\the\nosection.\the\nosubsection\ #1}
} 
\def\note #1{\advance\notenumber by 1
\footnote{$^{\the\notenumber}$}{\sevenrm #1}} 

\parindent=1em 
\newinsert\margin
\dimen\margin=\maxdimen
\count\margin=0 \skip\margin=0pt
\def\sslbl#1{\DefWarn#1%
\ifdraftmode{\hfill\escapechar-1{\rlap{\hskip-1mm%
\sevenrm\string#1}}}\fi%
\ifnum\nosection=0\if\prefix{}\xdef#1{}%
\edef\ewrite{\write\equa{{\string#1}}%
\write\equa{}}\ewrite%
\else
\xdef#1{\noexpand\hyperref{}{appendix}{\prefix}{\prefix}}%
\edef\ewrite{\write\equa{{\string#1},\prefix}%
\write\equa{}}\ewrite%
\writedef{#1\leftbracket#1}
\fi
\else%
\ifnum\nosubsection=0%
\xdef#1{\noexpand\hyperref{}{section}{\prefix\the\nosection}%
{\prefix\the\nosection}}%
\edef\ewrite{\write\equa{{\string#1},\prefix\the\nosection}%
\write\equa{}}\ewrite%
\writedef{#1\leftbracket#1}
\else%
\xdef#1{\noexpand\hyperref{}{subsection}{\prefix\the\nosection.%
\the\nosubsection}{\prefix\the\nosection.\the\nosubsection}}%
\writedef{#1\leftbracket#1}
\edef\ewrite{\write\equa{{\string#1},\prefix\the\nosection%
.\the\nosubsection}\write\equa{}}\ewrite\fi\fi}%

\newwrite\tfile \def\writetoca#1{}
\def\writetoc{\immediate\openout\tfile=\jobname.tab
\def\writetoca##1{{\edef\next{\write\tfile{\noindent ##1 \string\leaderfill%
\noexpand\number\pageno\par}}\next}}}

%
\def\nolabels{\def\wrlabeL##1{}\def\eqlabeL##1{}\def\reflabeL##1{}}
\def\writelabels{\def\wrlabeL##1{\leavevmode\vadjust{\rlap{\smash%
{\line{{\escapechar=` \hfill\rlap{\sevenrm\hskip.03in\string##1}}}}}}}%
\def\eqlabeL##1{{\escapechar-1\rlap{\sevenrm\hskip.05in\string##1}}}%
\def\reflabeL##1{\noexpand\llap{\noexpand\sevenrm\string\string\string##1}}}
\nolabels

\global\newcount\refno \global\refno=1
\newwrite\rfile
\def\ref{[\hyperref{}{reference}{\the\refno}{\the\refno}]\nref}
\def\nref#1{\DefWarn#1%
\xdef#1{[\noexpand\hyperref{}{reference}{\the\refno}{\the\refno}]}%
\writedef{#1\leftbracket#1}%
\ifnum\refno=1\immediate\openout\rfile=\jobname.ref\fi
\chardef\wfile=\rfile\immediate\write\rfile{\noexpand\item{[\noexpand\hyperdef%
\noexpand\hypernoname{reference}{\the\refno}{\the\refno}]\ }%
\reflabeL{#1\hskip.31in}\pctsign}\global\advance\refno by1\findarg}
\def\findarg#1#{\begingroup\obeylines\newlinechar=`\^^M\pass@rg}
{\obeylines\gdef\pass@rg#1{\writ@line\relax #1^^M\hbox{}^^M}%
\gdef\writ@line#1^^M{\expandafter\toks0\expandafter{\striprel@x #1}%
\edef\next{\the\toks0}\ifx\next\em@rk\let\next=\endgroup\else\ifx\next\empty%
\else\immediate\write\wfile{\the\toks0}\fi\let\next=\writ@line\fi\next\relax}}
\def\striprel@x#1{} \def\em@rk{\hbox{}}
\def\lref{\begingroup\obeylines\lr@f}
\def\lr@f#1#2{\DefWarn#1\gdef#1{\let#1=\UNd@FiNeD\ref#1{#2}}\endgroup\unskip}
\def\semi{;\hfil\break}
\def\addref#1{\immediate\write\rfile{\noexpand\item{}#1}} 
\def\listrefs{{}\vfill\supereject\immediate\closeout\rfile\writestoppt
\baselineskip=14pt\centerline{{\bf\iffrancmode R\'eferences\else References%
\fi}}
\bigskip{\parindent=20pt%
\frenchspacing\escapechar=` \input \jobname.ref\vfill\eject}\nonfrenchspacing}
\def\startrefs#1{\immediate\openout\rfile=\jobname.ref\refno=#1}
\def\xref{\expandafter\xr@f}\def\xr@f[#1]{#1}
\def\refs#1{\count255=1[\r@fs #1{\hbox{}}]}
\def\r@fs#1{\ifx\UNd@FiNeD#1\message{reflabel \string#1 is undefined.}%
\nref#1{need to supply reference \string#1.}\fi%
\vphantom{\hphantom{#1}}{\let\hyperref=\relax\xdef\next{#1}}%
\ifx\next\em@rk\def\next{}%
\else\ifx\next#1\ifodd\count255\relax\xref#1\count255=0\fi%
\else#1\count255=1\fi\let\next=\r@fs\fi\next}
%
\newwrite\lfile
{\escapechar-1\xdef\pctsign{\string\%}\xdef\leftbracket{\string\{}
\xdef\rightbracket{\string\}}\xdef\numbersign{\string\#}}
\def\writedefs{\immediate\openout\lfile=\jobname.def \def\writedef##1{%
{\let\hyperref=\relax\let\hyperdef=\relax\let\hypernoname=\relax
 \immediate\write\lfile{\string\def\string##1\rightbracket}}}}%
\def\writestop{\def\writestoppt{\immediate\write\lfile{\string\pageno%
\the\pageno\string\startrefs\leftbracket\the\refno\rightbracket%
\string\def\string\secsym\leftbracket\secsym\rightbracket%
\string\secno\the\secno\string\meqno\the\meqno}\immediate\closeout\lfile}}
\def\writestoppt{}\def\writedef#1{}
\writedefs
\def\biblio\par{\vskip0pt plus.1\vsize\penalty-100\vskip0pt plus-.1
\vsize\bigskip\vskip\parskip
\message{Bibliographie}
{\leftline{\bf \hyperdef\hypernoname{biblio}{bib}{Bibliographical Notes}}}
\nobreak\medskip\noindent\frenchspacing
\writetoca{\string\hyperref{}{biblio}{bib}{Bibliographical Notes}}}%

\def\biblionote{\iffrancmode Notes Bibliographiques\else Bibliographical Notes
\fi}
\def\beginbib\par{\vskip0pt plus.1\vsize\penalty-100\vskip0pt plus-.1
\vsize\bigskip\vskip\parskip
\message{Bibliographie}
{\leftline{\bf \hyperdef\hypernoname{biblio}{\the\nosection}%
{\biblionote}}}
\nobreak\medskip\noindent\frenchspacing
\writetoca{\string\hyperref{}{biblio}{\the\nosection}%
{\biblionote}}}%

\def\Exercises{\iffrancmode Exercices\else Exercises
\fi}
\def\exerc\par{\vskip0pt plus.1\vsize\penalty-100\vskip0pt plus-.1
\vsize\bigskip\vskip\parskip
\message{Exercises}
{\leftline{\bf \hyperdef\hypernoname{exercise}{\the\nosection}{\Exercises}}}
\nobreak\medskip\noindent\frenchspacing
\writetoca{\string\hyperref{}{exercise}{\the\nosection}{\Exercises}}
}
\def\eqnn{\global\advance\neqno by 1 \ifinner\relax\else%
\eqno\fi(\prefix\the\nosection.\the\neqno)}
%
\def\eqnd#1{\DefWarn#1%
\global\advance\neqno by 1 
{\xdef#1{($\noexpand\hyperref{}{equation}{\prefix\the\nosection.\the\neqno}%
{\prefix\the\nosection.\the\neqno}$)}}
\ifinner\relax\else\eqno\fi(\hyperdef\hypernoname{equation}{\prefix\the%
\nosection.\the\neqno}{\prefix\the\nosection.\the\neqno})
\writedef{#1\leftbracket#1}
\ifdraftmode{\escapechar-1{\rlap{\hskip.2mm\sevenrm\string#1}}}\fi
\edef\ewrite{\write\equa{{\string#1},(\prefix\the\nosection.\the\neqno)
{\noexpand\number\pageno}}\write\equa{}}\ewrite}
%
\def\checkm@de#1#2{\ifmmode{\def\f@rst##1{##1}\hyperdef\hypernoname{equation}%
{#1}{#2}}\else\hyperref{}{equation}{#1}{#2}\fi}
\def\f@rst#1{\c@t#1a\em@ark}\def\c@t#1#2\em@ark{#1}
\def\eqna#1{\DefWarn#1%
\global\advance\neqno by1\ifdraftmode{\hfill%
\escapechar-1{\rlap{\sevenrm\string#1}}}\fi%
\xdef #1##1{(\noexpand\relax\noexpand%
\checkm@de{\prefix\the\nosection.\the\neqno\noexpand\f@rst{##1}1}%
{\hbox{$\prefix\the\nosection.\the\neqno##1$}})}
\writedef{#1\numbersign1\leftbracket#1{\numbersign1}}%
} 
%

%
\def\em@rk{\hbox{}} 
\def\xeqn{\expandafter\xe@n}\def\xe@n(#1){#1}
\def\xeqna#1{\expandafter\xe@na#1}\def\xe@na\hbox#1{\xe@nap #1}
\def\xe@nap$(#1)${\hbox{$#1$}}
\def\eqns#1{(\e@ns #1{\hbox{}})}
\def\e@ns#1{\ifx\UNd@FiNeD#1\message{eqnlabel \string#1 is undefined.}%
\xdef#1{(?.?)}\fi{\let\hyperref=\relax\xdef\next{#1}}%
\ifx\next\em@rk\def\next{}%
\else\ifx\next#1\xeqn#1\else\def\n@xt{#1}\ifx\n@xt\next#1\else\xeqna#1\fi
\fi\let\next=\e@ns\fi\next}
\def\figure#1#2{\global\advance\nofigure by 1 \vglue#1%
\hyperdef\hypernoname{figure}{\the\nofigure}{}%
{\elevenpoint
\setbox1=\hbox{#2}
\ifdim\wd1=0pt\centerline{Fig.\ \the\nofigure\hskip0.5mm}%
\else\def\caption{Fig.\ \the\nofigure\quad#2\hskip0mm}
\setbox0=\hbox{\caption}
\ifdim\wd0>\hsize\noindent Fig.\ \the\nofigure\quad#2\else
                 \centerline{\caption}\fi\fi}\par}
\def\lfigure#1#2{\global\advance\nofigure by
1\vglue#1%
\hyperdef\hypernoname{figure}{\the\nofigure}{}%
\leftline{\elevenpoint\hskip10truemm  Fig.\
\the\nofigure\quad #2}} 
\def\figlbl#1{\ifdraftmode{\hfill\escapechar-1{\rlap{\hskip-1mm%
\sevenrm\string#1}}}\fi%
{\xdef#1{\noexpand\hyperref{}{figure}{\the\nofigure}%
{\the\nofigure}}}%
\edef\ewrite{\write\equa{{\string#1}\the\nofigure}%
\write\equa{}}\ewrite%
\writedef{#1\leftbracket#1}}
\def\tablbl#1{\global\advance\notable by
1\ifdraftmode{\hfill\escapechar-1{\rlap{\hskip-1mm%
\sevenrm\string#1}}}\fi%
{\xdef#1{\noexpand\hyperref{}{table}{\the\notable}%
{\the\notable}}}%
\hyperdef\hypernoname{table}{\the\notable}{}%
\edef\ewrite{\write\equa{{\string#1}\the\notable}%
\write\equa{}}\ewrite%
\writedef{#1\leftbracket#1}}

\catcode`@=12

\input epsf
\francmodefalse
\def\r{{\rm r}}
\def\Pib{ \Pi}
\def\phib{\phi}
\def\varphib{\varphi}
\def\psib{\psi}
\def\MS{$\overline{\rm MS}$}
\preprint{T00/055}

\title{Quantum Field Theory at Finite Temperature: An Introduction} 

\centerline{J.~ZINN-JUSTIN*}
\medskip
\centerline{\it CEA-Saclay, Service de Physique Th\'eorique**,}
\centerline{\it F-91191 Gif-sur-Yvette  Cedex, FRANCE.}
\smallskip
\centerline{\it and}
\smallskip
\centerline{\it University of Cergy-Pontoise}
\footnote{}{${}^*$email: zinn@spht.saclay.cea.fr}

\footnote{}{${}^{**}$Laboratoire de la Direction des
Sciences de la Mati\`ere du 
Commissariat \`a l'Energie Atomique}

\abstract
In these notes we review some properties of Statistical Quantum Field Theory
at equilibrium, i.e~Quantum Field Theory at finite temperature. We explain the
relation between finite temperature quantum field theory in $(d,1)$ dimensions
and statistical classical field theory in $d+1$ dimensions. This
identification allows to analyze the finite temperature QFT in terms of the
renormalization group and the theory of finite size effects of the classical
theory. We discuss in particular the limit of high temperature (HT) or the
situation of finite temperature phase transitions. There the concept of
dimensional reduction plays an essential role.        
Dimensional reduction in some sense reflects the known property that quantum
effects are not important at high temperature.  \par  
We illustrate these ideas with several standard examples, $\phi^4$ field
theory, the non-linear $\sigma $ model and the Gross--Neveu model, gauge
theories. We construct the corresponding effective reduced theories at
one-loop order, using the technique of mode expansion of fields in the
imaginary time variable. In models where the field is a vector with $N$
components, the large $N$ expansion provides another specially convenient tool
to study dimensional reduction.      
\endabstract
\vfill\eject
\listcontent
\vfill\eject
\nref\rKiLi{D.A. Kirznits {\it JETP Lett.} 15 (1972) 529\semi
D.A. Kirznits and A.D. Linde, {\it Phys. Lett.} B42 (1972) 471; {\ it Ann.
Phys.} 101 (1976) 195.} 
\nref\rWein{C. W. Bernard, {\it Phys. Rev.} D9 (1974) 3312\semi
S. Weinberg, {\it Phys. Rev.} D9 (1974) 3357.}
\nref\rDoJa{L. Dolan and R. Jackiw, {\it Phys. Rev.} D9 (1974) 3320.} 
\nref\rGroPisYa{D.J. Gross, R.D. Pisarski and L.G. Yaffe, {\it Rev. Mod.
Phys.} 53 (1981) 43.} 
\nref\rLWKLBM{N.P. Landsman and C. van Weert, {\it Phys. Rep.} 145 (1987)
141\semi 
J.~I.~Kapusta, {\it Finite Temperature Field Theory}, Cambridge Univ. Press
(Cambridge 1989)\semi 
M. Le Bellac, {\it Thermal Field Theory}, Cambridge Univ. Press (Cambridge
1996)\semi 
H. Meyer-Ortmanns, {\it Rev. Mod. Phys.} 68 (1996) 473.} 
\nref\rCTP{K.-C. Chou, Z.-B. Su, B.-L. Hao and L. Yu, {\it Phys. Rep.} 118
(1985) 1.} 
\nref\rbook{J. Zinn-Justin, 1989, {\it Quantum Field
Theory and Critical Phenomena}, in particular chap.~36 of third ed., Clarendon
Press (Oxford 1989, third ed. 1996).} 
\nref\rFSS{see for instance M.N. Barber in {\it Phase Transitions and Critical
Phenomena} vol. 8, C. Domb and J. Lebowitz eds. (Academic Press, New York
1983), and the papers collected in {\it Finite Size Scaling}, J.L. Cardy ed.
(North-Holland, Amsterdam 1988).} 
\nref\rFSSb{P. Hasenfratz, {\it Phys. Lett.} B141 (1984) 385\semi
E. Br\'ezin and J. Zinn-Justin, {\it Nucl. Phys.} B257 [FS14]
(1985) 867\semi 
J. Rudnick, H. Guo and D. Jasnow, {\it J. Stat. Phys.} 41 (1985) 353\semi
M. L\"uscher, {\it Phys. Lett.} 118B (1982) 391; {\it Nucl. Phys.} B219 (1983)
233.}
\nref\rKoMuRe{For a rigorous proof see C. Kopper, V.F. M\"uller and T. Reisz,
{\it Temperature Independent Renormalization of Finite Temperature Field
Theory} hep-th/0003254.}  
\nref\rShapo{S. Weinberg, {\it Phys. Lett.} B91 (1980) 51\semi
P. Ginsparg, {\it Nucl. Phys.} B170 (1980) 388\semi
T. Applequist and R. D. Pisarski, {\it Phys. Rev.} D23 (1981) 2305\semi
S. Nadkarni, {\it Phys. Rev.} D27 (1983) 917\semi
for a review see M.E. Shaposhnikov, {\it Proc. Int. School of Subnuclear Phys.} (Erice 1996), World Scientific, hep-ph/9610247.} 
\nref\rLand{N. P. Landsman, {\it Nucl. Phys.} B322  (1989) 498.}
\section Finite (and high) temperature field theory: General remarks

Study of quantum field theory (QFT) at finite temperature was initially motivated by cosmological problems \rKiLi~and more recently has gained additional attention in connection with high energy heavy ion collisions and speculations about possible phase transitions. References to initial papers like \refs{\rWein,\rDoJa} can be found in Gross {\it et al} \rGroPisYa. More recently several new review papers and textbooks have been published \rLWKLBM. Since we are  interested here only in {\it equilibrium physics}\/ the imaginary time formalism will be used throughout these notes. Non-equilibrium processes can be described either in the same formalism after analytic continuation to real time or alternatively by Schwinger's Closed Time Path formalism in the more convenient path integral formulation \rCTP.\par
In these notes we specially want to emphasize that additional physical intuition  about QFT at finite temperature at equilibrium can be gained by realizing that it can be also considered as an example of  classical statistical field theory in systems with finite sizes
\rbook.\par
Quantum field theory  at finite temperature  is the relativistic
generalization of finite temperature non-relativistic quantum
statistical mechanics. There it is known that quantum effects are only
important at low temperature. More precisely the important parameter
is the ratio of the thermal wave-length $\hbar/\sqrt{mT}$ and the
length scale which characterizes the variation of the potential (for
smooth potentials). 
Only when this ratio is large are  quantum effects important. Increasing
the temperature is at leading order equivalent to decrease $\hbar$.
Note that, from the point of view of the path integral representation of
quantum mechanics, the transition from quantum to classical 
statistical mechanics appears as a kind of dimensional reduction: in the
classical limit path (one dimensional system) integrals reduce to ordinary
(zero dimensional points) integrals.\par  
We discuss here quantum field theory  at finite temperature in $(d,1)$
dimensions, at equilibrium.  We want to explore its properties and
in particular study the relevance of quantum effects at high temperature. Note
that high temperature now means either that the theory contains massless
particles  or that one is working in the ultra-relativistic limit where the
temperature, in energy units, is much larger than the rest energy of massive
particles. In particular we want to understand the conditions under which
statistical properties of finite temperature QFT in $(d,1)$ dimension can be
described by an effective  classical statistical field theory in $d$
dimension. 
\subsection Finite temperature quantum field theory

The static properties of finite temperature QFT can be derived from the partition function ${\cal Z}=\tr\e^{-H/T}$, where $H$ is the hamiltonian of the quantum field theory and $T$ the
temperature. For a simple theory with boson fields $\phi$ and euclidean action ${\cal
S}(\phi)$, the partition function is given by the functional integral    
$${\cal Z}=\int[\d\phi]\exp\left[-{\cal S}(\phi)\right],\eqnd\eFTZp$$
where  ${\cal S}(\phi)$ is the integral of the
lagrangian density ${\cal L}(\phi)$
$${\cal S}(\phi)=\int_0^{1/T}\d\tau\int\d^d x\,{\cal L}(\phi), $$
and the field $\phi$ satisfies periodic boundary conditions  in
the (imaginary) time direction  
$$\phi(\tau=0,x)=\phi(\tau=1/T,x).$$
The quantum field theory may also involve fermions. Fermion fields
$\psi(\tau,x)$ instead satisfy anti-periodic boundary conditions 
$$ \psi(\tau=0,x)=-\psi(\tau=1/T,x).$$
\medskip
{\it Classical statistical field theory and renormalization group.}  The
quantum partition function \eFTZp~has also the interpretation of the partition
function of a classical statistical field theory in $d+1$ dimension. The zero
temperature limit of the quantum theory corresponds to the usual infinite
volume classical partition function. Correlation functions thus satisfy the
renormalization group (RG) equations of the corresponding $d+1$ dimensional
theory.\par      
In this interpretation finite temperature for the quantum partition function
\eFTZp~corresponds to a finite size $L=1/T$ in one direction for the classical
partition function. General results obtained in the study of finite size
effects \rFSS~also apply here \rbook.   
RG equations are only sensitive to short distance singularities, and therefore
finite size effects do not modify RG equations \refs{\rFSSb,\rKoMuRe}. Finite size effects affect
only the solution of the RG equations, because a new dimensionless, RG
invariant,  variable appears which can be written  as the product $Lm_L$, where
the correlation length $\xi_L=1/m_L$ characterizes the decay of correlation
functions in space directions.\par
For $L$ finite (non-vanishing temperature), we expect
a cross-over from a $d+1$-dimensional behaviour when the correlation
length $\xi_L$ is small compared to $L$, to the $d$-dimensional
behaviour when $\xi_L$ is large compared to $L$. This
regime can be described by an effective $d$-dimensional theory. Note
that in quantum field theory the initial microscopic scale
$\Lambda^{-1}$, where $\Lambda$ is the QFT cut-off, always appears. Therefore,
even at high temperature $L\to 0$, the product $L\Lambda$ remains large. 
\medskip
{\it Mode expansion.} As a consequence of periodicity, fields can
be expanded in eigenmodes in the time direction and the corresponding frequencies are quantized. For boson fields
$$\phib(x,t)=\sum_{\omega_n=2n\pi/L}\e^{i\omega_n t}\phib_n
(x).\eqnd\emodexp $$ 
In the case of fermions anti-periodic conditions lead to the expansion 
$$\psib(x,t)=\sum_{\omega_n=(2n+1)\pi/L}\e^{i\omega_n t}\psib_n
(x).\eqnd\efmodexp $$ 
When $T=L^{-1}\gg m$, where $m$ is
the zero-temperature physical mass of boson fields, a situation realized at
high temperature in the QFT sense, or when the mass vanishes, a non-trivial
physics exists for  momenta much smaller than the temperature $T$ or distances
much larger than $L$. In this limit one expects to be able to treat all
non-zero modes perturbatively: the perturbative integration over the non-zero
modes leads to an effective field theory  for the zero-mode, with a
$d$-dimensional action ${\cal S}_L$ \rShapo. \par      
Fermions instead, due to anti-periodic conditions, have no zero modes.  In the
same limit fermions can be completely integrated out. \par 
Apart from high temperature there is another situation where we expect this
mode integration to be useful, in the case of a  finite temperature second
order phase transition. Then it is the finite temperature correlation $\xi_L$
which diverges, and this induces a non-trivial long distance physics.   
\smallskip
{\it Remarks.} \par
(i) The mode expansion \eqns{\emodexp, \efmodexp} is well-suited to
simple situations where the field belongs to a linear space. In the case of
non-linear $\sigma $ models or gauge theories the separation of the zero-mode
will be a more complicated issue.  \par
(ii) More precisely the zero-mode has to be treated differently from other modes when the correlation length $\xi_L=1/m_L$ in the space directions is large compared to $L$, i.e.~$m_L\ll T$. This condition is equivalent to $m\ll T$ only at leading order in perturbation theory. 
\subsection  Dimensional reduction and effective field theory

To construct the effective $d$-dimensional theory, we thus keep the zero mode and integrate
perturbatively over all other modes. It is convenient to introduce some
notation  
$$\phi(x,t)=\varphi(x)+\chi(x,t), \eqnd\eFTmodfc $$
where $\varphi$ is the zero mode and $\chi$ the
sum of all other modes (equation \emodexp)
$$\chi(x,t)=\sum_{n\ne 0}\e^{i\omega_n t}\phib_n
(x), \quad \omega_n=2n\pi/L\,. \eqnn $$ 
The action ${\cal S}_L$ of the reduced theory is defined by 
$$\e^{-{\cal S}_L(\varphi)}=\int[\d\chi]\exp[-{\cal S}
(\varphib+\chi)]. \eqnd\eFTefact $$
At leading order in perturbation theory one simply finds
$${\cal S}_L(\varphi)=L \int\d^d x\,{\cal L}(\varphi) .\eqnn $$
We note that $L$ plays, in this leading approximation, the formal role of
$1/\hbar$, and the large $L$ expansion corresponds to a loopwise expansion.
The length $L$ is large with respect to $\Lambda^{-1}$. If $L\Lambda$ is
the relevant expansion parameter, which means that the perturbative expansion
is dominated by large momentum (UV) contributions, then the effective $d$
dimensional theory can still be studied with perturbation theory. This is
expected for large number of space dimensions where theories are non
renormalizable. However, another dimensionless combination can be found, $mL$,
which at high temperature is small. This may be the relevant expansion
parameter for theories which are dominated by small momentum (IR)
contributions,  a problem which arises at low dimension $d$. Then perturbation
theory is no longer possible or useful. \par    
An important parameter in the full effective theory is really $Lm_L$.
Therefore an important question is whether the integration over non-zero
modes, beyond leading order, generates a mass for the zero mode.  
\smallskip
{\it Loop corrections to the effective action.} After integration over
non-zero modes the effective action contains all possible interactions. In the
high temperature limit one can perform a {\it local expansion}\/ of the
effective action.  One expects, but this has to be checked carefully, that in general
higher order corrections coming from the mode integration will generate terms
which renormalize the terms already present at leading order, and additional
interactions suppressed by powers of $1/L$.  Exceptions are provided by gauge theories where new low dimensional interactions are generated by the breaking of $O(d,1)$ invariance. 
\smallskip
{\it Renormalization.} If the initial $d,1$ dimensional theory has been
renormalized, the complete theory is finite in the formal infinite cut-off
limit. However, as a consequence of the zero-mode subtraction, cut-off
dependent terms may remain in the reduced $d$-dimensional action. These terms
provide the necessary counter-terms which render the perturbative expansion
of the effective field theory finite.  The effective can thus be written
$${\cal S}_L(\varphi)={\cal S}^{(0)}_L(\varphi) +\ \hbox{\rm counter-terms}.
$$ 
Correlation functions have finite expressions in terms of the parameters of the effective action, in which the counter-terms have been omitted. The first part ${\cal S}^{(0)}_L(\varphi) $ thus satisfies the RG equations of the $d+1$ theory \refs{\rFSSb,\rLand}.\par
Finally the local expansion breaks down at momenta of
order $L^{-1}$. Actually the temperature $L^{-1}$ plays the role of an
intermediate cut-off. Determining the finite parts may involve some
careful calculations. 
\smallskip
{\it The finite temperature correlation length.}
As already stressed, a first and important problem is to understand the behaviour of the 
effective mass of the zero-mode generated  by integrating out the
non-zero modes. If this mass $m_L$ is of order of the QFT temperature
$T=L^{-1}$, the zero-mode is no longer different from other modes. The IR
problem disappears and one expects to again be able to use perturbation
theory. Actually  one should be able to rearrange the $d+1$   
perturbation theory to treat all modes in the same way.\par 
Conversely if the mass of the zero-mode remains much smaller
than the temperature, then perturbation theory may be
invalidated by IR contributions. However then one can use the {\it local
expansion}\/ of the effective action to study the non-trivial IR 
properties. \par
At high temperature the QFT remains with only one
explicit length scale $L$. The quantity $Lm_L$, where $m_L$ is the
physical mass of the complete theory, then only depends on
dimensionless ratios. If $Lm_L$ is of order one, the final zero mode
acquires a mass comparable to the other modes. Note that this is
what  happens in theories with non-trivial IR fixed points.
\section The example of the $\phi^4_{d,1}$ quantum field theory

We first study the example of a simple scalar field theory. The scalar   field  $\phib$ is a $N$-component vector and the hamiltonian $ {\cal H}(\Pib,\phib) $ is $O(N)$ symmetric \sslbl\ssFTfiv  
$$ {\cal H}(\Pib, \phib)=\ud \int \d ^d x \, \Pib^2(x)+\Sigma (\phib)\,, \eqnn $$
with
$$\Sigma (\phib)= \int \d ^d x \left\lbrace\ud
\left[ \nabla \phib (x) \right]^{2}+\ud (r_c+r)\phib
^{2} (x)+{1 \over 4!}u\bigl(\phib^2(x)\bigr)^2 
 \right\rbrace . \eqnd\eactred $$
A cut-off $\Lambda$ as usual is implied, to render the field theory UV finite.
The quantity $r_c(u)$ has the form of a mass renormalization. It is defined by the condition that  at zero temperature, $T=0$, when $r$ vanishes the physical mass $m$ of the field $\phib$
vanishes. At $r=0$ a  transition occurs between a symmetric phase, $r>0$, and
a broken phase, $r<0$. We recall that  the field theory is meaningful only if
the physical mass $m$ is much smaller than the cut-off $\Lambda$. This implies
either (the famous {\it fine tuning}\/ problem) $|r| \ll \Lambda^2  $
or, for $N\ne 1$,  $r<0$ which corresponds to a spontaneously broken symmetry
with massless Goldstone modes. This latter situation will be examined in
section \label{\ssFTnls} within the more suitable formalism of the non-linear
$\sigma $-model. \par      
Note that we sometimes will set
$$u=\Lambda^{3-d}g\,, \eqnd\eFTuLg $$
 where $g$ is dimensionless.\par 
The finite temperature quantum partition function reads
$${\cal Z}=\int[\d\phi]\exp[-{\cal S}(\phi)]\,, \eqnn $$
with periodic boundary conditions in the time direction,  and
$${\cal S}(\phi)=\int_0^L \d t\left[\int\d^d x\ud(\d_t\phib)^2
+\Sigma (\phib)\right], \eqnn $$  
where $T\equiv 1/L$ is the temperature. \par
We now construct the effective $d$-dimensional theory and discuss its
validity. Note, however, that this construction is useful only if the IR
divergences are strong enough to invalidate perturbation
theory. Therefore we do not expect the construction to be very useful if
the initial theory has a dimension $d+1>4$, because the reduced
$d$-dimensional theory has a finite perturbation expansion  even in the massless limit. This is a property we will check by discussing the dimension $d=4$. 
\subsection  Renormalization group at finite temperature

As already explained, some useful information can be obtained from
renormalization group analysis. One important quantity is the product $Lm_L$,
where $\xi_L=m_L^{-1}$ is the finite temperature (finite size) correlation
length, and $m_L$ therefore the mass of the zero-mode in the effective theory.
\par    
The zero temperature theory satisfies the RG equations of a $d+1$ dimensional
field theory in infinite volume. The dimension $d=3$ is special, since then
the $\phi^4_{d+1}$ theory is just renormalizable.   
\smallskip
{\it Dimensions $d>3$}. For $d>3$ the theory is non-renormalizable, which
means that the gaussian fixed point $u=0$ is stable. The coupling constant
$u=g\Lambda^{3-d}$ is small in the physical range, and perturbation theory is
applicable. At zero temperature the physical mass in the symmetric phase
scales like in the free theory $$m\propto r^{1/2}.$$
The leading corrections to the two-point function due to finite temperature
effects are of order $u$. Therefore in the symmetric phase, for dimensional
reasons,  
$$m_L=1/\xi_L\propto ( r+{\rm const.}\ g\Lambda^{3-d} L^{1-d})^{1/2}.$$
This expression has several consequences. \par
If at zero temperature the symmetry is broken, then a phase transition occurs
at a temperature $T_c$ which scales like
 $$T_c=1/L_c\propto \Lambda \left(-r/\Lambda^2 \right)^{1/(d-1)},$$
which means high temperature, since the physical mass scale is $(-r)^{1/2}$.
In particular in the case $N=1$, the critical temperature is large with
respect to the initial physical mass $m\propto (-r)^{1/2}$ 
$$T_c\propto m^{2/(d-1)}\Lambda^{(d-3)/(d-1)} \gg m\,.$$
At high temperature or in the massless theory ($r=0$) the effective mass $m_L$
increases like   
$$m_L  L\propto (L\Lambda)^{(3-d)/2}\ll 1\,.$$
The behaviour $L m_L \ll 1 $ implies the validity of
the mode expansion. 
\smallskip
{\it Dimension $d=3$.} The theory is just renormalizable and logarithmic
deviations from naive scaling appear. RG equations take the form  
$$ \left[ \Lambda{ \partial \over \partial \Lambda} +\beta(g){\partial \over
\partial g}-{n \over 2}\eta (g)-\eta_{2}(g)r{\partial \over \partial r}
\right] \Gamma^{(n)} \left(p_{i};r,g,\Lambda \right)=0\,. \eqnd\eganTRG $$
The product $Lm_L=F(L\Lambda , g,rL^2 )$  is a dimensionless RG invariant,
thus it satisfies
$$ \left[ \Lambda{ \partial \over \partial \Lambda} +\beta(g){\partial \over
\partial g}-\eta_{2}(g)r{\partial \over \partial r}
\right] F=0\,.$$
The solution can be written
$$ Lm_L=F(\Lambda L, L^2 r, g)=F\bigl(\lambda \Lambda L,g(\lambda ), L^2
r(\lambda)\bigr),\eqnd\eFTRGmass $$ 
where $\lambda $ is a scale parameter, and $g(\lambda ), r(\lambda )$ the
corresponding running parameters (or effective parameters at scale $\lambda $)
$$\lambda {\d g(\lambda )\over \d\lambda }=\beta\bigl(g(\lambda )\bigr), \quad
\lambda {\d r(\lambda )\over \d\lambda }=-r(\lambda )\eta_2\bigl(g(\lambda
)\bigr).$$ 
The form of the RG $\beta$-function 
$$\beta(g)={(N+8) \over 48\pi^2}g^2+O\left(g^{3} \right), \eqnn $$
implies that the theory is IR free, i.e.~that $g(\lambda )\to 0$ for $\lambda
\to 0$. The effective coupling constant at the physical scale is
logarithmically small. For example to reach the  scale $T=1/L$ we have to
choose $\lambda =1/\Lambda L\ll1$, and thus   
$$g(1/\Lambda L)\sim {48 \pi^2 \over (N+8)\ln(\Lambda L)}. \eqnd\eFTgrun $$
From 
$$\eta_2(g)=-{N+2 \over 48\pi^2} g+O\left(g^2 \right), $$ 
one also finds
$$r(1/\Lambda L)\propto {r\over (\ln \Lambda L)^{(N+2)/(N+8)}}.
\eqnd\eRGivtrun$$ 
Therefore RG improved perturbation theory can be used, and we expect results
to be qualitatively similar to those of $d>3$, except that powers of $\Lambda
$ are replaced by powers of $\ln \Lambda $.  
\smallskip
{\it Dimensions $d=2$.} 
The three-dimensional classical theory has an IR fixed point $g^*$. Then finite size scaling (equation \eFTRGmass) predicts, in the symmetric phase, 
$$L m_L= f(rL^{1/\nu}),$$
where is $\nu$ the exponent of the three-dimensional system. Therefore
$m_L$ remains of order $T$ and the zero-mode is special only if the function
$f(x)$ is small (compared to 1). \par
This happens at a phase transition, but in an effective two-dimensional theory a phase transition is possible only for $N=1$. Then if $f(x)$ vanishes at $x=x_0$, for $|x-x_0|\ll 1$, i.e.~when the temperature $T$ is close to the critical temperature $T_c$,  
$$T_c=x_0^{-\nu}(-r)^\nu\propto m\,, \quad |T-T_c|\ll (-r)^\nu\propto m\,,$$
the IR properties are described by an effective two-dimensional theory.
If the system has a finite temperature phase transition, it is at zero temperature in a broken phase.\par
Again the situation of a broken phase at zero temperature for $N\ne 1$
will be examined separately. 
\subsection One-loop effective action
 
{\it Mode expansion and effective action at leading order.} 
To construct the effective field theory in $d$ dimensions one expands the
field in eigenmodes in the time direction (equation \emodexp). 
One then calculates the effective action \eFTefact~by integrating
perturbatively over all non-zero modes. In the notation \eactred~the result
at leading order simply is 
$${\cal S}_L(\varphib)=L \,\Sigma (\varphib) .\eqnn $$
Note that if we rescale $\varphib$ into $\varphib L^{-1/2}$ the
coupling constant is changed into $u/L$
$$ {\cal S}_L(\varphib)= \int \d ^d x \left\lbrace\ud \left[ \nabla \varphib
(x) \right]^{2}+\ud r\varphib^2 (x)+{1 \over
4!}(u/L)\bigl(\varphib^2(x)\bigr)^2  \right\rbrace .\eqnd\eFTfivSLi $$
At this order $r_c$ vanishes and therefore has been omitted.
The action \eFTfivSLi~generates a perturbation theory. In terms of a
dimensionless coupling $g=u\Lambda^{d-3}$, the expansion parameter is 
$(\Lambda/m_L )^{4-d}g/\Lambda L$. For $d\ge 4$ it is always small because
$\Lambda L$ is large. For $d=3$ the expansion parameter is of order
$g/L m_L$. Since dimensional reduction is useful only for $Lm_L $ small, the
situation is subtle  because the running coupling constant $g(1/\Lambda L)$ at
scale $L^{-1}\ll \Lambda$, renormalized by higher modes, is also small:
$g(1/\Lambda L)=O(1/\ln (L\Lambda ))$. A more detailed discussion of the
situation requires a one-loop calculation.\par     
For $d<3$ IR singularities are always present both in the initial and the
reduced theory, and the small coupling regime can never be reached for
interesting situations. The $\varepsilon=3-d$ expansion can be useful in some
limits, otherwise the problem has to be studied by non-perturbative methods. 
\medskip
{\it One-loop calculation.} 
The one-loop contribution takes the form ($\ln \det=\tr\ln$)
$$\delta {\cal S}_L=\ud\tr\ln \left[(-\d_t^2-\Delta_d +r +
\frac{1}{6}u\varphib^2)\delta_{ij}  
+\frac{1}{3}u\varphi_i\varphi_j\right]-\ (\varphib=0). $$
The situation of interest here is when the correlation length $\xi=1/m$ of
zero-temperature or the infinite volume  $d+1$ system is at least of the order
of the size $L$.  
In this situation we expect to be able to make a local expansion in
$\varphi$.\par
The leading order in the derivative expansion can be obtained by treating
$\varphi(x)$ as a constant. To calculate the  one-loop contribution to the
reduced $\delta {\cal S}_L$  it is then convenient to use the identity    
$$\tr\ln A-\tr\ln B=\tr\int_0^\infty {\d s \over
s}\left(\e^{-Bs}-\e^{-As}\right). $$ 
This leads to Schwinger's representation of the one-loop contribution
$$\eqalign{\delta {\cal S}_L&
=-\ud\int \d^d x\int_0^\infty {\d s\over s}\int {\d^d p\over
(2\pi)^d}\sum_{n\ne 0} 
\left[(N-1)\e^{-s(p^2+\omega_n^2+r+u\varphib^2(x)/6)}\right.\cr &\quad\left.+
\e^{-s(p^2+\omega_n^2+r+u\varphib^2(x)/2)} 
-(\varphib \equiv 0)\right]. \cr}$$
This expression has UV divergences which appear as divergences at $s=0$. A possible regularization, which we will adopt here, consists in cutting the $s$ integration at
$s=1/\Lambda^2$.   
The integral over momenta is simple. The sum over the frequencies $\omega_n$
can be expressed in terms of  the function $\vartheta_0(s) $ (equation
\eqns{\eJacobi}) which satisfies   
$$\vartheta_0(s) = s^{-1/2} \vartheta_0(1/s). $$
One obtains
$$\eqalignno{\delta {\cal S}_L(\varphib)&
=-\ud{1\over (4\pi)^{d/2}} \int \d^d x\int^\infty_{1/\Lambda^2}  {\d s\over
s^{1+d/2}}\e^{-sr} \bigl(\vartheta_0(4\pi s/L^2)-1\bigr) \cr
&\quad\times
\left[(N-1)\e^{-us\varphib^2/6}+\e^{-us\varphib^2/2}-N\right].&\eqnd\eFTfivol
\cr}$$  
After the change of variables $4\pi  s/L^2 \mapsto s$  the expression becomes
$$\eqalignno{\delta {\cal S}_L(\varphib)&
=-\ud L^{-d} \int \d^d x\int_{s_0}^\infty {\d s\over
s^{1+d/2}}\e^{-sL^2r/4\pi} \bigl(\vartheta_0(s)-1\bigr) \cr
&\quad\times
\left[(N-1)\e^{-usL^2\varphib^2/24\pi}+\e^{-usL^2\varphib^2/8\pi}-N\right].
&\eqnd\eFTfivol \cr}$$
with $s_0=4\pi /L^2\Lambda^2$. 
We now perform a small $\varphib$ expansion. At this order this expansion
makes sense only if $L^2 r>-4\pi^2$, a condition which more generally
involves the dimensionless scale invariant ratio $L^2  m^2$, and implies
being not too far in the ordered region. This is not surprising since an expansion around $\varphib=0$ makes sense only if the field expectation value is small.
\medskip
{\it Order $\varphib^2$.} Note that for the quadratic term the local
approximation was not needed because the corresponding one-loop diagram is a
constant  
$$\delta {\cal S}^{(2)}_L=\frac{1}{12}(N+2)uG_2\int\d^d x\,\varphib^2(x)\,,
\eqnn $$
where the constant $G_2$ is given by
$$G_2(r,L)={L^{2-d}\over 4\pi}\int_{s_0}^\infty{\d s\over
s^{d/2}}\e^{-rL^2s/4\pi} 
\left(\vartheta_0(s)-1\right). \eqnd\eGinttwo    $$
\medskip
{\it Order $\varphib^4$.} The quartic term  is proportional to the initial
interaction 
$$\delta {\cal S}^{(4)}_L=-{1\over 144}(N+8)u^2G_4\int\d^d
x\,(\varphib^2(x))^2 \eqnn$$ 
with 
$$G_4={L^{4-d}\over (4\pi)^2}\int_{s_0}^\infty{\d s\over
s^{d/2-1}}\e^{-rL^2s/4\pi} 
\bigl(\vartheta_0(s)-1\bigr)=-{\partial \over \partial r} G_2 \,.
\eqnd\eGintfour  $$ 
\medskip
{\it The one-loop reduced action.} We first keep only the terms already present in the tree approximation. The value of $r_c$
correspond to the mass renormalization which renders the $T=0$ theory massless
at one-loop order. Thus $G_2$ has to be replaced by $[G_2]_\r$. Using the same
regularization one obtains ($d>1$) 
$$[G_2]_\r=G_2-{L\over(2\pi)^{d+1}}\int {\d^{d+1} k\over
k^2}=G_2-{2L\Lambda^{d-1} \over(d-1)(4\pi)^{(d+1)/2}}.$$ 
After the rescaling $\varphi \mapsto \varphi L^{-1/2}$ 
the effective action can be written 
$$ {\cal S}_L(\varphi)=\int \d ^d x \left\lbrace\ud 
\left[ \nabla \varphib (x) \right]^{2}+\ud \sigma_2 \varphib
^{2} (x)+{1 \over 4!} \sigma_4 \bigl(\varphib^2(x)\bigr)^2
\right\rbrace ,\eqnd\eFTfivSol $$
with
$$\sigma_2= r +\frac{1}{6}(N+2)u[G_2]_\r/L \,,\quad
\sigma_4=u/L -\frac{1}{6}(N+8) u^2 G_4 /L^2.$$
\medskip
{\it Other interactions.} For space dimensions $d< 5$ the coefficients of the
other interaction terms are no longer UV divergent. Since the zero-mode
contribution has been subtracted no IR divergence is generated even in the
massless limit. In this limit the coefficients are thus proportional to powers of $L$ obtained by
dimensional analysis (in the normalization \eFTfivSLi)    
$$\delta {\cal S}^{(2n)}_L \propto g^n (\Lambda L)^{-n(d-3)} L^{n(d-2)-d}
\int\d^d x\,(\varphib^{2}(x))^{n} ,\eqnn$$ 
and therefore increasingly negligible at high temperature at least for $d\ge
3$.\par 
The local expansion of the one-loop determinant also generates monomials with
derivatives. No term  proportional to $(\partial_\mu \varphi)^2$ is generated
at one-loop order. All other terms with derivative are finite for $d<5$, and
thus the structure of the coefficients again is given by dimensional analysis.
To $2k$ derivatives corresponds an additional factor  $ L^{2k}$.   \par 
Finally for $r\ne 0$ but $rL^2\ll 1$, we can expand in powers of $r$ and the previous arguments immediately generalize. 
\section High temperature and critical limits
 
We now examine two interesting situations. First we discuss $r\to 0$ which
corresponds to a massless theory at zero temperature in the QFT context (and
to the critical temperature of  the $d+1$ dimensional statistical field
theory). This will give the leading contributions in the high temperature
limit. It will prove useful to also keep terms linear in $r\varphi^2$.\par    
Then we consider another situation, which, from the classical
statistical point of view, corresponds to choose the parameter $r$ as a
function of $L$ to remain at the critical point, and from the QFT
point of view to choose $T=1/L$, the temperature, to take a critical value
$T_c$, if possible. Then the correlation length diverges and the effective
field theory for the zero-mode is indeed a $d$-dimensional field theory for a
massless scalar field. We again   
find a situation of dimensional reduction.
\medskip
{\it The massless limit.} 
The constants  $[G_2]_\r$ and $G_4$ for $r=0$ become
$$\eqalignno{[G_2]_\r&= {1\over
2\pi^{(d+1)/2}}\Gamma\bigl((d-1)/2\bigr)\zeta(d-1)L^{2-d} 
-{2\Lambda^{d-2}\over (d-2)(4\pi)^{d/2}}, &\eqnd\eGtwoze \cr
G_4&={\pi^{d/2-4}\over 8}\Gamma(2-d/2)\zeta(4-d)L^{4-d}+{2\Lambda^{d-4}\over
(d-4)(4\pi)^{d/2}} +{2L\Lambda^{d-3} \over (d-3)(4\pi)^{(d+1)/2}},\cr 
&&\eqnd\eFTGfour \cr}$$
where the results of appendices \label{\appthree}~have been used.
The expression for $[G_2]_\r$ is the sum of two terms, a renormalized mass
term for the zero-mode, and the one-loop counterterm which renders the
two-point function one-loop finite in the reduced theory.
The expression for $G_4$ contains a finite temperature contribution, a
zero-temperature renormalization for $d\ge 3$ and a counter-term for the
reduced theory for $d \ge 4$.\par  
Finally from the value of $G_4$ and the relation \eGintfour~we immediately obtain the term linear in $r$ in $G_2$ 
$$G_2(r,L)=G_2(0,L)-r G_4(0,L)+O(r^2). \eqnn $$
\subsection Dimension $d=4$

For $d>3$ the coupling constant $u$ which is of order $\Lambda^{3-d}$ is
very small. The renormalized mass generated for the zero-mode is of order
$L^{-1}(L\Lambda)^{(3-d)/2}$ and thus small in the temperature scale,
justifying a mode expansion.  \sslbl\ssFTfiviv \par 
Let us examine more precisely the $d=4$ case, keeping the contribution of
order $r$ in $G_2$. Then 
$$\eqalign{[G_2]_\r&={1\over4\pi^2 L^2}\zeta(3)-{\Lambda^2\over
16 \pi^2} +r\left[ -{\gamma\over 16\pi^2}-{2\over (4\pi)^{5/2}}\Lambda L
 +{1\over 8\pi^2 }\ln(\Lambda L) \right]+O(r^2)\cr
G_4&={\gamma\over 16\pi^2}+{2\over (4\pi)^{5/2}}\Lambda L
 -{1\over 8\pi^2 }\ln(\Lambda L) , \cr} $$
where $\gamma $ is Euler's constant, $\gamma =-\psi(1)$.
The  infinite volume terms proportional to $\Lambda L$ which induces a finite
renormalization $g\mapsto g_\r   $ of the dimensionless $\phi^4$ coupling  $g$,
and $r\mapsto r_\r  $   
$$ u=g/\Lambda \,,\quad g_\r   =g-{1\over (4\pi)^{5/2}} {N+8 \over 3} g^2,\quad  
r_\r  =r- {1\over (4\pi)^{5/2}} {N+2 \over 3}    gr\,.$$
The remaining cut-off dependent terms of $ [G_2]_\r$ and $G_4$ will render the
effective $d=4$ theory one-loop finite. Using  expression \eFTfivSol~and
introducing the small dimensionless (effective) coupling constant $\lambda _L$
$$\lambda _L=g_\r   /(\Lambda L), $$ 
we can write the effective action at one-loop order
$${\cal S}_L(\varphib)=\int \d ^4 x \left\{ \ud 
\left[ \nabla \varphib (x) \right]^{2}+\ud r_L \varphib
^{2} (x)+{1 \over 4!} g_L \bigl(\varphib^2(x)\bigr)^2
 \right\}  +  \delta {\cal S}_{L,\Lambda }(\varphib),\eqnn $$
where $ \delta {\cal S}_{L,\Lambda }$ is the sum of one loop counter-terms
$$ \delta {\cal S}_{L,\Lambda }(\varphib)=\int \d ^4 x \left[ \ud \delta r_L \varphib
^2 (x)+{1 \over 4!} \delta g_L \bigl(\varphib^2(x)\bigr)^2
 \right] $$
and the various coefficients are
$$\deqalignno{r_L &=r_\r  +
\frac{1}{6}(N+2){\zeta(3)\over4\pi^2}L^{-2}  \lambda _L, &\delta r_L&={N+2 \over 96\pi^2}\bigl(-\Lambda ^2+2r_\r \ln(\Lambda L)-\gamma r_\r \bigr)\lambda_L  \,,\cr
g_L &=\lambda_L , &\delta g_L&={N+8\over 96\pi^2}\bigl(2 \ln(\Lambda L)-\gamma \bigr)
\lambda_L^2 \,.\cr}$$
\smallskip
{\it Other local interactions.} In the same normalization an interaction term
with $2n$ fields and $2k$ derivatives is proportional to $g_L^n L^{2n-4+2k}$
and thus negligible in the situations under study for $n>2$ or $n=2$, $k>0$. 
\smallskip
{\it One-loop calculation with reduced action.} To check that indeed one-loop
counter-terms have been provided, we calculate the one-loop contributions  
to the two-point function $\Gamma^{(2)}$ and the four-point function
$\Gamma^{(4)}$ at zero momentum.\par 
For the one-loop contribution $r_L=r_\r=r$ because the differences are of order $\lambda
_L$. Taking into account the counter-terms one finds
$$\eqalign{\Gamma^{(2)}_{\rm one~loop}&={N+2 \over 6}{\lambda_L \over
(2\pi)^4} \int^\Lambda {\d^4 k\over k^2+r}-{N+2 \over96\pi^2}\Lambda ^2
\lambda_L+{N+2\over 96\pi^2} \bigl(2\ln(\Lambda L)-\gamma \bigl)r\lambda _L \cr  
&={N+2\over 96\pi^2}\bigr(\ln(rL^2)-1\bigr)\lambda _L r\,.\cr}
$$
Therefore the complete two-point function at one-loop order is
$$\Gamma^{(2)}(p)=p^2+r_\r  +\frac{1}{6}(N+2){\zeta(3)\over4\pi^2}L^{-2}  \lambda
_L +{N+2\over 96\pi^2}\bigr(\ln(r_\r  L^2)-1\bigr)\lambda _L r_\r  \,.
\eqnd\eFTfiviipt $$ 
For the four-point function  at zero momentum we find
$$\eqalign{\Gamma^{(4)}_{\rm one~loop}&=-{N+8\over
6}{1\over(2\pi)^4}\int^\Lambda {\d^4 k\over( k^2+r)^2}\lambda_L^2+ {N+8\over
96\pi^2}\bigl(2\ln(\Lambda L)-\gamma \bigr) \lambda_L^2  
\cr
&={N+8\over 96\pi^2}\ln(r_\r  L^2) \lambda _L^2 .\cr}$$
Thus the complete four-point function reads
$$\Gamma^{(4)}(p_i=0)=\lambda_L+{N+8\over 96\pi^2}\ln(r_\r  L^2) \lambda _L^2
+O(\lambda_L^3).$$ 
We note that $L^{-1}$ plays the role of the cut-off in the reduced theory. We
also find large  logarithms which can be summed by the RG of the
four-dimensional reduced theory. However, because the initial coupling constant
$\lambda_L$ is very small, it has no time to run.    
\smallskip
{\it The massless theory.} 
At leading order  in the massless theory $r=0$ we find an effective mass 
$$Lm_L =\left[\frac{1}{24\pi^2}(N+2)\zeta(3) \lambda _L\right]^{1/2}\ll 1\,.$$
Although the induced mass remains small,
because the effective four-dimensional theory has at most logarithmic
IR singularities, and the effective coupling is of order $1/(\Lambda L)$, the
reduced theory can still be solved by perturbation theory. \par
For $r\ne0$ but still such that $rL^2$ is small (high temperature) the product
$Lm_L$ remains small, the term $rL^2=L^2m^2$ becoming dominant when $mL\gg
(\Lambda L)^{-1/2}$.   
\smallskip
{\it The critical temperature.} 
We now calculate the critical temperature. Note that for dimensions $d\ge 3$
we can study the effective theory by perturbation theory and renormalization
group. If we start from four or lower dimensions perturbative methods are no
longer applicable, because the massless theory is IR divergent. \par
Using the expression  \eFTfiviipt~one finds at leading order
$$r +{(N+2)\zeta(3) \over 24\pi^2 L^2}\lambda_L =0\,.$$ 
This equation justifies a small $r$ expansion,
and shows in particular that a phase transition is  possible only
if at $T=0$ (zero QFT temperature) the system is in the ordered phase.  
The critical temperature $T_c$ has the form
$$T_c \sim \bigl((N+2)\zeta(3)\bigr)^{-1/3}(2\pi)^{2/3}
|\left<\phi\right>|^{2/3}\gg m\,,\eqnd\eFTTcvi $$ 
where $ |\left<\phi\right>|$ is the zero temperature field expectation value and $m\propto \sqrt{-r}$ the physical mass-scale.
This result confirms that the critical temperature is in the high temperature
region. 
\subsection Dimension $d\le 3$

{\it Dimension $d=3$.} For $d=3$ and at order $r$ we now obtain \sslbl\ssFTfiviii
$$\eqalignno{[G_2]_\r&={1\over12 L}-{2\Lambda\over
(4\pi)^{3/2}}-{L\over 8\pi^2}\left[\ln(\Lambda L)-\ud
\gamma-\ln(4\pi)\right]r.&\eqnn  \cr 
G_4 &={L\over 8\pi^2}\left[\ln(\Lambda L)-\ud \gamma-\ln(4\pi)\right].&
\eqnn \cr} $$
The coupling constant $u$ is dimensionless $u\equiv g$.
Then $G_4$ just yields the one-loop contribution to the perturbative expansion of the running
coupling $g(1/\Lambda L)$, 
$$g(1/L\Lambda ) = g-{N+8\over 48\pi^2}\left[\ln(\Lambda L)-\ud
\gamma-\ln(4\pi)\right] g^2. $$ 
In fact we know from RG arguments that all quantities will only
depend on the running coupling constant. \par
In the same way $G_2$ contains a one-loop contribution to perturbative expansion of the running $\phi^2$ coefficient $r(1/\Lambda L)$
$$r(1/\Lambda L)/r=1-{N+2\over 48\pi^2}\left[\ln(\Lambda L)-\ud
\gamma-\ln(4\pi)\right] g\,.$$ 
The $d=3$ effective theory is super-renormalizable, and thus requires only a
mass renormalization. In $G_2$ we find two terms, one cut-off dependent which
is a one-loop counter-term, and the second which gives a mass  to the
zero-mode. The one-loop effective action takes the form 
$$\eqalignno{ {\cal S}_L(\varphi)&=\int \d ^3 x \left\lbrace\ud 
\left[ \nabla \varphib (x) \right]^{2}+\ud r_L \varphib
^{2} (x)+{1 \over 4!} g_L\bigl(\varphib^2(x)\bigr)^2
 \right\rbrace\cr &\quad +\hbox{one loop counter-terms},&\eqnn \cr}$$
with
$$r_L=r(1/\Lambda L)+{N+2  \over 72 }{g(1/\Lambda L)\over L^2} \,, \quad g_L={g(1/\Lambda L) \over L}\,.$$
Expanding in powers of $g$ one can check explicitly that, as in the case $d=4$, the term proportional to $\Lambda $ in $[G_2]_\r$ is indeed the one-loop counter-term and one is left with a one-loop contribution to $\Gamma^{(2)}$:   
$$-{N+2\over 48\pi}g{\sqrt{r}\over L}\,.$$ 
Again we note that $L^{-1}$ now plays the role of the cut-off. 
\nref\rParw{R. Parwani, {\it Phys. Rev.} D51 (1995) 4518.}
\nref\rmatcha{E. Braaten and A. Nieto, {\it Phys. Rev.} D51 (1995) 6990.}
\nref\rChHa{See for instance 
F. Karsch, A. Patkos and P. Petreczky, {\it Phys. Lett.} B401 (1997) 69\semi
S. Chiku, T. Hatsuda, {\it  Phys. Rev.} D58 (1998) 76001, hep-ph/9803226 which use a method proposed in
R. Seznec and J. Zinn-Justin, {\it J. Math. Phys.} 20 (1979) 1398\semi
H. Nakkagawa and H. Yokata, hep-ph/9902422 
.}
\nref\rbookii{see for instance reference \rbook, chap.~28.}
\nref\rZJLG{{\it Large Order Behaviour of Perturbation Theory} - Current Physics Sources and Comments, Vol. 7 (Elsevier Science, 1990)  J.C. Le Guillou and J. Zinn-Justin eds.}
\smallskip
{\it The massless theory.} For $r=0$
$$(L m_L)^2= {N+2 \over 72} g(L\Lambda ) .$$ 
The solution to the RG equation tells us that when $L\Lambda \gg 1$
$g(L\Lambda ) $ goes to zero as $1/\ln(L\Lambda)$ (equation \eFTgrun). 
Therefore
$$(L m_L)^2 \sim{2\pi^2 (N+2) \over 3(N+8)
}{1\over \ln(\Lambda L)}\, . \eqnd\eFTmiii $$ 
The mass of the zero-mode is smaller, though only logarithmically
smaller, than the other modes, justifying the mode expansion. Moreover
the perturbative expansion of the three dimensional effective theory is, for small momenta, an expansion in $g(\Lambda L)/L$ divided by the mass which is of order $\sqrt{g(\Lambda L)}/L$. The expansion parameter thus is $\sqrt{g(\Lambda L)}$ which is small, due to the IR freedom of the four-dimensional theory.  
Higher order calculations have been performed \refs{\rParw, \rmatcha}.
The convergence, however, is expected to be extremely slow and therefore summation techniques have been proposed \rChHa. General summation methods, which have been used in the calculation of 3D critical exponents, should be useful here also \refs{\rbookii,
\rZJLG}. \par
The situation $r=0$ is representative of a regime, in which $|L^2r(1/\Lambda L)|=L^2m^2$ is small compared to, or of the order of $ g(L\Lambda )$, where the physical mass scale $m$ is for $r>0$ also proportional to the physical mass.
\smallskip
{\it The critical temperature.} If at zero temperature the system is in an ordered phase
($r<0$), at higher temperature a phase transition occurs at  a critical temperature $T_c=1/L$, which at leading order is solution of the equation
$$ r_L=r(1/\Lambda L)+{N+2  \over 72 }{g(1/\Lambda L)\over L^2} =0\,. $$
This relation can be rewritten in various forms, for example
$$\sqrt{(N+2)/12} \,T_c\sim  |\left<\phi\right>|\propto m\sqrt{\ln (\Lambda/m)}\propto
(-r)^{1/2}\bigl(\ln(-r)\bigr)^{3/(N+8)}  , \eqnd\eFTTciii $$
where $m$ and $\left<\phi\right>$ are  the physical mass scale and field expectation value resp.~at zero temperature.  We note that the critical temperature is large compared to the mass scale $m$ and thus belongs to the high temperature regime. The critical theory, which  can no longer be studied by perturbative methods, is the theory relevant to a large class of phase 
transitions  in statistical physics. It has been studied by a number of different methods.
\smallskip
{\it Other local interactions.} In the same normalization an interaction term
with $2n$ fields and $2k$ derivatives is proportional to $g^n L^{n-3+2k}$ and
thus negligible in the situations under study for $n>2$ or $n=2$, $k>0$,
because even the zero-mode mass is large.  
\medskip
{\it Dimension $d<3$.} 
Renormalization group arguments imply that the finite $L$
correlation length $\xi_L$ is proportional to $L$ for $d<3$. Since $Lm_L$ is
of order unity, there is no justification anymore to treat the zero mode
separately. To calculate correlation functions at momenta small compared to
the temperature, and for small field expectation value, a local expansion of
the type of chiral perturbation theory still makes physical sense, but it is
necessary to modify the perturbative expansion. For example in the $d+1$ field
theory one can add and subtract a mass of order $1/L$ a mass term for the zero
mode. This temperature dependent mass renormalization modifies the propagator
and introduces an IR cut-off. One then determines the mass term by demanding
cancellation of the one-loop correction to the mass. \par    
An alternative strategy is to work in $d=3-\varepsilon$ dimension and  use the
$\varepsilon$-expansion. Then the zero-mode effective mass is formally small
of order $\sqrt{\varepsilon}/L$, and the expansion parameter is
$\sqrt{\varepsilon}$. 
\smallskip
{\it The critical temperature.} For $d=2$, a problem which arises only for
$N=1$, RG equations lead to the scaling relation  
$$T_c\propto |\left<\phi\right>|^{2/(1+\eta)}\propto m \,,$$
where $\eta$ is the 3D Ising model exponent $\eta\approx 0.03$. 
%
\nref\rPBZB{A.M. Polyakov, {\it Phys. Lett.} 59B (1975) 79\semi
E. Br\'ezin and J. Zinn-Justin, {\it Phys. Rev. Lett.} 36 (1976) 691\semi E. Br\'ezin, J. Zinn-Justin and J.C. Le Guillou, {\it  Phys. Rev.} D14 (1976) 2615\semi
W.A. Bardeen, B.W. Lee and R.E. Shrock, {\it Phys. Rev.} D14 (1976) 985.}
\nref\rBZLG{E. Br\'ezin and J. Zinn-Justin,  {\it Phys. Rev.} B14 (1976) 3110.}

\section The non-linear $\sigma$ model in the large $N$ limit

We now discuss another, related, example: the non-linear $\sigma$
model because the presence of Goldstone modes introduces some new
aspects in the analysis. Moreover, due the non-linear character of the
group representation, one is confronted with difficulties which also
appear in non-abelian gauge theories. Actually the non-linear $\sigma $ model
and non-abelian theories share another property: both are asymptotically free
in the dimensions in which they are renormalizable \refs{\rPBZB,\rBZLG}. \sslbl\ssFTnls\par     
Before dealing with the non-linear $\sigma$ in the perturbative
framework, we discuss the finite temperature properties in the large $N$
limit. Large $N$ methods are particularly well suited to study finite
temperature QFT because one is confronted with a problem of crossover
between different dimensions. We recall that it has been proven within the
framework of the $1/N$ expansion that the non-linear $\sigma $ model is
equivalent to the $\left((\phib^2)\right)^2$  field theory, both quantum field
theories generating  two different perturbative expansions of the same
physical model \rBZLG.    
\medskip
{\it The non-linear $\sigma $ model.}  The non-linear $\sigma $ model is an
$O(N)$ symmetric quantum field theory, with an $N$-component scalar field
${\bf S}(x)$ which belongs to a sphere, i.e.~satisfies the constraint ${\bf
S}^2(x)=1$. To study the model in the large $N$ limit, it is convenient to
enforce the constraint by a $\delta $-function in its Fourier representation.
We thus write the partition function of the non-linear $\sigma$ model:  
$${\cal Z}= \int \left[\d{\bf S}(x)\d\lambda(x)\right]
\exp\left[-{\cal S}({\bf S},\lambda)\right] ,\eqnn  $$ 
with:
$${\cal S}({\bf S},\lambda) = {1 \over 2t}\int \d^{d+1}x \left[ \bigl(
\partial_{\mu}{\bf S}(x) \bigr)^2 + \lambda(x) \left({\bf S}^2(x) -1
\right)\right], \eqnd\eactsigla $$
where the $\lambda$ integration runs along the imaginary axis. The parameter
$t$ is the coupling constant of the quantum model as well as the temperature
of the corresponding  classical theory in  $d+1$ dimensions.\par  
Note that to compare the expectation value of $\bf S$ with the expectation of
the field $\phib$ of the $\phi^4$ field theory one must set ${\bf
S}=t^{1/2}\phib$.  
\nref\rONgen{for a recent review see for example J. Zinn-Justin, {\it Vector models in the large N limit: a few applications}, 11th Spring 	school in particle and fields, Taipei, 1997, hep-th/9810198.}  

\subsection The large $N$ limit at zero temperature 

We briefly recall the solution of the $\sigma$-model  in the large $N$
limit at zero temperature. Integrating over $N-1$ components of ${\bf S}$
and calling $\sigma$ the remaining component, we obtain \refs{\rBZLG, \rONgen}:
$${\cal Z}= \int \left[\d\sigma(x)\d\lambda(x)\right]
\exp\left[-{\cal S}_N(\sigma,\lambda)\right] , \eqnn $$  
with:
$${\cal S}_N \left(\sigma,\lambda  \right)= 
 {1 \over 2t}\int \left[ \left(\partial_{\mu}\sigma \right)^{2}+
\left(\sigma^{2} (x)-1\right) \lambda (x)
\right] \d^{d+1}x +{1 \over 2} (N-1 ) \tr\ln \left[
-\Delta +\lambda (\cdot) \right] .  \eqnn $$
The large $N$ limit is here taken at $tN$ fixed. The functional integral can then be calculated by steepest descent. At leading order we replace
$N-1$ by $N$. The saddle point equations are:
\eqna\emgNsig
$$\eqalignno{m^2\sigma &=0\, ,& \emgNsig{a} \cr
\sigma^{2}& = 1 - {N t \over (2\pi)^{d+1}} \int^{\Lambda}{\d^{d+1}k
\over k^2+ m^2} \,,& \emgNsig{b}\cr} $$
where we have set $\left<\lambda(x)\right>=m^2$. For $t$ small the field expectation value $\sigma $ is different from zero, the $O(N)$ symmetry is broken and thus $m$, which is the mass of the $\pi$-field,  vanishes. Equation \emgNsig{b} gives the field expectation value: 
$$\sigma^{2} = 1 - t/t_{c}\,, \eqnd\emgTNsig $$
where we have introduced $t_c$, the critical coupling constant where $\sigma$ vanishes
$${1 \over t_{c}} = {N \over (2\pi)^{d+1}} \int^{\Lambda}{\d^{d+1}k \over
k^{2}} .\eqnn $$
Above $t_{c}$, $\sigma$ instead vanishes, the symmetry is unbroken, and $m$ which is now the common mass of the $\pi$- and $\sigma$-field is given by the gap equation
$$ {1 \over (2\pi)^{d+1}} \int^{\Lambda}{\d^{d+1}k
\over k^2+ m^2}={1\over N t}\,.$$ 
Depending on space dimension we thus find:\par
\noindent (i) For $d>3$:
$$ m\propto \sqrt{t-t_c}\,.\eqnd\eNsigmv$$
(iii) For $d=3$
$${1 \over t_{c}}- {1 \over t} \sim {N \over 8\pi^2} m^2 \ln (\Lambda /m).
\eqnd\eNsigmiv $$ 
(iv) For $d=2$
$$ m={4\pi \over N}\left({1 \over t_{c}}- {1 \over t}\right).
\eqnd\eNsigmiii $$
(v) For $d=1$
$$m \propto \Lambda \e^{-2\pi/N t}.\eqnd\eNsigmii $$
The physical  domain then corresponds to $t$ small, $t=O(1/\ln(\Lambda /m))$. 
\nref\rBGMOPS{G. Baym and G. Grinstein, {\it Phys. Rev.} D15 (1977) 2897\semi
E. Br\'ezin, {\it J. Physique (Paris)} 43 (1982) 15\semi
H. Meyer-Ortmanns, H.J. Pirner and B.J. Schaefer, {\it Phys. Lett.} B311
(1993) 213\semi 
M. Reuter, N. Tetradis and C. Wetterich, {\it Nucl. Phys.} B401 (1993) 567, hep-ph/9301271\semi
see also G. N. J. Añaños, A. P. C. Malbouisson, N. F. Svaiter,  {\it Nucl.
Phys.} B547 (1999) 221, hep-th/9806027.} 

\subsection Finite temperature

As we have already explained, finite temperature $T$ corresponds to finite
size $L=1/T$ in the corresponding $d+1$ dimensional classical theory.  \par 
The saddle point or gap equation \emgNsig{b}, in the symmetric phase $\sigma=0$,
becomes \rBGMOPS 
$$\eqalignno{1&= Nt {1\over (2\pi)^{d}L}\int\d^{d} k \sum_{n}{1\over
m_L^2+k^2+(2\pi n/L)^2}\,, &\eqnn \cr
&=Nt L^{1-d}{1\over 4\pi}\int_{s_0}^\infty{\d s\over
s^{d/2}}\e^{-m_L^2 L^2s/4\pi} \vartheta_0(s), &\eqnd\esaddNTf  \cr} $$
with $s_0=4\pi /L^2\Lambda^2$. \par
Here $\xi_L=m_L^{-1}$ has the meaning of a correlation length in the space
directions. \par
A phase transition is possible only if the integral is finite for $m_L=0$. IR divergences can come only from  the contribution of the zero-mode: since the integral is $d$-dimensional, a 
phase transition is  possible only for $d>2$. This is already an example of {\it dimensional reduction}\/ $d+1\mapsto d$. \par
We have seen that from the point of view of perturbation theory a crossover between different dimensions is a source of technical difficulties because IR divergences are more severe in lower dimensions. Instead the large $N$ expansion is particularly well suited
to the study of this problem because it exists for any dimension.
\medskip
{\it Dimension $d=1$}. Let us first examine the case $d=1$. This
corresponds to a situation where even at zero temperature, $L=\infty$,  the
phase always is symmetric and the mass is given by equation \eNsigmii. 
Using the gap equation for zero temperature with the same cut-off, we write
the finite  temperature  gap equation  
$$\int_0^\infty \d s\,s^{-1/2}\e^{- m_L^2L^2 s/4\pi} \left[ \vartheta_0 ( s)
-s^{-1/2}\right]+\int_{s_0}^{s_1} {\d s\over s}\e^{- z^2 s/4\pi}=0\,, $$ 
with   $s_1=s_0 m^2/m_L^2$. It follows
$$\ln(m_L/ m)={1\over2}\int_0^\infty \d
s\,s^{-1/2}\e^{- m_L^2L^2 s/4\pi} \left[ \vartheta_0 ( s) -s^{-1/2}\right] .$$
High temperature corresponds to $m \ll 1/L$ and thus we also
expect to $m_L\gg m$. The integral diverges only for $m_L L\to0$, and it is then
dominated by the contribution of the zero-mode, 
$${1\over4\pi}\int_0^\infty \d
s\,s^{-1/2}\e^{- z^2 s/4\pi} \left[ \vartheta_0 ( s) -s^{-1/2}\right] ={1\over
2z}+{1 \over2\pi} \bigl(\ln z+\gamma-\ln(4\pi)\bigr)+O(z), $$
and therefore
$$ Lm_L  =-\pi/ \ln(m_L/m)\sim - \pi/ \ln( mL). \eqnd\eFTNnlsi$$ 
As we will see later, the logarithmic decrease at high temperature of the
product $Lm_L$ corresponds to the UV asymptotic freedom of the classical non-linear $\sigma
$ model in two dimensions. 
\medskip
{\it Dimensions $d>1$.} In higher dimensions the system can be in either phase
at zero temperature depending on the value of the coupling constant $t$.
Introducing the critical coupling constant $t_c$ we can then rewrite the gap
equation   
$$\eqalignno{G_\Lambda (L m_L)&=L^{d-1}\left({1\over Nt}-{1\over Nt_c}\right),
 &\eqnd\eFTnlsgap\cr
G_\Lambda (z)&\equiv {1\over 4\pi}\int_{s_0}^\infty \d s\,s^{-d/2}\left[\e^{-
z^2 s/4\pi} \vartheta _0( s) -s^{-1/2}\right].& \eqnd\eFTGzd \cr}$$
\smallskip
{\it Dimension $d=2$.} At finite temperature the phase is always symmetric
because  no phase transition is possible in two dimensions. The function
$G_\Lambda $ has a limit for large cut-off $G_\infty $ and the gap equation
thus has a scaling form for $d=2$ as predicted by finite size RG arguments.
For $t>t_c$ (and $t-t_c$ small)  the r.h.s.~involves the product of the mass
$m$ at  zero temperature (equation \eNsigmiii) by $L$     
$$Lm=4\pi G_\infty (L m_L). $$
with
$$ G_\infty (z)= -{1\over 2\pi}\ln\bigl(2\sinh(z/2)\bigr),\eqnd\eFSSGz  $$
For $t=t_c$, a situation relevant to high temperature QFT, we find the
equation $G_\infty (z)=0$. The solution is
$$m_L L= 2\ln\bigl((1+\sqrt{5})/2\bigr)\ , $$
and therefore the mass of the zero-mode is proportional to the mass of the other modes.\par
The zero-mode instead dominates if $Lm_L$ is small and this can arise only in
the situation $t<t_c$, i.e.~when the symmetry is broken at zero temperature.
We then have to examine the  behaviour of $G_\infty (z)$ for $z$ small. From the
explicit expression \eFSSGz~we obtain   
$$G_\infty (z)\mathop{=}_{z\to 0}-{1\over 2\pi}\ln z+O(z^2)\,\quad\Rightarrow\
L m_L= \exp\left[-{2\pi L\over N}\left({1\over t}-{1\over t_c}\right) \right] 
.\eqnd\eFTnlsmLii $$ 
Dimensional reduction makes sense only for $Lm_L$ small. On the other hand the
physical scale in the broken phase is $m\propto 1/t-1/t_c$. Therefore $Lm_L$
is small only for  $Lm$ large, i.e.~at low but non-zero temperature, a
somewhat surprising situation, and a precursor of the zero temperature phase transition. Another possibility corresponds to $t<t_c$
fixed and thus a physical scale of order $\Lambda $: this is the situation
of chiral perturbation theory, and corresponds to the deep  IR (perturbative) region
where only Goldstone
particles propagate. Then $Lm_L $ is small even at high temperature. Note that
the mass $m_L$ has, when the coupling constant $t$  goes to zero or
$L\to\infty$, the exponential behaviour characteristic of the dimension
two.\par 
For $t<t_c$ the equation can also be written
$$L m_L= \e^{- 2\pi L\left(\left<\phi\right>\right)^2/N},  $$
where $\left<\phi\right>$ is the field expectation value in the normalization of the $\phi^4$
 field theory. 
\smallskip
{\it Dimension $d=3$.}
For $d=3$ the situation is different because a phase transition is possible in
a three-dimensional classical theory. This is consistent with the existence of the
quantity $G_\infty (0)>0$ which appears in the relation between coupling
constant and temperature at the critical point $m_L=0$:
$${1\over t}-{1\over t_c}={N\over 12 L^2}\,.\eqnn $$
For a coupling constant $t$ which corresponds to a phase of broken symmetry
at zero temperature ($t<t_c$), one now finds a transition temperature
$$T_c=L_c^{-1}\sim (12/N)^{1/2} |\left<\phib\right>|,$$
a result consistent with equation \eFTTciii.  \par
Studying more generally the saddle point
equations one can derive all other properties of this system. 
Another limit of interest is the high temperature QFT. For $z\ne 0$ the
coefficient of $z^2$ in expression \eFTnlsgap~ has a cut-off dependence
$$G_\Lambda (z)= {1\over12 }-{z\over4\pi}+{z^2\over
16\pi^2}\left[-2\ln(\Lambda L)-\gamma+2\ln(4\pi)\right] +O(z^3).   $$
At $t=t_c$ we find that $(m_L L)^2$ is of order $1/\ln(\Lambda L)$. Thus
at leading order
$$(m_L L)^2={2\pi^2\over 3\ln(\Lambda L)}, $$
in agreement with equation \eFTmiii.  
\smallskip
{\it Dimension $d=4$.} From $G_\infty (0)=\zeta(3)/4\pi^2$  and the simple
relation 
$${\partial G_\Lambda (d=4,z) \over \partial z^2}=-{1\over 4\pi}G_\infty
(d=2,z) -{L\Lambda \over 16\pi^{5/2}}, $$
we find:\par
(i) The critical temperature $T_c$ for $d=4$ 
$$T_c=L_c^{-1}\sim
(2\pi)^{2/3}\bigl(N\zeta(3)\bigr)^{-1/3}|\left<\phib\right>|^{2/3}, $$ 
again consistent with equation \eFTTcvi.\par
(ii) In the massless limit $G(z)=0$ leads to
$$L^2 m_L^2\sim \pi^{-1/2}\zeta(3)/L\Lambda \,,\eqnd\eFTnlsmiv$$
in agreement with the behaviour found in section \ssFTfiviv.
\smallskip
{\it The $((\phib)^2)^2$ field theory at large $N$.} To compare with  the
situation in the $\phi^4$ theory of section \ssFTfiviii~it is interesting to
also write the corresponding gap equation  for $d=4$ in the large $N$ limit.
One finds   
$$L^2m_L^2={N g \over 6\Lambda L}G_\Lambda (Lm_L). $$
For $Lm_L$ small one expands
$$L^2m_L^2={N g \over 6\Lambda L}\left[{\zeta(3) \over4\pi^2}-{L\Lambda \over
4\pi^{3/2}} m_L^2L^2+{1\over 8\pi^2} m_L^2L^2\ln(Lm_L)+O(m_L^2L^2)\right]. $$ 
At leading order one finds
$$L^2m_L^2={N \over 6}{\zeta(3) \over4\pi^2}g_L \,,\qquad g_L={1\over \Lambda
L}{g \over (1+Ng/3(4\pi)^{5/2})}\,.$$ 
This behaviour is consistent  with the behaviour found in section
\ssFTfiviv~and the behaviour \eFTnlsmiv. 

\def\pib{\pi}
\section The non-linear $\sigma$ model: Dimensional reduction

We want now to derive the reduced effective action for the non-linear $\sigma
$ model. Because the space of fields is a sphere, a simple mode expansion
destroys the geometry of the model. Several strategies then are available. We
explore here two of them, and mention a third one. \par
A first possibility, which we will not discuss because we can do better,  is
based on solving the constraint ${\bf S}^2(x)=1$ by parametrizing the field
${\bf S}(x)$  
$${\bf S}(x) = \{\sigma(x) , \pib(x)\} , $$
and eliminating locally the field $\sigma(x)$ by:
$$ \sigma(x) = \left(1-\pib^2(x)\right)^{1/2}.$$
One then performs a mode expansion on $\pi(x)$, and integrates perturbatively
over the non zero-modes. 
This mode expansion somewhat butchers the geometry and this is the main source
of complications. Otherwise, provided one uses dimensional regularization (or
lattice regularization, but the calculations are much more difficult) to deal
with the functional measure,  this strategy is possible.\par   
However one can find other methods in which the geometric properties are obvious. One convenient method involves parametrizing the zero-mode in terms of a time-dependent rotation matrix which rotates the field zero-mode to a standard direction in the spirit of section 36.5 of \rbook.
Here instead we describe a method based on the introduction of an auxiliary field. This will allow us to use a more physical cut-off regularization of
Pauli--Villars type.    
\subsection Linearized formalism. Renormalization group
 
We again start from the action in the form \eactsigla. 
$${\cal S}({\bf S},\lambda) = {\Lambda^{d-1} \over 2t}\int \d^{d+1}x \left[
\bigl( \partial_{\mu}{\bf S}(x) \bigr)^2 + \lambda(x) \left({\bf S}^2(x) -1
\right)\right], \eqnn $$
but we have rescaled the coupling constant in such a way that $t$ now is
dimensionless. The correlation functions of the $\bf S$ field satisfy RG equations
$$\left[ \Lambda{ \partial \over \partial
\Lambda} +\beta (t){\partial \over \partial t}-{n \over 2}\zeta
(t)\right] \Gamma^{(n)} \left(p_{i};t,L,\Lambda \right)=0\, ,
\eqnd\eFTRGsig  $$
with 
$$\beta(t)=(d-1)t+O(t^2).$$
The solution can be written
$$ \Gamma^{(n)} \left(p_{i};t,L,\Lambda \right)=m^{d+1}(t) M_{0}^{-n}
(t)F^{(n)} \bigl(p_i/m(t) , Lm(t) \bigr). \eqnd\eFTRGsolve $$
with 
$$ \eqalignno{ M_0 (t) & = \exp\left[-{1 \over 2} \int^{t}
_{0}{\zeta \left(t' \right) \over \beta \left(t'
\right)} \d  t'\right] , & \eqnd\eMzero \cr m(t)={1\over \xi (t)} & =
\Lambda t^{-1/(d-1)}\exp\left[-\int^{t}_{0} \left({1 \over
\beta \left(t' \right)}-{1 \over (d-1) t'} \right)
\d  t'\right] . & \eqnd\enlsmdef  \cr}
$$
The RG functions are related to properties of the zero temperature theory.
The function $m(t)$ has the nature of a physical mass. In the broken phase it
is a crossover scale between the large momentum critical behaviour and the
small momentum perturbative behaviour. The function $M_0(t)$ is proportional
to the field expectation value.\par 
For $d=1$ 
$$\beta(t)=-{N-2 \over 2\pi}t^2 +O(t^3)\,,\quad\zeta (t)={N-1 \over 2\pi}t
+O(t^2), \eqnd\enlsRGf $$ 
and the definition \enlsmdef~has to be modified
$$m(t)\propto \Lambda \exp\left[-\int^{t} {\d t' \over \beta (t')} \right] \
\Rightarrow\ \ln(m/\Lambda )=-{2\pi \over (N-2)t}+O(\ln t).\eqnd\enlsmdii $$
Another way to express the solution of RG equations at finite temperature is
to introduce the coupling $t_L$ at scale $L$ 
$$ \ln(\Lambda L)=\int^{t}_{t_L}{\d t' \over \beta(t')}\,, \eqnd\etLxi $$
where $t_L$ is a function of $t$ and $L$ only through
the combination $L m(t)$
$$\ln Lm(t)=-\int^{t_L}{\d t'\over \beta (t')}. \eqnd\eFTtL$$
For $d>1$ and $t<t_c$ fixed, the equation \etLxi~implies that $t_L$ approaches
the IR fixed point $ t=0 $ at fixed temperature 
$$t_L\sim1/\bigl(Lm(t) \bigr)^{d-1}.\eqnd\etLGold $$ 
In the mass scale $m(t)$ which is of order $\Lambda $, this is a low temperature regime, where finite temperature effects can be calculated from perturbation
theory and renormalization group. \par 
At $t_c$, and more generally in the critical domain, techniques based on an $\varepsilon =d-1$ expansion can be used. Since $t_c$ is a RG fixed point $t_L(t_c)=t_c$. 
\par
Finally in  two dimensions ($d=1$) we see from equation \eFTtL~that $t_L$ goes
to zero  for $Lm(t)$ small, i.e.~at high temperature, because $t=0$ then is a
UV fixed point,  
$$t_L\sim {2\pi\over(N-2)\ln(m(t)L)},\eqnd \eFStLde $$
and this is the limit in which the two-dimensional perturbation theory is
useful. 
\subsection Dimensional reduction

We expand the fields in eigenmodes in the time dimension, and
keep the tree and one loop contributions. We call $\varphib,\rho$ the zero
momentum modes and ${\cal S}_L(\varphib,\rho)$ the reduced $d$ dimensional
action.  At leading order we find  
$$ {\cal S}_L(\varphib,\rho)=L{\cal S}(\varphib,\rho) . \eqnn $$ 
The one-loop contribution now  is
$$\eqalignno{\delta {\cal S}_L&=\ud N\tr\ln(-\Delta+\rho)+\ud
\tr\ln\left[\varphib(-\Delta+\rho)^{-1}
\varphib\right] \cr
&=\ud (N-1)\tr\ln(-\Delta+\rho)+\ud
\tr\ln\left[\varphib(-\Delta+\rho)^{-1}
\varphib(-\Delta+\rho) \right]. &\eqnd\eFTnlsfila  \cr}$$
The form of the last term may surprise, until one remembers that the perturbative
expansion is performed around a non-vanishing value of $\varphib$.\par
We use the identity, obtained after some commutations,
$$\eqalign{\varphib(-\Delta+\rho)^{-1}\varphib(-\Delta+\rho)&
=\varphib\cdot\varphib+\varphib(-\Delta+\rho)^{-1}[\Delta,\varphib]\cr
&=\varphib\cdot\varphib+\varphib(-\Delta+\rho)^{-1}
\left[(\Delta\varphib)+2\partial_\mu\varphib \partial_\mu\right]. \cr}$$
At this order $\varphib\cdot\varphib=1$ and we expect that $\rho $ can be
neglected because it yields an interaction of higher dimension , which is therefore negligible in the long
distance limit.   
If we then expand $\tr\ln$ we see that the first term yields a term
with two derivatives and higher orders yield additional derivatives
which also are sub-leading in the long distance limit. The first term yields
$$\tr\varphib(-\Delta+\rho)^{-1}\left[(\Delta\varphib)+2\partial_\mu\varphib
\partial_\mu\right] \sim \tr(\partial_\mu\varphib)^2(-\Delta)^{-1},
 $$
where the relations
$$\varphib\cdot\partial_\mu\varphib=0\,,\quad(\partial_\mu\varphib)^2
+\varphib\cdot \Delta\varphib=0\,,$$
valid at leading order, have been used.\par
In the same way we expand the first term in \eFTnlsfila~in powers of the field
$\rho $. At leading order only one term is relevant, and we thus obtain 
$$\delta {\cal S}_L=\ud G_2\int\d^d
x\left[(\partial_\mu\varphib(x))^2+(N-1)\rho(x) 
\right] , $$
where the constant $G_2$ defined in \eGinttwo~has to taken at $r=0$. One finds
(appendix \label{\appthree}) 
$$\eqalignno{G_2&={1\over
2\pi^{(d+1)/2}}\Gamma\bigl((d-1)/2\bigr)\zeta(d-1)L^{2-d} 
-{2\Lambda^{d-2}\over (d-2)(4\pi)^{d/2}}+{2L\Lambda^{d-1}
\over(d-1)(4\pi)^{(d+1)/2}}. ,\cr 
&\quad&\eqnd\eGtwozen \cr}$$
We conclude that at one-loop order
$${\cal S}_L(\varphi,\rho) ={L\Lambda^{d-1}\over 2t} \int\d^d x\left[
(Z_\varphi/Z_t)(\partial_\mu\varphib(x))^2+\rho(x) 
\left(\varphi^2(x)-Z^{-1}_\varphi\right)\right], $$
with
$$\eqalign{Z_t&=1+(N-2)\Lambda^{1-d}L^{-1}G_2 t+O(t^2) \cr
Z_\varphi&=1+(N-1)\Lambda^{1-d}L^{-1}G_2 t+O(t^2) .\cr} $$
\nref\rSHN{S. Chakravarty, B.I. Halperin and D.R. Nelson,
{\it Phys. Rev.} B39 (1989) 2344.}
\nref\rHaNie{R. Hasenfratz and F. Niedermayer, {\it Phys. Lett.} B268 (1991)
231.} 
\smallskip
{\it Dimension $d=2$}\/ \refs{\rSHN,\rHaNie}. 
For $d=2$ the constant $G_2$ in equation \eGtwozen~has a UV contribution
which is three dimensional of order $\Lambda $, and a two-dimensional 
contribution of order $\ln(\Lambda L)$, corresponding to the omitted 
zero-mode
$$G_2=-{1\over 2\pi}\bigl(\ln(\Lambda
L)-\ud\gamma\bigr)+{2\over(4\pi)^{3/2}}\Lambda L\,. $$ 
The term proportional to $\Lambda L$ generates a finite renormalization of $t$
$$t_\r =t+ 2(4\pi)^{-3/2}(N-2)t^2, $$
and of the field $\varphi$
$$\varphi=\left[1- (4\pi)^{-3/2}(N-1)t\right]\varphi_\r\,.$$
We now introduce the effective coupling constant $g$
$$g_L=t_\r /(\Lambda L)\,. $$
The effective action becomes
$${\cal S}_L(\varphi_\r,\rho_\r) ={1\over 2g_L} \int\d^2 x\left[ (\tilde
Z_\varphi/Z_g)(\partial_\mu\varphib_\r(x))^2+\rho_\r(x) 
\left(\varphi_\r^2(x)-\tilde Z^{-1}_\varphi\right)\right]. $$
We verify that the remaining factors $ Z_g, \tilde Z_\varphi$ render the
reduced theory one-loop finite 
$$\eqalign{ Z_g&=1-{N-2\over 2\pi} \bigl(\ln(\Lambda L)-\ud\gamma\bigr) g_L
+O(g_L^2) \cr 
\tilde Z_\varphi&=1-{N-1\over 2\pi} \bigl(\ln(\Lambda L)-\ud\gamma\bigr) g_L
+O(g_L^2).\cr} $$ 
The solution of the two-dimensional non-linear $\sigma $ model then requires
non-pertur\-bative techniques, but the two-dimensional RG tells us 
$$\ln(m_L L)\propto -{2\pi  \over(N-2)g_L}= -{2\pi L\Lambda
\over(N-2)t}=-{2\pi \over N-2}L m(t), $$  
where the last equation involves the three-dimensional RG. The result is
consistent with equation \eFTnlsmLii. 
\smallskip
{\it Dimension $d=1$.} Then
$$G_2={L\over 2\pi}\left[\ud\gamma-\ln(\pi)+\ln(\Lambda L)\right]. $$
The reduced one-dimensional theory is of course finite. Therefore $Z_\varphi$
and $Z_t$ are the renormalization factors which are associated with the change
from the scale $\Lambda $ to the temperature scale $1/L$. We set
\eqna\eFTnlsDi
$$\eqalignno{\varphi_\r &=Z_\varphi^{1/2}\varphi=\left[1+(N-1)G_2
t/2L\right]\varphi & \eFTnlsDi{a}
\cr   {1\over g}&={1 \over t Z_t}={1\over t}
-(N-2)G_2/L+O(t) \,.&\eFTnlsDi{b}
 \cr}$$ 
Both quantities $Z_\varphi$ and $g$ satisfy the RG equations of the zero
temperature field theory 
$$\Lambda {\partial Z_{\varphi}\over \partial \Lambda }+\beta (t){\partial
Z_{\varphi}\over \partial t}-\zeta (t)Z_{\varphi}=0\,, \quad 
\Lambda {\partial g\over \partial \Lambda }+\beta (t){\partial g\over \partial
t}=0 \,, $$ 
where the RG functions at this order are given in \enlsRGf.\par 
The one-dimensional non-linear $\sigma $ model again cannot be solved by
perturbation theory, but since it corresponds to a simple angular momentum
squared hamiltonian, it can be solved exactly. The difference between the
energies of the ground state energy and first excited state is   
$$ m_L =\ud (N-1)g /L\,.$$
Expressing $t$ in terms of the mass scale \enlsmdii, which is proportional to
the physical mass, we obtain 
$${1\over m_L L}\sim -{1\over \pi}{N-2 \over N-1}\ln(mL),\eqnd\eFTmLdi $$
a result consistent with equation \eFTNnlsi.
The result reflects the UV asymptotic freedom of the non-linear $\sigma $ model in
two dimensions; the effective coupling constant decreases at high temperature
where $mL\to 0$.  
\nref\rmatchb{K. Kajantie, M. Laine, K. Rummukainen, M. Shaposhnikov, {\it
Nucl. Phys.} B458 (1996) 90, hep-ph/9508379.} 

\subsection Matching conditions

If the explicit form of the reduced theory can be guessed, another strategy is
available, based on matching conditions. The idea is to calculate some physical observables in $d+1$ dimensions and to expand them for high temperature thus $L$ small. One then
calculates the same quantities in the guessed reduced theory in $d$
dimensions. Identifying the two set of results, one obtains the relations
between the parameters of the initial and reduced action
\refs{\rHaNie,\rmatcha,\rmatchb}. One advantage of the 
method is the possibility to check the ansatz of dimensional reduction by
calculating more quantities than needed, and requiring consistency. In addition
one has a better control of the correspondence for what concerns large
momentum effects. The main drawback is that one is often led to calculate
detailed expressions, here the two-point correlation function in an external
field, where the most part is not useful (related to IR properties).
Contributions of the zero-mode have to be separated for each diagram.   
\par
To guess the reduced theory the main guiding principles are power counting and symmetries, as usual for effective low energy field theories.\par 
In what follows dimensional regularization is used to avoid the functional
measure problem: in the absence of a Lagrange multiplier, a Pauli--Villars
cut-off does not regularize the $O(N)$ invariant measure. The more
``physical'' lattice regularization is also available, but explicit
calculations are more difficult.\par   
Finally, in the dimensions of interest, it is necessary to add an explicit
symmetry breaking linear in the field,  to avoid IR divergences. Once the
correspondence between the parameters of the finite temperature and the
reduced theories has been determined one can take the symmetric limit.\par   
We consider the finite temperature $d+1$ dimensional theory
$${\cal S}({\bf S})={\Lambda^{d-1}Z_S \over 2t Z_t} \int\d^{d+1}
x\bigl(\partial_\mu {\bf S}(x)\bigr)^2-{\Lambda^{d-1}\over t}\int\d^{d+1}
x\,{\bf h}\cdot {\bf S}(x), \eqnn $$  
where the \MS~scheme is used to define renormalization constants, and
$${\bf S}^2(x)=Z_S^{-1}. \eqnn $$
Therefore $t$ is the effective coupling constant at scale $\Lambda $.\par
RG equations in an external field $h$ take the form
$$\left[ \Lambda{ \partial \over \partial
\Lambda} +\beta (t){\partial \over \partial t}-{n \over 2}\zeta
(t)+ \rho(t) h{\partial \over \partial h}
\right] \Gamma^{(n)} \left(p_{i};t,h,L,\Lambda \right)=0\, ,
\eqnd\eRGsigma  $$
where the new  RG function is not independent:
$$ \rho (t)=1-d+\ud \zeta(t)+\beta(t)/t\,.  \eqnn $$
\medskip
{\it Dimensional reduction.} 
We compare it with the zero temperature $d$ dimensional theory
$${\cal S}(\varphi)={L^{2-d}Z_\varphi  \over 2 g Z_g} \int\d^d
x\bigl(\partial_\mu \varphi(x)\bigr)^2-{L^{2-d}\over g}\int\d^d x\,{\bf
h}\cdot \varphi(x), \eqnn $$  
where the \MS~scheme is used to define renormalization constants, and
$$\varphi^2(x)=Z_\varphi ^{-1}. \eqnn $$
The coupling constant $g$ instead is the effective coupling at the temperature scale
$L^{-1}$.\par 
We expect that between the two fields ${\bf S}$ and $\varphi$ some
renormalization will be required. 
\midinsert
\epsfxsize=115.4mm
\epsfysize=23.4mm
\centerline{\epsfbox{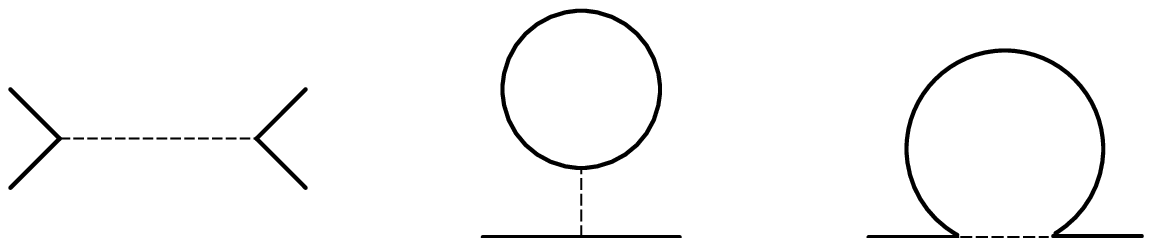}}
\vskip3mm
\centerline{\elevenpoint
$V^{(4)}=\frac{1}{8}\delta_{i_1i_2}\delta_{i_3i_4}[(p_1+p_2)^2+h]
\hskip4mm\ud(N-1)h I(h)\hskip28mm p^2 I(h)$\hskip19mm}
\medskip
\figure{0mm}{One-loop diagrams: the dotted lines do not correspond to 
propagators but are used only to represent faithfully the flow of group
indices.}
\figlbl\figFTi
\endinsert
The one-loop diagrams are listed in figure \figFTi. 
In the reduced model, at one-loop order the two-point function is
$$  \Gamma_d^{(2)} (p)={L^{2-d} \over g}
\left(p^2 Z_\varphi/Z_g +h Z^{1/2}_\varphi\right)+\left[ p^2+\ud (N-1) h
\right] I(h)+O(g)\,,$$   
where $h=|{\bf h}|$ and 
$$I(h)={1\over (2\pi)^d} \int{\d^d p\over p^2+h} \,. \eqnd\etadeph $$ 
\par
At finite temperature, in the $d+1$ theory, one finds instead  
$$\eqalignno{\Gamma_{d+1}^{(2)} (p_0=0,p)&={\Lambda^{d-1}L \over
t}\left(p^2Z_{\bf S}/Z_t +h Z_{\bf S}^{1/2}\right) 
+ \left[ p^2+\ud (N-1)h\right] \cr &\quad \times\left[G_2(h,L)+I(h)\right]  +O
(t) ,& \eqnn \cr} $$ 
where the contribution $I(h)$ of the zero-mode has been separated explicitly
and the function $G_2(r=h,L)$ is defined in \eGinttwo. In  the limit $h=0$ 
$$ G_2(0,L)= N_d {\pi \over \sin (\pi
d/2)}(2\pi)^{d-2}\zeta(2-d)L^{2-d} ,\eqnd\eGiiDii $$ 
where $N_d$ is the usual loop factor
$$N_d={2\over(4\pi)^{d/2}\Gamma(d/2)}={1\over2\pi}+O(d-2). \eqnd\eloopfac $$
\smallskip
{\it Dimension $d=2$.}  For $d\to 2$  the renormalization constants at
one-loop in the \MS~scheme are 
$$Z_g= 1+(N-2){N_d \over d-2}g \,,\quad
Z_\varphi=1+(N-1){N_d  \over d-2}g \,. \eqnd\enlsZZti $$
In particular
$$Z/Z_g=1+ {N_d \over d-2}g_\r\,. $$
Therefore the renormalized $d$-dimensional two-point function reads
$$  \Gamma_d^{(2)} (p)={1 \over g}
(p^2+h)+\left[ p^2+\ud (N-1) h \right] I_\r(h)\,,$$  
with
$$I_\r(h)=\lim_{d\to2}I(h)+{N_d \over d-2}L^{2-d}=-{1\over 4\pi}\ln(hL^2) \,.
\eqnn $$  
In the finite temperature theory no renormalization is required because the
theory is non-renormalizable, and dimensional regularization cancels all power
divergences.  
Thus 
$$\Gamma_{d+1}^{(2)} (p)\mathop{=}_{d\to2}{\Lambda L \over t}\left(p^2 +h
\right) + \left[ p^2+\ud (N-1)h\right]\bigl(G_2(h,L)+I(h)\bigr)+O (t) , \eqnn
$$ 
with
$$G_2(0,L)={N_d \over d-2} -{1\over 2\pi}\ln L+O(d-2)\,.$$
We note that at this order no field renormalization is required to compare the
two functions and then 
$${1\over g}={\Lambda L\over t}+O(t).$$
\smallskip
{\it Dimension $d=1$.} In $d=1$ dimension the reduced theory has no
divergences and the one-loop expression reads 
$$\Gamma_d^{(2)} (p)={L \over g}(p^2+h)
+ \left[ p^{2}+\ud (N-1)h\right]I(h)+O (g).$$
We compare this expression with the finite temperature two-point
function, calculated in the \MS~scheme (with renormalization scale $\Lambda$).
For this purpose we have to subtract to expression \eGiiDii~the
\MS~counterterm. For $d\to 1$ we find
$$[G_2]_\r(0,L)={L\over 2\pi}\bigl(\ln(\Lambda L)+\gamma-\ln(4\pi)\bigr),
\eqnd\eFTGiirdi $$ 
and therefore
$$\eqalignno{\Gamma_{d+1}^{(2)} (p)&={L \over t}\left(p^2 +h\right)
+ \left[ p^2+\ud (N-1)h\right] \left[I(h)+{L\over 2\pi}\bigl(\ln(\Lambda
L)+\gamma-\ln(4\pi)\bigr)\right] \cr &\quad+O (t) .&\eqnn \cr}$$ 
This time we have also to take into account the field renormalization. 
We set 
$$\varphi(x)={\bf S}(x)\sqrt{Z_{\varphi{\bf S}}}\,  \qquad Z_{{\bf
S}\varphi}=1+ (N-1) \bigl(\ln(\Lambda L)+\gamma-\ln(4\pi)\bigr){t \over 2\pi}
,$$  
and 
$${1\over t}={1 \over g Z_{tg}}\,,\qquad Z_{gt}=1-\bigl(\ln(\Lambda
L)+\gamma-\ln(4\pi)\bigr){t \over 2\pi}, $$ 
or inverting the relation 
$${1 \over g}=  {1\over t}-{(N-2)\over
2\pi}\bigl(\ln(\Lambda L)+\gamma-\ln(4\pi) \bigr) +O(t), $$
a result which can also be obtained by the method of section 36.5 of \rbook.
The results  for $Z_{gt}$ and $g$ are consistent with the equations \eFTnlsDi{}
\section The Gross--Neveu in the large $N$ expansion

To gain some intuition about the role of fermions at finite temperature we now
examine a simple model of self-interacting fermions, the Gross--Neveu (GN)
model.  The GN model is described in terms of a $U(N)$ symmetric action for a
set of $N$ massless Dirac fermions $\{\psi^i, \bar\psi^i \}$:\sslbl\ssGNNmod 
$${\cal S}\left(\bar \psib, \psib\right)= -\int \d^{d+1} x \left[
\bar\psib\cdot \sla{\partial} \psib +\ud G\left(\bar\psib\cdot \psib \right)^2
\right].\eqnd\eGNact $$ 
The GN model has in all dimensions a discrete symmetry 
$${\bf x}=\{x_1,x_2,\ldots, x_d\}\mapsto\tilde {\bf
x}=\{-x_1,x_2,\ldots,x_d\},
\quad \cases{\psi(x)\mapsto \gamma_1 \psi(\tilde x), \cr
\bar\psi(x)\mapsto -\bar\psi(\tilde x)\gamma_1 \cr},$$
which prevents the addition of a mass term.  
In even dimensions it implies a discrete chiral symmetry, and in odd
dimensions it corresponds to space reflection. Below, to simplify, we will speak about chiral
symmetry, irrespective of dimensions. \par  
The GN model is renormalizable in $d=1$ dimension, where it is asymptotically
free and the chiral symmetry is always broken at zero temperature. \par  
We recall that within the $1/N$ expansion it can be proven that the GN model
is equivalent to the GNY (Y for Yukawa) model, a model with the same symmetry,
but with an elementary scalar particle  coupled to fermions through a
Yukawa-like interaction, which is renormalizable in four dimensions \ref\rZJGN{J. Zinn-Justin,
{\it Nucl. Phys.} B367 (1991)	105.}. This
equivalence provides a simple interpretation to some of the results that
follow.\par     
Since fermions at finite temperature have no zero modes, limited insight about
the physics of the model can be gained from perturbation theory; all fermions
are simply integrated out. Therefore we study here the GN model within the
framework of the $1/N$ expansion.   
\nref\rRWP{B. Rosenstein, Brian Warr, S.H. Park,  {\it Phys. Rep.} 205 (1991) 59.} 

\subsection The GN model at zero temperature, in the large $N$ limit

We first recall the properties of the GN model at zero temperature \refs{\rRWP,\rZJGN} in the
large $N$ limit. To generate the large $N$ expansion one introduces an 
auxiliary field $\sigma$, replaces the action \eGNact~by an equivalent action:
$${\cal S}(\left(\bar \psi, \psi,\sigma\right)=\int \d^{d+1} x\left[ -\bar\psi
\cdot (\sla{\partial}+\sigma) \psi +{1\over2 G} \sigma^2\right], $$ 
and integrates over $N-1$ fermions. One finds ($\psi\equiv \psi_1$)
$${\cal S}_N\left(\bar \psi, \psi,\sigma\right)=\int \d^{d+1} x\left[
-\bar\psi  (\sla{\partial}+\sigma) \psi +{1\over2 G}
\sigma^2\right] -(N-1)\tr \ln\left(\sla{\partial}+\sigma \right). \eqnn $$
For $G=O(1/N)$ the corresponding partition function can be calculated by the
steepest descent method. The saddle point (or gap) equation obtained by
differentiating with respect to $\sigma$ has the trivial solution $\sigma=0$
and (at leading order for $N\to\infty$ we can replace $N-1$ by $N$) 
$${1\over G}={N'\over (2\pi)^{d+1}}\int^\Lambda {\d^{d+1} k \over
k^2+\sigma^2} ,\eqnd\eGNNsadd $$ 
where $N'=N\tr{\bf 1}$ is the total number of fermions. \par
Note that  at leading order for  $N$ large the scalar field
expectation value $\sigma=\left<\sigma \right>$ is also the fermion mass
$m_\psi=\left<\sigma \right>$.  
\par
For $d=1$ the chiral symmetry is spontaneously broken for all $G>0$, and one
finds 
$$\sigma=m_\psi \propto \Lambda \e^{-\pi/NG}.$$
\par
For $d>1$ a phase transition occurs at a value $G_c$ such that
$${1\over G_c}={N'\over (2\pi)^{d+1}}\int^\Lambda {\d^{d+1} k \over k^2} .$$
For $G<G_c$ the saddle point is $\sigma =0$ and the chiral symmetry is
preserved. For $G>G_c$ the chiral symmetry is broken, and for $d<3$ 
$$\sigma \propto (G-G_c)^{1/(d-1)}, $$
which implies that the physical region corresponds to $|G-G_c|$ small.\par
For $d=3$ logarithmic corrections appear and one finds instead
$$\sigma^2\ln(\Lambda /\sigma )\sim {8\pi^2\over N'}\left({1\over G_c}-{1\over
G}\right)\,,$$ 
a reflection of the IR triviality of the effective renormalizable GNY
model.\par  
In higher dimensions the model is equivalent to a weakly interacting GNY
model, with an IR stable  gaussian fixed point and 
$$\sigma \propto (G-G_c)^{1/2}. $$
In the broken phase  the $\sigma $-propagator is given by
$$\Delta^{-1}_\sigma (p)= {N'
\over 2 (2\pi)^{d+1} }\left(p^2+4\sigma^2 \right)\int^\Lambda{\d^{d+1} k \over
\left(k^2+\sigma^2 \right) \left[(p+k)^2 +\sigma^2 \right]},\eqnn $$
 where the saddle point equation has been used. 
The mass of the scalar field $m_\sigma =2\left<\sigma\right>$, is such at
leading order the $\sigma $ particle is a fermion bound state at threshold
($m_\sigma =2m_\psi$). \par  
In the chiral symmetric phase $G<G_c$ instead one finds 
$$\Delta^{-1}_\sigma (p)={1\over G}-{1\over G_c}+ {N'
\over 2 (2\pi)^{d+1} } p^2 \int^\Lambda{\d^{d+1} k \over k^2 (p+k)^2},\eqnn $$
a reflection of the property that the $\sigma $ particle now is a resonance
which can decay into a fermion pair. 
\nref\rFTGN{
U. Wolff, {\it Phys. Lett.} B157 (1985) 303\semi
K. Klimenko, {\it Z. Phys.} C37 (1988) 457\semi
T.F.  Treml,  {\it Phys. Rev.} D39 (1989) 679\semi 
T. Inagaki, T. Kouno, T. Muta, {\it Int. J. Mod. Phys.} A10 (1995) 2241, hep-ph/9409413\semi 
 A. Barducci, R. Casalbuoni, R. Gatto, M. Modugno and G. Pettini, {\it Phys. Rev.} D51 (1995) 3042, hep-th/9406117.} 

\subsection The  GN model at finite temperature

Due to the anti-periodic boundary conditions fermions have no zero modes, and
at high temperature can be integrated out, yielding an effective action 
for the periodic scalar field $\sigma $. In the situations in which the
$\sigma$ mass is small compared with the temperature, one can perform a mode
expansion of the $\sigma $ field, integrate over the non-zero modes and obtain
a local $d$-dimensional action for the zero-mode.  It is important to realize
that, since the reduced action is local and symmetric in $\sigma \mapsto
-\sigma $, it describes the physics of the Ising transition  with short range
interactions (unlike what happens at zero temperature).      
The question which then arises is the possibility of a breaking of  this
remaining symmetry  of Ising type. If a transition exists and is continuous,
the $\sigma $-mass vanishes at the transition and a potential IR problem
appears.\par  
Additional effects due to the addition of a chemical potential will
not be considered in these notes  \rFTGN.\par  
After integration over all fermions we obtain a non-local action ${\cal S}_N$
for the field $\sigma $,  
$$ {\cal S}_N\left(\sigma\right)={1\over2 G}\int_0^L\d\tau \int \d^d x \,
\sigma^2 -N\tr \ln\left(\sla{\partial}+\sigma \right), \eqnd\eGNNactb $$
where $L$ is the inverse temperature $T=1/L$, and the $\sigma $ field
satisfies periodic boundary conditions in the time direction.  
As we have seen  a non-trivial perturbation theory is obtained by expanding
for large $N$. 
\medskip
{\it The gap equation.} The gap equation at finite temperature again splits
into two equations $\sigma =0$ and  
$$\eqalignno{{L\over G}&=N' {\cal G}_2(\sigma ,L) ,&\eqnd\eFTGNNsad\cr
{\cal G}_2(\sigma ,L)&={1\over (2\pi)^d}\sum_n \int^\Lambda {\d^d k \over
k^2+\omega_n^2 +\sigma^2}\,,\qquad \omega _n =(2n+1)\pi/L\,.&\cr}  $$ 
Using Schwinger's representation, and the corresponding regularization, the
function ${\cal G}_2$  can be expressed in terms of another function
$\vartheta_1(s)$, of elliptic type (equation~\eqns{\eAellfer}),   
$${\cal G}_2(\sigma ,L)={L^{2-d}\over 4\pi} \int_{s_0}{\d s \over
s^{d/2}}\e^{-sL^2\sigma^2/(4\pi)}\vartheta_1 (s), \eqnn $$ 
with $s_0=4\pi/(\Lambda L)^2$. From equation \eqns{\ePoissonfer} we learn that
$\vartheta_1 (s)=1/\sqrt{s}$ for $s\to 0$, up to exponentially small
corrections. The function ${\cal G}_2(\sigma ,L)$ has a regular small $\sigma
$ expansion. The two first terms are 
$${\cal G}_2(\sigma ,L)={\cal G}_2(0 ,L)+\sigma^2{\cal G}_4 +O(\sigma^4), $$
with  for $d<5$ (equations \eqns{\eFTGiif,\eFTGivf})
$$\eqalignno{{\cal G}_2(0,L)&= {4L^{2-d}\over
(4\pi)^{(d+1)/2}}(1-2^{d-2})\Gamma\bigl((d-1)/2\bigr)\zeta(d-1) +{1\over
d-1}{2L\Lambda^{d-1}\over (4\pi)^{(d+1)/2}} \hskip6mm &\eqnn   \cr  
{\cal G}_4&=L^{4-d}{\Gamma(2-d/2)\over
8\pi^{4-d/2}}\left(2^{4-d}-1\right)\zeta(4-d)+{1\over d-3}{2L
\Lambda^{d-3}\over (4\pi)^{(d+1)/2} } .&\eqnn \cr} $$  
Finally  the propagator $\Delta_\sigma (p)\equiv \Delta_\sigma (\omega =0,p)$
of the $\sigma$ zero-mode, in the broken phase, is given by (after use of the
gap equation \eFTGNNsad)  
$$\Delta^{-1}_\sigma (p)= {N'
\over 2 (2\pi)^{d}L }\left(p^2+4\sigma^2 \right)\sum_n \int^\Lambda{\d^{d} k
\over \left( k^2+\omega_n^2+ \sigma^2 \right) \left[(p+k)^2
+\omega_n^2+\sigma^2 \right]}.\eqnn $$ 
We again find that the $\sigma$-mass is $2\sigma$. In the symmetric phase
instead the propagator of the zero mode is given by 
$${1 \over N'\Delta_\sigma (p)}={1\over N'G}-{{\cal G}_2(0,L)\over L}+ {p^2
\over 2 (2\pi)^d L}  \sum_n\int^\Lambda{\d^{d} k \over (k^2+\omega_n^2)
[(p+k)^2+\omega_n^2] }.\eqnn $$ 
\subsection Phase structure for $d>1$

For $d>1$ we introduce the critical value $G_c$ where $\sigma$
vanishes at  zero temperature ($L=\infty$),
$$[{\cal G}_2]_\r(\sigma ,L)={\cal G}_2-{L\over N'G_c}={L^{2-d}\over 4\pi}
\int_{s_0}{\d s \over s^{d/2}}\left[\e^{-sL^2\sigma^2/(4\pi)}\vartheta_1
(s)-s^{-1/2}\right], \eqnd\eFTGiifr $$  
The function $[{\cal G}_2]_\r(\sigma ,L)$ is a decreasing function of $\sigma
$, thus 
$$[{\cal G}_2]_\r(\sigma ,L)\le [{\cal G}_2]_\r(0 ,L)=L^{2-d}{\cal I}_2(d),
\eqnn $$ 
with
$${\cal I}_2(d)={4\over (4\pi)^{(d+1)/2}}\left(1-2^{d-2}\right)
\Gamma\bigl((d-1)/2\bigr)\zeta(d-1) . \eqnn $$ 
The integral is always negative and therefore the gap equation \eFTGNNsad~has
a solution only for $G>G_c$, i.e.~when at zero temperature chiral symmetry is
broken.   
For $d<3$ the integral \eFTGiifr ~converges at $s=0$  and the gap equation
\eFTGNNsad~takes a scaling form. For $d=2$ it can be expressed in terms of the
fermion mass at zero temperature $m_\psi$  
$$Lm_\psi=-\int_0{\d s \over s}\left(\e^{-sL^2\sigma^2/(4\pi)}\vartheta
_1(s)-s^{-1/2}\right) .$$  
\medskip
{\it The phase transition.} 
A phase transition between the two Ising phases takes place at a temperature
$T_c=L_c^{-1}$ where $\sigma $ solution to the equation \eFTGNNsad~vanishes: 
$$L_c^{d-1}\left({1 \over G}-{1\over G_c} \right)=N' {\cal I}_2(d). $$ 
Since the r.s.h.~of the gap equation \eFTGNNsad~is a decreasing function of
$\sigma $,  the $\sigma\to -\sigma$ Ising symmetry is broken for $T<T_c$ and
restored for $T>T_c$.\par  
It is interesting to express the critical temperature in terms of the fermion
mass $m_\psi$. For $d>3$ one finds 
$$T_c=(L_c)^{-1}\propto m_\psi (\Lambda /m_\psi)^{(d-3)/(d-1)}\gg m_\psi\,.$$
Therefore the critical temperature is a high temperature in the scale of the
particle masses.\par 
For $d=3$ the critical temperature is given by
$$L_c^2\left({1\over G_c}-{1\over G} \right)={N' \over48}\,. $$
Therefore
$$T_c \sim {\sqrt{6}\over \pi}m_\psi \sqrt{\ln (\Lambda /m_\psi)}\propto\sqrt{
G-G_c},$$ 
which again corresponds to a high temperature regime.   \par 
Finally for $d=2$ the critical temperature is proportional to the fermion
mass: 
$$T_c={1\over 2\ln 2}m_\psi\,.$$
\medskip
{\it Local expansion.} When the $\sigma $ mass or expectation value are small
compared to $1/L$ we can perform a local expansion of the action \eGNNactb,
and study it to all orders in the $1/N$ expansion. Consistency requires that
one also performs a mode expansion of the field $\sigma $ and retains only the
zero mode. In the reduced theory $1/L$ plays the role of a large momentum
cut-off. \par     
The first terms of the effective $d$ dimensional action are
$${\cal S}_d (\sigma )=\int\d^d x \left[\ud Z_\sigma (\partial_\mu \sigma
)^2+\ud r \sigma ^2 +{1\over 4!}u \sigma^4\right], \eqnd\eFTGNloc $$
where terms of order $\sigma^6$ and $\partial ^2 \sigma ^4 $ and higher have
been neglected, and the three parameters are given by 
$$Z_\sigma =\ud N' {\cal G}_4\,, \quad r={L\over G}-N' {\cal G}_2(0,L) \,,
\quad u=6 N' {\cal G}_4\,, $$
where  ${\cal G}_4=L^{4-d}{\cal I}_4(d)$. For $d>1$ after the shift of the
coupling constant one finds  
$$r={L\over G}-{L\over G_c}-N'L^{2-d}{\cal I}_2(d)=N'{\cal
I}_2(d)L\left(L_c^{1-d}-L^{1-d} \right). \eqnn $$
Though, after rescaling of the field we observe that the effective $\sigma^4$
coupling is logarithmically small, close enough to the critical temperature
the effective theory cannot be solved by perturbative methods. \par  
For $d=2$ the integrals are UV finite after the shift of $G$. The effective
theory describes the physics of the two-dimensional Ising model. 
\medskip
{\it  Dimensional reduction and $\sigma$ mass.} 
For $d>3$, in the symmetric phase $L<L_c$, the $\sigma$ mass behaves like
$$m_\sigma^2 \propto L^{-2}(\Lambda  L)^{3-d}\left[ 1-(L/L_c)^{d-1}\right], $$
and thus is small with respect to $L$, justifying dimensional reduction.
Moreover, after rescaling of the field $\sigma Z^{1/2}_\sigma \mapsto \sigma $
one sees that the effective $\sigma^4$ coupling is of order   
$$u/Z_\sigma^2\propto L^{d-4}(\Lambda L)^{3-d} .$$
For $d \ge 4$ the coupling is small, the physics  perturbative, and no
additional analysis is required. \par
For the mathematical case $3<d<4$, the situation is more subtle. For
dimensional reasons the true expansion parameter is 
$$m_\sigma^{d-4}u/Z_\sigma^2\propto (m_\sigma L)^{d-4} (\Lambda L)^{3-d}. $$
At high temperature, i.e.~for $1-L/L_c$ positive and finite, $Lm_\sigma
\propto (\Lambda L)^{(3-d)/2}$ and one finds  
$$m_\sigma^{d-4}u/Z_\sigma^2\propto (\Lambda L)^{(3-d)(d-2)/2}, $$
which is small. On the other hand for $ |T-T_c|\propto L-L_c$  small enough
perturbation theory is no longer useful.  
\par
For $d=3$ dimensional reduction is justified near the critical temperature,
but the reduced model is non-perturbative. Since the coefficient ${\cal G}_4$
has still a logarithmic UV contribution, in the symmetric phase at high
temperature one finds   
$$L m_\sigma \propto 1/\sqrt{\ln(\Lambda L)}, $$
which also justifies a local expansion. The effective coupling 
$$ {u \over m_\sigma Z_\sigma^2}\propto {1\over\sqrt{\ln(\Lambda L)}}, $$
and thus the reduced model can be solved using perturbation theory. This is a
situation we have already met in the example of the $\phi ^4$ field theory,
and which reflects the IR triviality of the GNY model.  \par  
In dimension $d=2$, near the critical temperature the $\sigma $ mass is small
and a local and mode expansions can be performed. The reduced theory cannot be
solved by perturbation theory.  
\par
In general the model obeys simple scaling relations (a reflection of the
existence of a non-trivial IR fixed point in the GNY model). At high
temperature $\sigma =0$, and the $\sigma $ mass if of order $1/L$. No local
expansion is justified and necessary. 
\subsection Dimension $d=1$

The situation $d=1$ is doubly special, since at zero temperature chiral
symmetry is always broken and at finite temperature the Ising symmetry is
never broken.  The GN model is renormalizable and UV free  
$$\beta(G)=-{N'-2 \over 2\pi}G^2+O(G^3). $$
The RG invariant mass scale $\Lambda(G)$, to which all masses at zero
temperature of the rich GN spectrum are proportional, has the form 
$$\Lambda(G) \propto\Lambda \exp\left[-\int^G{\d G'\over \beta (G')}\right] .
$$ 
Physical masses are small with respect to $\Lambda $ when $G$ is small
$$\ln (\Lambda /m_\psi)={2\pi \over( N'-2)G}+O(\ln G).$$
At finite temperature all masses, in the sense of inverse of the correlation
length in the space direction, have a scaling property. For example the
$\sigma $ mass has the form   
$$Lm_\sigma=f(Lm_\psi). $$ 
One can also express the scaling properties by introducing a temperature
dependent coupling constant $G_L$ defined by 
$$\int_G^{G_L}{\d G'\over \beta(G')}=-\ln(L \Lambda ).$$
At high temperature $G_L$ decreases
$$G_L  \sim {2\pi \over (N'-2)\ln(Lm_\psi)}.$$
We therefore expect a trivial high temperature physics with weakly interacting
fermions. 
\par
At finite temperature the large $N$ gap equation becomes
$${2\pi \over N' G}=\ln(\Lambda L)+{1 \over 2}
\int_0{\d s \over s^{1/2}}\left[\e^{-L^2\sigma^2
s/(4\pi)}\vartheta_1 (s)-s^{-1/2} \theta(1-s)\right], $$
which, in terms of the zero-temperature mass scale
$$m_\psi=\Lambda \e^{-2\pi/N'G}, $$
can be rewritten:
$$\eqalign{\ln(m_\psi L)&= F(\sigma L), \cr
F(z)&=-{1 \over 2}
\int_0{\d s \over s^{1/2}}\left[\e^{-z^2 s/(4\pi)}\vartheta_1 (s)-s^{-1/2}
\theta(1-s)\right]. \cr} $$
The function $F(z)$ is an increasing function, with a finite
limit for $z\to 0$ and which behaves  like $\ln z$ for $z\to\infty$, in such a
way that at low temperature one recovers $m_\sigma =2m_\psi$. Again we find a
phase transition at a temperature $L_c^{-1}\propto m_\psi$. 
This mean-field-like prediction of a phase transition contradicts the
well-known property of the absence of phase transitions in an Ising-type model in $d=1$ dimension. In the high temperature phase $\sigma =0$ and one finds  
$${1\over N'\Delta_\sigma  (p)}={1\over 2\pi}\left[-\ln(Lm_\psi)-\gamma+\ud
\ln\pi\right] +\sum_{n\ge 0} {p^2\over L \omega_n (p^2+4\omega_n^2)}, $$  
from which one derives a scaling relation
$$Lm_\sigma=f(Lm_\psi ).$$ 
\par
If we assume that dimensional reduction is justified we can return to the
action \eFTGNloc. At leading order we find a simple model in quantum
mechanics: the quartic anharmonic oscillator. Straightforward considerations
show that the correlation length, inverse of the $\sigma $ mass parameter,
becomes small only when the coefficient of $\sigma^2$ is large and negative.
This happens only at low temperature, where the two lowest eigenvalues of the
hamiltonian corresponding to space direction are almost degenerate, a
precursor of the zero temperature phase transition. One then finds       
$$Lm_\sigma \propto (\ln m_\psi L)^{5/4}\e^{-{\rm const.}(\ln m_\psi
L)^{3/2}}. $$  
\section Abelian gauge theories

We  first discuss the abelian case which is much simpler,
because the mode decomposition is consistent with the gauge structure. Some additional problems arising in non-abelian gauge theories will be considered in next section. 
Because the gauge field has a number of components which depends on the number
of space dimensions,  the mode expansion have some new properties and affects 
gauge transformations. The simplest non-trivial example of a gauge theory is QED, a theory which is IR free in four dimensions, and therefore from the RG point of view has properties similar to the scalar $\phi^4$ field theory. Another example is provided by the  abelian Higgs model but since it has a first order phase transition, it has a more limited validity.   
\subsection Mode expansion and gauge transformations

We decompose a general gauge field $A_\mu(t,x)$ into the sum of a zero mode
$B_\mu(x)$ and the sum of all non-zero modes $Q_\mu(t,x)$ 
$$A_\mu(t,x)=B_\mu(x)+Q_\mu(t,x). $$
At finite temperature $T=L^{-1}$, $Q_\mu(t,x)$ thus satisfies
$$\int_0^L\d t\,Q_\mu(t,x)=0\,.\eqnd\eFTabgaug $$
With this decomposition gauge transformations 
$$\delta A_\mu(t,x)=\partial_\mu \varphi (t,x),$$
become
$$\delta B_\mu(x)=\partial_\mu \varphi_0(t,x),\quad
\delta Q_\mu(t,x)=\partial_\mu \varphi_1(t,x)\,,\quad \varphi=\varphi_0+\varphi_1\,. \eqnd\eFTgaugetr $$
Since $\delta B_\mu$ does not depend on $t$ we conclude that $\varphi_0(t,x)$
must have the special form 
$$\varphi_0(t,x)=F(x)+\Omega t\,, \eqnd\eFTAom $$
where $\Omega $ is a constant. The space components $B_i$ transform as the
components of a $d$-dimensional gauge field;  the time component $B_0$ is a
$d$-dimensional scalar field which is translated by a constant  
$$\delta B_0(x)=\Omega \,.\eqnd\eFTBom $$
Symmetry which respect to the translation \eFTBom~implies  that the scalar field $B_0$ is massless. 
\par
The condition \eFTabgaug~then implies
$$\partial_i \int_0^L\d t\, \varphi_1(t,x)=0\,, \quad \int_0^L\d t\,
\partial_t \varphi_1(t,x)=0\ \Rightarrow\ \varphi_1(0,x)=\varphi_1(L,x). $$ 
The  transformations of the gauge field $Q_\mu$ are thus specified by periodic functions $\varphi_1(t,x)$ with a constant zero-mode,
which can be set to zero.  \par 
Finally we verify that the function $\varphi =\varphi_1+\varphi_0$ is such
that $\partial_\mu \varphi $ is periodic as it should, since $A_\mu$ is
periodic.  
\medskip   
{\it Matter fields.} We now couple the gauge field to matter, for instance charged fermions
$\psi(t,x),\bar\psi(t,x)$. At finite temperature fermion fields satisfy
anti-periodic boundary conditions. To the gauge transformation
\eFTgaugetr~corresponds for the fermions  
$$\psi(t,x)=\e^{i\varphi (t,x)}\psi'(t,x),\quad\bar \psi(t,x)=\e^{-i\varphi
(t,x)}\bar \psi'(t,x). $$  
Anti-periodicity implies that
$$\varphi(L,x)=\varphi(0,x) \pmod {2\pi}\,.$$ 
Since $\varphi_1$ is periodic, the condition implies for the constant $\Omega $ in
\eFTAom
$$\Omega =2n \pi/L\,.\eqnd\eFTgausc $$
This restriction on the transformation \eFTBom~of the scalar component $B_0$ has
important consequences. As a result of quantum corrections generated by the
interactions with charged matter,  the scalar field $B_0$ does not remain
massless. Instead the thermodynamic potential is a periodic function of $B_0$
with period $2\pi/L$.    
\nref\rFTSch{A.V. Smilga, {\it Phys. Lett.} B278 (1992) 371\semi 
A. Fayyazuddin, T.H. Hansson, M.A. Nowak, J.J.M. Verbaarschot, I. Zahed, {\it Nucl. Phys.} B425 (1994) 553, hep-ph/9312362\semi
J.V. Steele, A. Subramanian, I. Zahed, {\it Nucl. Phys.} B452 (1995) 545, hep-th/9503220\semi 
S. Durr, A. Wipf, {\it Ann. Phys. (N.Y.)} 255 (1997) 333, hep-th/9610241.}
\subsection  Gauge field coupled to  fermions: quantization
 
We now consider a gauge field coupled to an $N$-component massless charged
fermion 
$$ {\cal S} \left(\bar \psib ,\psib,A _{\mu} \right) = \int \d^{d+1}x
\left[{\textstyle{ 1 \over 4e^2}}  F ^{2} _{\mu \nu}(x) -
\bar \psib (x) \cdot\left(\sla{\partial} + i\Abar \right) 
\psib(x) \right] . \eqnd\eactQEDN $$
The theory has RG properties which bear some similarities with the $\phi^4$ theory; it is renormalizable for $d=3$ and IR free (trivial). It can be solved in the large $N$ limit. Finally in dimension $d=1$ it reduces to the massless Schwinger model which can be solved exactly even at finite temperature \rFTSch, because bosonization methods still work.
\medskip
{\it The temporal gauge.} To calculate the partition function we first quantize in the temporal gauge $A_0(x,t)=0$ because the corresponding hamiltonian formalism is simple. The action becomes
$$ {\cal S} \left(\bar \psib ,\psib,A _{\mu} \right) = \int \d^dx\d t
\left[{\textstyle{ 1 \over 4e^2}}\left(2\dot A_i^2+  F ^{2} _{ij}(t,x)
\right)- \bar \psib (t,x) \cdot\left(\sla{\partial} + i\Abar \right) 
\psib(t,x) \right] . \eqnd\eactQEDAze $$
In calculating $\tr\e^{-L H}$ we have to remember  that Gauss's law has still to be imposed. This means that the trace has to be taken in the subspace of wave functionals invariant under space-dependent gauge transformations. To project onto this subspace  we impose periodic conditions in the time direction up to a gauge
transformation:
$$\eqalign{A_i(L,x)&=A_i(0,x)-L\partial_i \varphi(x), \cr
\psib(L,x)&=\e^{i L \varphi(x)}\psi(0,x),\cr} 
$$
and integrate over the gauge transformation $ \varphi(x)$. We then set
$$\eqalign{A_i(t,x)&=A_i'(t,x)-t\partial_i \varphi(x), \cr
\psib(t,x)&=\e^{i t \varphi(x)}\psi'(t,x).\cr} $$
where the fields $A'_i$, $\psib',\bar\psib'$ now are periodic and anti-periodic resp..
This induces two modifications in the action 
$$\eqalign{\int\d x\d t\,(\partial_t A_i)^2\ &\mapsto\ \int\d x\d
t\,(\partial_t A_i)^2 + L\int\d x\,(\partial_i \varphi(x))^2 \cr
\int\d x\d t\,\bar \psib (t,x)\gamma_0 \partial_ t \psib(t,x) &
\mapsto \int\d x\d t\,\bar \psib (t,x)\gamma_0( \partial_ t +i\varphi
(x))\psib(t,x) . \cr}$$
Therefore $\varphi(x)$ is simply the residual zero mode of the $A_0$
component
$$\varphi(x)\equiv B_0(x). $$
Its presence is a direct consequence of enforcing Gauss's law.\par
The field theory has a $d$-dimensional gauge invariance with the zero mode $B_i(x)$ of $A_i(t,x)$ as gauge field. In addition it contains $d$ families of neutral vector fields with masses $2\pi n/L$, $n\ne 0$, quantized in a unitary, and thus non-renormalizable gauge.  
\par
Note that from the technical point of view the usual difficulties which appear in perturbation calculations with the temporal gauge (the gauge field propagator is singular) reduce to the need for quantizing the remaining zero-mode, and to the non explicit renormalizability. The latter problem can be solved with the help of dimensional regularization for example (for gauge invariant observables). An alternative possibility of course is to introduce a renormalizable gauge.
The change of gauges can be performed by the standard zero-temperature method \rbook. 
\medskip
{\it Covariant gauge.}  To change to the covariant gauge we introduce a time component $ A_0$ for the gauge field (periodic in time) and multiply the functional measure by the corresponding $\delta$-function
$$1=\int[\d Q_0(t,x)]\prod_{t,x}\delta (Q_0). $$
The action can then be written in a gauge invariant form. 
We then introduce a second identity in the functional integral
$$1=\det(-\partial^2)\int[\d\varphi_1]\delta\bigl(\partial _\mu Q_\mu +\partial^2\varphi_1-h(t,x)\bigr),
\eqnd\eFTFPa $$
where $\varphi_1$ and $h(t,x)$ are two periodic functions without zero-mode.   We  perform the gauge transformation 
$$Q_\mu +\partial_\mu \varphi_1\mapsto  Q_\mu \,.$$
The $\varphi_1 $ dependence remains only in $\delta (Q_0-\partial_t \varphi_1)$, and the integration over $\varphi_1$ yields a constant. Integration over $h(t,x)$ with a gaussian weight yields the standard covariant gauge
$${\cal S}_{\rm gauge}={L\over 2\xi}\int\d x\bigl(\partial_i B_i(x)\bigr)^2+
{1\over 2\xi}\int\d t\d x(\partial_\mu Q_\mu)^2\equiv
{1\over 2\xi}\int\d t\d x(\partial_\mu A_\mu)^2. $$
since
$$ \partial_\mu A_\mu=\partial_t Q_0(t,x) +\partial_i B_i(x)
+\partial_i Q_i(t,x ).$$
We conculde that the gauge fixing term is just obtained by substituting the mode decomposition into the 
gauge fixing term  of the zero-temperature action. From the point of view of the $B$ gauge field this corresponds to a quantization in the covariant gauge in $d$ dimensions.
\par
Note that the transformation from the temporal gauge to the covariant gauge generates a  determinant (equation \eFTFPa) which is field-independent, but contributes to the free energy.
\nref\rQECD{P. Arnold and C. Zhai, {\it Phys. Rev.} D51 (1995) 1906.}
\nref\rBIP{J.-P. Blaizot, E. Iancu, R.R. Parwani, {\it Phys. Rev.} D52  (1995) 2543, hep-ph/9504408.}
\nref\rQED{J.O. Andersen, {\it Phys. Rev.} D53 (1996) 7286.}
\subsection  Dimensional reduction

At finite temperature, to generate the effective action for the gauge field zero-modes, we have  to integrate over all fermion modes (anti-periodic boundary conditions) and over the non-zero modes $Q_\mu(t,x)$ of the gauge field.  At leading order one finds a free theory containing a gauge field $B_i$ and a massless scalar $B_0$.
At one-loop order only fermion modes  contribute. Replacing the gauge field $A_\mu$ by its zero mode $B_\mu$ and performing the fermion integration explicitly we find
the effective action 
$${\cal S}_L=L\int\d^d x\left[{1\over 2 e^2}(\partial_i B_0)^2+{1\over 4 e^2}
F_{ij}^2(B)\right] -N \tr \ln \left(\sla{\partial} + i\Bbar \right) , \eqnn $$
where latin indices means space indices. \par
An important issue is the behaviour of induced mass of the time component
$B_0=\varphi$ of the gauge field. We thus first calculate the effective potential for constant $\varphi$.
\medskip
{\it The effective potential.} The effective potential for constant field
$\varphi $ is then given by 
$$V(\varphi)=-\ud N'\sum_n{1\over (2\pi)^d}\int\d^d k\,\ln\left[k^2
+\bigl(\varphi+(2n+1)\pi/L\bigr)^2\right], $$
where $N'=N\tr{\bf 1}$ is the total number of fermion degrees of freedom.\par 
Its evaluation involves the new function 
$$\sum_n \e^{-t[(2n+1)\pi/L+\varphi]^2}\equiv \vartheta _2(4\pi t/L^2;\nu,0)=
{L\over 2\sqrt{\pi t}}\sum_n(-1)^n \cos(nL\varphi)\e^{-n^2 L^2/4t}, $$
where $\nu=\ud+L\varphi/2\pi$, and the general Poisson's formula
\eqns{\eJacobig} has been used. Then
$$V(\varphi)=\ud N'{1\over(4\pi)^{d/2}}\int{\d t \over t^{1+d/2}}
\left[\vartheta _2(4\pi t/L^2;\nu,0) -\vartheta _2(4\pi
t/L^2;1/2,0)\right],\eqnn  $$  
where $V(\varphi)$ has been shifted so as to vanish for $\varphi=0$.
After the usual change of variables $t=L^2 s/4\pi$ we find
$$V(\varphi)=\ud N' L^{-d}\int_{s_0}^\infty{\d s \over s^{1+d/2}}
\left[\vartheta _2(s;\nu,0) -\vartheta _2(s;1/2,0)\right],\eqnn  $$ 
We note 
$$\vartheta _2(s;\nu,0) -\vartheta _2(s;1/2,0)=s^{-1/2}\sum_n(-1)^{n+1}
\left[1-\cos(nL  \varphi)\right]\e^{-n^2 \pi/ s}, $$  
which shows that the $n=0$ term cancels and thus
the potential is UV finite.
This is not too surprising since in the zero temperature, large $L$, limit no
gauge field mass or quartic potential are generated. \par
The potential has an extremum at $\varphi=0$. 
The coefficient of $\varphi^2$ is:
$$\eqalign{V(\varphi)&=\ud L^{2-d}\varphi^2 K_2(d)+O(\varphi^4) \cr
K_2(d)&=N' {8\over
(4\pi)^{d/2+1/2}}\Gamma\bigl((d+1)/2\bigr)\left(2^{d-2}
-1\right)\zeta(d-1) .\cr}$$
The constant $K_2(d)$ is positive showing that $\varphi=0$ is a minimum.
More generally the potential is $$V(\varphi)=K(d)L^{-d}{\cal V}(L\varphi)\,,\quad
K(d)= N'{ 1\over \pi^{d/2+1/2}}\Gamma\bigl((d+1)/2\bigr)
,  $$
where the function ${\cal V}(z)$ is given by the sum
$${\cal V}(z)=\sum_{n=1}(-1)^{n+1}{\bigl(1-\cos(n z)\bigr) \over
n^{d+1}} ={1-\cos z\over 2\Gamma (d+1)}\int_0^\infty \d \tau {\tau^d
\tanh(\tau/2) \over \cosh \tau +\cos z}\,  .$$ 
Its derivative has the sign of $\sin z$, it is negative for $-\pi<z<0$ and
positive for $0<z<\pi$. Therefore $z=0$ is the unique minimum in the interval
$-\pi<z<\pi$.\par  
A special case is $d=1$ for which one finds
$${\cal V}'(z)=\ud z\ {\rm for}\ |z|<\pi\quad{\rm and\ thus}\ {\cal
V}(z)=\frac{1}{4} z^2. $$ 
To obtain the mass of $\varphi $ field at this order it is necessary to also
calculate the coefficient of $(\partial_i\varphi )^2$ generated by the
determinant. Expanding to order $p^2$ the Fourier transform of the coefficient of $\varphi\varphi/2$ in ${\cal S}_L$ we find
$${\cal S}_L^{(2)}(p)={L p^2\over e^2}+ N'\sum_n \int {\d^d k  \over (2\pi)^d}
\left[{\omega_n^2-k^2 \over (k^2+\omega_n^2)^2} +{p^2 \over 6}{3k^2-\omega_n^2
\over (k^2+  \omega_n^2)^3}\right]+O(p^4). $$  
The coefficient of $(\partial_i\varphi )^2$ is renormalized and for $d<3$ becomes
$${L\over 2 e^2}(\partial_i\varphi )^2\left[1+N'L^{3-d}K_3(d) e^2\right], $$
with
$$K_3(d)={1\over(4\pi)^{d/2}}{d-1 \over 3}(1-2^{d-4})\Gamma(2-d/2)\zeta(4-d),$$ 
(which vanishes  for $d=1$ as expected from the exact solution of the Schwinger model). 
For $d\ge 3$ one has to add to $K_3(d)$ a cut-off dependent contribution
generated for instance by a Pauli--Villars  regulator field, a fermion of large mass $\Lambda $,  
$$  {1\over 3(4\pi)^{(d+1)/2}}\Gamma\bigl((3-d)/2\bigr)(\Lambda L)^{d-3}.$$ 
These contributions generate in dimensions $d\ge 3$ a renormalization
$e\mapsto e_\r$ of the coupling constant. 
\smallskip
{\it Discussion.} We thus obtain a mass term which is
proportional to $e_\r L^{(1-d)/2}$. If $e$ is generic, i.e.~of order 1 at the microscopic scale $1/\Lambda $, then $e\propto \Lambda^{(3-d)/2}$ and the scalar mass is proportional to $(\Lambda L)^{(3-d)/2}/L$. It is
thus large with respect to the vector masses for $d<3$ and small for $d>3$.
We conclude that for $d<3$ the scalar field can be
completely integrated out, but for $d> 3$ it survives. Note that for $d>3$
additional corrections are even smaller because UV divergences cannot
compensate the small coupling constant.  For $d=3$  QED is IR free, 
$$\beta_{e^2}={N\over 6\pi^2} e^4 +O(e^6),$$
$e_\r$ has to be replaced by the effective coupling constant $e(1/\Lambda L)$ with is logarithmically small 
$$e^2(1/\Lambda L)\sim {6\pi^2\over N\ln(\Lambda L)}, $$
and the scalar mass thus is still small
$$m^2_\varphi\propto {1\over L^2\ln(\Lambda L)}.$$
The separation between zero and non-zero modes remains justified. The situation is completely analogous to high temperature $\phi^4$ field theory, and perturbation for the same reason remains applicable.\par
Finally if one is interested  in IR physics only, one can in a second step integrate over the massive scalar field $\varphi$. 
\par
For more details and more systematic QED calculations see
\refs{\rQECD,\rBIP,\rQED}. 
\nref\rFKRS{K. Farakos, K. Kajantie, K. Rummukainen, M. Shaposhnikov, {\it Nucl. Phys.} B425 (1994) 67, hep-ph/9404201, hep-ph/9704416.} 
\nref\rAlHiggs{
A.D. Linde {\it Rep. Prog. Phys.} 42 (1979) 389\semi
A. Hebecker, {\it Z. Phys.} C60 (1993) 271\semi
J.O. Andersen, {\it Phys. Rev.} D59 (1999) 65015, hep-ph/9709418.}
\subsection The abelian Higgs model

The abelian gauge fields interacting with charged scalar fields has also been investigated  \refs{\rFKRS,\rAlHiggs,\rBIP} as  a toy model to study properties of the electro-weak phase transition at finite temperature. The gauge action reads
$${\cal S}(A_\mu, \phi)=\int\d t\d^d x\left[{1\over 4 e^2}F_{\mu\nu}^2+ |{\rm D}_\mu\phi|^2+ U(|\phi|^2)\right], \eqnn $$
where the quartic potential $U(|\phi|^2)$ 
$$U(z)=rz +\frac{1}{6} g z^2, \eqnd\eFTUabHig $$
is such that the $U(1)$ symmetry is broken at zero temperature. \par
The model can directly be quantized in the unitary (non-renormalizable) gauge and calculations of gauge-independent observables can be performed with dimensional regularization. Below we use instead the temporal gauge because the unitary gauge becomes singular near the phase transition. \par
One limitation of the model is that RG shows that in $3+1$ dimensions the hypothesis of second order phase transition is inconsistent, and therefore the transition is most likely first order. Indeed, in a more general model with $N$ charged scalars  for $d=3$ the RG $\beta $-functions are
$$\eqalign{\beta_g&={1\over 24\pi^2}\left[(N+4)g^2-18ge^2+54 e^4\right] \cr
\beta_{e2}&= {1\over 24\pi^2}Ne^4.\cr} $$
The origin $e^2=g=0$ is a stable IR fixed point only for $N\ge 183$. For $N$ small, 
the continuum model remains meaningful if initially the coupling constants are small enough, in such a way that by the time the running coupling contants reach the physical scale, they have not yet reached the region of instability. The transition then is weak first order.\par
Presumably the same result applies for small values of $N$ to the three-dimen\-sional classical statistical field theory, which is also the Landau--Ginzburg model of superconductivity. \par
Neglecting all non-zero modes we obtain one massive vector field degenerated 
in mass with a scalar field, and the Higgs field. We expect the degeneracy between vector and scalar masses to be lifted by the integration over non-zero modes. 
To construct the reduced action we quantize in the temporal gauge $A_0=B_0$ and we set
$$A_i =B_i + Q_i\,,\quad \phi=\varphi+\chi\,.$$
The reduced action at leading order is simply
$${\cal S}_L(B_\mu, \varphi)=L\int\d^d x\left[{1\over 2e^2}(\partial_i B_0)^2+ {1\over 4 e^2}F_{ij}^2(B)+ |{\rm D}_i\varphi|^2+|\varphi|^2 B_0^2+U(|\varphi|^2)\right], \eqnn $$
where the covariant derivative now refers to the gauge field $B_i$
$${\rm D}_i=\partial_i +iB_i\,.$$ 
At one-loop order we need the terms quadratic in $Q_\mu,\chi$. 
In the gaussian integration over $Q_\mu,\chi $ at high temperature the leading effects come from the shift in the masses.   We can therefore take $B_0$ and $\varphi$ constant.
The quadratic terms in the action relevant for the $\varphi$ mass shift are
$${\cal S}_2=\int\d t\d^d x\left[ {1\over 2e^2}(\partial_t Q_i)^2+ {1\over 4 e^2}F_{ij}^2(Q)+ |\partial_t\chi|^2+ |\partial_i \chi +iQ_i\varphi|^2 
+ \frac{2}{3}g|\varphi|^2|\chi|^2 \right], $$
where we have omitted the term proportional to $r|\chi|^2$, a high temperature approximation. The integration over $\chi$ yields a term
$$e^2|\varphi|^2\sum_\omega \int\d^d k \,Q_i(\omega ,k)\left(\delta_{ij}-{k_i k_j\over k^2+\omega^2}  \right)Q_j(-\omega ,-k)\,, $$
and thus finally a contribution $\delta r$ to the coefficient of $|\varphi|^2$
$$\delta r=G_2(0,L)(d e^2+2g/3)/L\,,$$
where $G_2$ is defined by equation \eGinttwo. For $d=3$ we have found $G_2=1/12L$ 
(section \ssFTfiviii) and therefore
$$\delta r={1\over 36 L^2}(9e^2+2g).$$
For the contribution to the $B_0$ mass the relevant quadratic action is
$${\cal S}_2=\int\d t\d^d x  |{\rm D}_\mu \chi|^2 . $$
In the limit of constant $B_\mu$, the space components $B_i$ can be eliminated by a gauge transformation. The remaining $B_0$ component cannot be eliminated  because 
$\chi$ satisfies periodic boundary conditions in the time direction. Instead, as we have already discussed, the mode integration generates a periodic potential for $B_0$
$$\int\d^d x\sum_\omega \int\d^d k\,\ln\left[\bigl(\omega +B_0(x)\bigr)^2+k^2\right]
.$$
Expanding to order $B_0^2$ we obtain the mass term
$$(d-1)G_2\int\d^d x\, B_0^2(x), $$
where the identity, valid  in dimensional regularization, $${1\over (2\pi)^d}\sum_{\omega }\int{\d^d k\, \omega^2 \over (k^2+\omega ^2)^2}=(1-d/2)G_2\,,$$
has been used.
Thus for $d=3$
$$m^2_{B_0}={e^2\over 3 L^2}.$$
As we have discussed several times, in three dimensions additional UV contributions transform the parameters $e^2, g,r$ into the one-loop expansion of the running parameters at scale $1/\Lambda L$.  Finally, for completeness, let us point out that the coefficient of $|\varphi|^2B_0^2$ gets renormalized. At one-loop one finds
$$1\mapsto 1+{e^2+g\over 12\pi^2}.$$ 
\smallskip
{\it Discussion.} We assume that at zero temperature the $U(1)$ symmetry is broken which implies that the coefficient $r$ in the potential $U$ (equation \eFTUabHig) is sufficiently negative.  When the temperature increases, the coefficient of $|\varphi^2|$ increases until  a critical temperature is reached where the $U(1)$ symmetry is restored. Near the transition the scalar field $B_0$ is massive and therefore the effective theory relevant for the phase transition is simply the classical $U(1)$ gauge model
$$\tilde {\cal S}_L(B_i, \varphi)=L\int\d^d x\left[{1\over 4 \tilde e^2}F_{ij}^2+ |{\rm D}_i\varphi|^2+ \tilde U(|\varphi|^2)\right], \eqnn $$
where the parameters $\tilde e^2$ and in $\tilde U$ can be obtained by integrating the reduced action also over the heavy field $B_0$.
\nref\rQCD{C. Zhai and B. Kastening, {\it Phys. Rev.} D52 (1995) 7232.}
\nref\rNAbHiggs{Z. Fodor and A. Hebecker, {\it Nucl. Phys.} B432 (1994) 127\semi
P. Arnold and O. Espinosa, {\it Phys. Rev.} D47 (1993) 3546.}
\nref\rABS{ J.P. Blaizot, E. Iancu, A. Rebhan, {\it Phys. Rev. Lett.} 83  (1999) 2906, hep-ph/9906340; {\it Phys. Lett.} B470 (1999) 181, hep-ph/9910309\semi 
J.O. Andersen, E. Braaten, M. Strickland. {\it Phys. Rev.} D61 (2000) 74016, 
hep-ph/9908323; hep-ph/9905337.} 
\section Non-abelian gauge theories

Pure non-abelian gauge theories, as well as non-abelian gauge coupled to a
small number of fermions, are UV asymptotically free in four dimensions. From the
RG point of view, we expect some similarities with the non-linear $\sigma $ model
(in different dimensions).    
In particular in $d=3$ dimensions the effective coupling constant $g(L)$ decreases at high temperature, $g(L)\propto 1/\ln(mL)$, where $m$ is the RG invariant mass scale of
the gauge theory. \sslbl\ssFTnag\par  
\medskip
{\it Notation.}
We consider fields ${\bf A}_{\mu}$ written as matrices in the adjoint
representation of some compact group $G$. 
Gauge transformations take the form
$${\bf A}'_{\mu}(x) = {\bf g}(x) {\bf A}_{\mu}(x) {\bf
g}^{-1}(x) + {\bf g}(x) \partial_{\mu} {\bf g}^{-1}(x), \eqnd\eAmutr $$
where $\bf g$ is a group element.\par
Covariant derivatives $ {\bf D}_{\mu}$ acting on fields $\varphib$ belonging the adjoint
representation are 
$$ {\bf D}_{\mu}\varphib = \partial_{\mu}\varphib  + [{\bf A}_{\mu},\varphib]
\,. \eqnd\eAdcovder 
$$
The corresponding curvature $ {\bf F}_{\mu\nu}(x) $ tensor is
$$ {\bf F}_{\mu\nu}(x) = \left[ {\bf D}_{\mu},{\bf
D}_{\nu}\right] = \partial_{\mu}{\bf  A}_{\nu} -  \partial_{\nu}{\bf
A}_{\mu}  + \left[ {\bf A}_{\mu},{\bf A}_{\nu}\right], \eqnn $$
and the pure gauge action
$${\cal S}({\bf A})=-{1\over 4g^2}\tr\int\d^{d+1} x\,{\bf F}_{\mu \nu
}^2.\eqnd\eFTSnag $$ 
\medskip
{\it Temporal gauge.}
The situation is here more complicated because the mode decomposition is not gauge invariant. Thus we first quantize, choosing the temporal gauge. Then the space
components $A_i$ must be periodic up a gauge transformation which again  enforces Gauss's law:
$${\bf A}_i(L,x) = {\bf g}(x) {\bf A}_i(0,x) {\bf
g}^{-1}(x) + {\bf g}(x) \partial_i  {\bf g}^{-1}(x),  $$
We parametrize the group element $\bf g$ in terms of an element $\varphib$ of the Lie algebra:
$${\bf g}(x)=\e^{L\varphib(x)}, $$
and introduce
$${\bf g}(t,x)=\e^{t\varphib(x)}, $$
After the gauge transformation
$${\bf A}_i(t,x)={\bf g}(t,x){\bf A}'_i(t,x){\bf g}^{-1}(t,x)+{\bf g}(t,x)
\partial_i {\bf g}^{-1}(t,x) , $$ 
the new field ${\bf A}'_i$ becomes strictly periodic. The gauge component ${\bf
A}_0$ which vanishes before the transformation becomes 
$$0={\bf g}(t,x) {\bf A}'_0(t,x) {\bf g}^{-1}(t,x)+{\bf g}(t,x) \partial_t
{\bf g}^{-1}(t,x) , $$ 
and therefore
$${\bf A}'_0(t,x) =\varphib(x).$$
Again the temporal gauge reduces the time-component ${\bf A}_0$ to its zero-mode,
the field $\varphib(x)$. \par 
Let us express the pure gauge action \eFTSnag~ in terms of the new fields 
$$\eqalign{{\cal S}({\bf A},\varphib )&=-{1\over2 g^2} \tr\int\d t\d^d
x\,\bigl(\partial_i\varphib(x) -\partial_t {\bf A}_i+[{\bf A}_i,\varphib]
\bigr)^2 -{1\over 4g^2}\tr\int\d t\d^{d} x\,{\bf F}_{ij}^2 \cr  
& =-{1\over2 g^2}\tr\int\d t\d^d x\,({\bf D}_i \varphib -\partial_t {\bf
A}_i)^2 -{1\over 4g^2}\tr\int\d t\d^{d} x\,{\bf F}_{ij}^2 .\cr }$$
\medskip
{\it Dimensional reduction.} 
We now separate the zero-modes of the space components of the gauge field
$${\bf A}_i(t,x)={\bf B}_i(x)+{\bf Q}_i(t,x), $$
with
$${\bf Q}_i(t,x)=\sum_{n\ne 0}\e^{2i\pi nt/L}{\bf Q}_{n,i}(x).$$
Then
$${\bf D}_i (A)\varphib= {\bf D}_i (B)\varphib+[{\bf Q}_i,\varphib]. $$ 
In the same way
$${\bf F}_{ij}(A)={\bf F}_{ij}(B)+{\bf D}_i {\bf Q}_j-{\bf D}_j {\bf
Q}_i+[{\bf Q}_i,{\bf Q}_j],$$ 
where the covariant derivative now refers to the gauge field ${\bf B}$.\par
We see that the resulting action is gauge invariant with respect to time-independent gauge  transformations with gauge field ${\bf B}$. The gauge field
is coupled to a massless scalar $\varphib$  and massive vector fields with
masses $4\pi^2n^2/L^2$, all belonging to the adjoint representation. \par  
The problem of quantization then reduces to the quantization of the field
${\bf B}$. If we choose a covariant gauge then, not only is the theory
quantized, but the vector fields do not have the kind of singular propagator
typical of the temporal gauge. A question remains, massive vector fields lead
to non-renormalizable theories.    
A way to solve this problem is to go over to a covariant gauge.
We introduce a time component ${\bf A}_0$ for the gauge field (periodic in
time) and multiply the functional measure by the corresponding
$\delta$-function. The action is a function only of the sum
${\bf A}_0(t,x)+\varphib(x)$. Let us thus temporarily call $\tilde {\bf
A}_\mu$ the field 
$$\tilde{\bf A}_i={\bf A}_i\,, \quad \tilde{\bf A}_0={\bf
A}_0(t,x)+\varphib(x).$$ 
The $\delta$-function then becomes $\delta(\tilde{\bf A}_0-\varphib)$.
We then perform the standard manipulations to pass to the covariant gauge   
with gauge function $\partial_\mu \tilde{\bf A}_\mu$. Eventually we shall get
the gauge average of the constraint $\delta(\tilde{\bf A}_0-\varphib)$. This
gives a determinant which only depends on $\varphib$, while $\varphib$ appears
nowhere else. The integral over $\varphib$ factorizes and gives a constant
factor. Of course in the process we have introduced ghost fields
which satisfy periodic boundary conditions, unlike ordinary fermions.
\medskip
{\it One-loop calculation of the effective $\varphi$ potential.} We expect
that quantum corrections generated by the integration over non-zero modes give
a mass to the scalar field $\varphib$,  as in the abelian example.\par  
For $\varphib$ constant, and omitting the massless gauge field we find a
simplified action. 
It is convenient here to introduce the generators of the Lie algebra of the
compact group $G$, in the form of hermitian matrices $\tau^\alpha $ 
$$\tr \tau^\alpha \tau^\beta =\delta_{\alpha \beta}\,,\quad[ \tau^\alpha,
\tau^\beta]=i f_{\alpha \beta \gamma }  \tau^\gamma \,,$$ 
where the structure constants $f_{\alpha \beta \gamma }$ are chosen
antisymmetric. We then set
$${\bf Q}_i=i Q^\alpha_i \tau^\alpha \,,\quad \varphib=i \varphi^\alpha
\tau^\alpha .$$  
Then the relevant $Q$ action is
$${\cal S}_2(Q)={1\over2 g^2}\int\d t \d^d x\left[\ud\left(\partial_i Q^\alpha_j
-\partial_j Q^\alpha_i\right)^2 
+\left(\partial_t Q^\alpha_i +f_{\alpha \beta \gamma }Q^\beta_i \varphi^\gamma
\right)^2 \right].     $$ 
The integration yields a determinant which generates an additive contribution
to the effective action 
$$ \eqalignno{ V(\varphi )&=\ud \sum_{n\ne 0} \tr\ln
\left[\left(k^2\delta_{ij}-k_ik_j +\omega_n^2 
\delta_{ij} \right)\delta_{\alpha \beta }+\delta_{ij}\left(2i\omega_n
f_{\alpha \beta \gamma }\varphi^\gamma  + 
f_{\alpha \gamma \delta }f_{\beta \epsilon \delta }\varphi ^\gamma \varphi
^\epsilon \right)\right] \cr 
&\quad-(\varphi =0), &\eqnn \cr}$$ 
with $\omega_n=2\pi n/L$. The result can be written as the sum of two
contributions, along $k$ and transverse. 
The first contribution is
$$\eqalignno{V_1(\varphi )&=\ud\sum_x \sum_{n\ne 0} \tr\ln \left[\omega_n^2
\delta_{\alpha \beta }+\left(2i\omega_n f_{\alpha \beta \gamma }\varphi^\gamma
+f_{\alpha \gamma \delta }f_{\beta \epsilon \delta }\varphi ^\gamma \varphi
^\epsilon \right)\right]-(\varphi =0) \cr     
& =\sum_x \sum_{n\ne 0} \tr\ln \left(\delta_{\alpha \beta }+if_{\alpha \beta
\gamma }\varphi^\gamma / \omega_n \right). \cr 
&=\ln\det\left[\prod_x \,2\Phi^{-1}L^{-1}\sinh( L\Phi/2)\right],&
\eqnd\eFTgrmeas \cr}$$ 
where we have introduced the matrix $\Phi$ with elements
$$\Phi^{\alpha \beta }=f_{\alpha \beta \gamma }\varphi^\gamma .$$ 
This term contributes to the $\varphi(x)$ integration measure
and yields a factor at each point $x$
$$\prod_x \d\varphi(x) \det{L\Phi(x) /2  \over \sinh\bigl(L\Phi(x)
/2\bigr)},$$ 
which cancels the invariant group measure in the $\varphi$ parametrization
(see Appendix \label{\appGrmea}).\par 
The second term, after division by the space volume, then reads
$$V_2(\varphi )=\ud(d-1)\sum_{n\ne 0}{1\over(2\pi)^d}\int\d^d
k\ln\det\left[k^2\delta_{\alpha \beta }+(\omega_n +i\Phi)^2_{\alpha \beta }
\right] -(\varphi =0).$$   
Using Schwinger's parametrization we obtain
$$V_2(\varphi )=\ud(d-1){1\over(4\pi)^{d/2}}\sum_{n\ne 0}\int_0^\infty {\d t
\over t^{1+d/2}}\tr \left[\e^{-\omega_n^2t}-\e^{(\omega_n+i\Phi)^2t}\right],$$
and therefore
$$\eqalignno{V_2(\varphi )&=-\ud(d-1){1\over(4\pi)^{d/2}}\int_0^\infty {\d t
\over t^{1+d/2}}\tr \Bigl[\vartheta_2 (4\pi t/L^2;\nu,0) -\vartheta_2 (4\pi
t/L^2;0,0) \cr  &\left.\quad -\e^{t\Phi^2}+1\right],&\eqnn  \cr}$$ 
where $\nu$  now is a matrix $\nu=iL\Phi/2\pi$. We change variables $4\pi
t/L^2=s$ and find 
$$V_2(\varphi )=-\ud(d-1)L^{-d}\int_0^\infty {\d s \over s^{1+d/2}}\tr
\left[\vartheta_2 (s;\nu,0) -\vartheta_2 (s;0,0) -\e^{L^2\Phi^2
s/4\pi}+1\right].\eqnn  $$ 
The end of the calculation is somewhat similar to the abelian case. We
complete the calculation for the $SU(2)$ group and obtain 
$$V_2(\varphi)=-(d-1) L^{-d}\int{\d s \over s^{1+d/2}}
\left[\vartheta_2 (s;\nu,0) -\vartheta_2
(s;0,0)-\e^{-sL^2\varphi^2/4\pi}+1\right],\eqnn  $$  
where $\varphi$ now is a three component vector and $\nu=L\varphi/2\pi$. Using
the Poisson formula \ePoissonfer,  
$$\eqalign{\vartheta_2 (s;\nu,0) -\vartheta_2 (s;0,0)&={1\over\sqrt{s}}\sum_n
\left(\e^{2i\pi \nu n}-1\right)\e^{-\pi n^2/s} \cr 
&=-{1\over\sqrt{s}}\sum_n \bigl(1-\cos(2\pi\nu)\bigr)\e^{-\pi n^2/s} , \cr} $$
we can rewrite the expression
$$\eqalign{V_2(\varphi)&=(d-1)L^{-d}\int{\d s \over s^{3/2+d/2}}\sum_n\e^{-\pi
n^2/s}\left[1-\cos(nL\varphi)\right]+{(d-1) \Gamma(-d/2)
\over(4\pi)^{d/2}}\varphi^d \cr  
&=2(d-1)L^{-d}{\Gamma\bigl((d+1)/2\bigr)\over \pi^{(d+1)/2}}  
\sum_{n=1}{1-\cos(nL\varphi)\over n^{d+1}}+{(d-1) \Gamma(-d/2)
\over(4\pi)^{d/2}}\varphi^d .\cr } $$
where dimensional regularization has been used. The last term, which comes
from the zero mode, is there to provide a counter-term to the $\varphi $
four-point function for $d=4$. The sum can be replaced by another integral   
$$\eqalign{\sum_{n=1}{1-\cos(nL\varphi)\over n^{d+1}}&={1\over2\Gamma (d+1)}
\int_0\d s\, s^d \left({2\over e^{s}-1}-{1\over \e^{s+iL\varphi}-1}-{1\over
\e^{s-iL\varphi}-1} \right) \cr
&={1-\cos(L\varphi) \over2\Gamma (d+1)}\int_0{\d s \, s^d\over
\tanh(s/2)\left[\cosh s-\cos(L\varphi)\right]}.  \cr} 
$$
\par
As in the QED case $\varphi=0$ is the minimum of the potential (which
is also periodic in $\varphi$), and the generated mass is UV finite
$$V_2(\varphi)=(d-1){\Gamma\bigl((d+1)/2\bigr) \over \pi^{(d+1)/2}}\zeta(d-1)
L^{2-d}\varphi^2+O(\varphi^4), $$ 
because gauge invariance ensures the absence of mass terms for gauge
fields.\par
For $d>3$ again the mass is small. For $d<3 $ it is large and the
scalar field can be integrated out. For $d=3$ the situation is different
from the QED case because the theory is UV asymptotically free. We expect a
situation similar to the non-linear $\sigma $ model in two dimensions: the
effective coupling constant at high temperature is logarithmically  small,  
$g(L)\propto 1/\ln(mL)$, $m$ being the RG invariant mass scale of the gauge theory
(related to the $\beta$-function). Thus
we can trust the effective reduced field theory. However the effective theory  most likely
cannot  be solved by perturbative methods.
\smallskip
{\it Remarks} Detailed calculations have been performed for models of physical interest like QCD \refs{\rQECD,\rQCD} with the problem of the quark-gluon plasma phase and the Higgs sector of the Standard Model with the question of the $SU(2)\times U(1)$ symmetry restoration \rNAbHiggs. In QCD the problem of slow convergence also arises and various summation schemes have been proposed \rABS.
\bigskip
{\bf Acknowledgments.} The hospitality of the Center for Theoretical Physics
at MIT, where a first draft of these notes has been written, is gratefully
acknowledged. The author has also benefited from useful conversations with
U.J.~Wiese, J.P.~Blaizot and R.~Guida.   
\vfill\eject
\appendix{Feynman diagrams at finite temperature}

In this appendix we summarize a few definitions and identities useful
for general one-loop calculations. But first a short section is devoted to a
few reminders on group measures, useful for gauge theories. 
\section One-loop calculations

We give here some technical details about explicit calculations of one-loop
diagrams. 
 \sslbl\appFTiloop 
\subsection One-loop diagrams: general remarks

Let us add a few remarks concerning the calculation of Feynman diagrams
in finite temperature field theory. In the review we have used techniques
which have been developed for the more general finite size problems.  
It is based on the introduction of Schwinger's parameters
and write the momentum space propagator $\Delta(p)$, 
$$\Delta(p)={1\over p^2+\mu^2}=\int_0^\infty\d t\,\e^{-t(p^2+\mu^2)},$$
a method also used at zero temperature.
After this transformation some zero temperature gaussian integrals over
momenta are replaced by discrete sum over integers which can no longer
be calculated exactly. However, dimensional continuation can still be defined.
In the review we have mostly considered simple one-loop diagrams of the form  
$D_\gamma$ which can be written$$\eqalign{D_\gamma&\equiv {1\over (2\pi)^d
L}\sum_n\int{\d^d p \over \left(p^2+\omega_n^2+\mu^2\right)^\sigma}\cr & =   
{1\over (2\pi)^d \Gamma(\sigma)}{1\over L}\int_0^\infty\d t\,
t^{\sigma-1}\sum_n\int\d^d p\, \e^{-t(p^2+\omega_n^2+\mu^2)}, \cr}$$
with $\omega_n=2\pi n/L$.\par  
In terms of the function $\vartheta_0 (s)$ defined by \eqns{\eJacobi},  
the sums can be written 
$$\eqalign{D_\gamma&={1\over(4\pi)^{d/2}\Gamma(\sigma)}{1\over L}\int_0^\infty\d t\,
t^{\sigma-d/2-1}\e^{-t\mu^2}\vartheta_0 (4t \pi /L^2)\cr
&={1\over(4\pi)^\sigma \Gamma(\sigma)}L^{2\sigma -d-1}\int_0^\infty\d s\,s^{\sigma -d/2-1}
\e^{-s\mu^2 L^2/4\pi}\vartheta_0(s).\cr}$$  
The identity (equation \eqns{\ePoisson})
$$\vartheta_0(s) = (1/s)^{1/2} \vartheta_0 \left(1/s\right) $$
shows, in particular, that the zero temperature limit is approached
exponentially when the mass $\mu$ is finite. Indeed from
$$\vartheta_0(s)-{1\over 2\pi}\int\d \omega \,\e^{-\pi s \omega ^2}\sim 2 s^{-1/2}\e^{-\pi /s}$$
one concludes
$$ D_\gamma(L)-D_\gamma(L=\infty) \mathop{\sim}_{L\to\infty}{L^{2\sigma -d-1}
\over(4\pi)^{\sigma-1} \Gamma(\sigma)}{\e^{-\mu L} \over L\mu}\,. $$
\medskip
{\it Other analytic techniques.} We just mention here the more traditional and
more specific techniques also available in finite temperature quantum field
theory. The idea is the following: in the mixed $d$-momentum, time
representation  the propagator is the two-point function $\Delta(t,p)$ of the
harmonic oscillator with frequency $\omega (p)=\sqrt{p^2+m^2}$ and time
interval $L$:  
$${1\over p^2+\omega^2+m^2}\mapsto  \Delta(t,p)={1\over 2 \omega(p)}
{\cosh\left[(L/2-|t|)\omega (p)\right]\over \sinh\bigl(L\omega (p)/2\bigr)}.
$$ 
Summing over all frequencies is equivalent to set $t=0$. For the simple
one-loop diagram one finds 
$${1\over 2\pi }\int{\d\omega \over \omega^2 +p^2+m^2}\mapsto {1\over 2
\omega(p)} {\cosh\bigl(\omega (p)L/2)\bigr)\over \sinh\bigl(\omega
(p)L/2\bigr)}, $$ 
This again allows a simple separation into IR and UV contributions:
$${1\over 2\omega(p)\tanh(L\omega(p) /2)}={1\over 2\omega (p)}+{1\over \omega
(p)(1-\e^{-L\omega (p)})}, $$ 
where the first term is the zero-temperature result, and the second term,
which involves the relativistic Bose statistical factor,  decreases
exponentially at large momentum.\par  
Finally we calculate the massless propagator, with excluded zero-mode.
\medskip
{\it The massless propagator.} The Fourier representation 
$$\Delta(t,x)={1\over (2\pi)^d L}\sum_{n\ne 0}\int{\d^d p\e^{-ipx-i\omega_n t}
\over p^2+\omega_n^2}, \eqnn $$
with $\omega_n=2\pi n/L$ can be transformed into an infinite sum 
$$\eqalignno{\Delta(t,x)&={\Gamma\bigl((d-1)/2\bigr)\over 4 \pi^{(d+1)/2}}
\sum_n\left[x^2+(t+nL)^2\right]^{(1-d)/2} \cr
&\quad-{\Gamma(d/2-1)\over L 4 \pi^{d/2}}x^{2-d}.&\eqnd\ePropLze \cr}$$
\subsection  Some one-loop calculations

Many one-loop results which have been used  can be derived from the simple
integral \sslbl\appthree  
$$G_2(r,L)={L^{2-d}\over 4\pi}\int_{s_0}^\infty{\d s\over
s^{d/2}}\e^{-rL^2s/4\pi} \left(\vartheta_0(s)-1\right), \eqnn  $$
with $s_0=4\pi /L^2\Lambda^2$. We first note that the function for dimension
$d$ is related to the function  for the dimension $d-2$. 
$${\partial G_2(d)\over \partial r}=-G_4(d)=-{L^2 \over 4\pi}G_2(d-2). $$
The expansion of the function for $r$ small has been several times needed. We
first calculate the value at $r=0$ 
$$G_2(0,L)= {L^{2-d}\over 4\pi}\int_{s_0}^\infty{\d s\over
s^{d/2}}\left(\vartheta_0(s)-1\right). \eqnn  $$
A method which can be used quite often in these one-loop calculations is the following: one
calculates explicitly the integral in dimensions $d$ in which it converges at $s=0$ and subtracts the integral on $[0,s_0]$. One then proceeds by analytic continuation
$$\eqalign{G_2&= {L^{2-d}\over 4\pi}\int_0^\infty{\d s\over 
s^{d/2}}\left(\vartheta_0(s)-1\right)- {L^{2-d}\over 4\pi}\int_0^{s_0}{\d
s\over s^{d/2}}\left(\vartheta_0(s)-1\right) \cr
&=\ud \pi^{d/2-2}L^{2-d}\Gamma(1-d/2)\zeta(2-d) +{2L\Lambda^{d-1}\over
(d-1)(4\pi)^{(d+1)/2}} -{2\Lambda^{d-2}\over(d-2)(4\pi)^{d/2}}. \cr}$$ 
where the first integral is given by equation \eqns{\eFTintoa} and in the
second, for small $s$, 
$\vartheta_0(s)$ has been replaced by $1/\sqrt{s}$.
Note that from the identity \eqns{\ezetaref} the constant can also be written
$$\ud \pi^{d/2-2}\Gamma(1-d/2)\zeta(2-d)=\ud
\pi^{-(d+1)/2}\Gamma\bigl((d-1)/2\bigr) \zeta(d-1)=N_d\Gamma(d-1)\zeta(d-1)  .
$$ 
Additional terms in the small $r$ expansion can be obtained by Mellin
transform 
$$\eqalign{M( \lambda )&=\int_0^\infty \d r\,r^{\lambda -1}G_2(r,L)\cr
&={L^{2-d}\over 4\pi}\int_{s_0}^\infty{\d s\over
s^{d/2}}\bigl(\vartheta_0(s)-1\bigr) \int_0^\infty \d r\,r^{\lambda
-1}\e^{-rL^2s/4\pi} \cr &=(4\pi)^{\lambda -1}L^{2-d-2\lambda }\Gamma (\lambda
)\int_{s_0}^\infty{\d s\over s^{d/2+\lambda
}}\left(\vartheta_0(s)-1\right).\cr}$$ 
The last integral can then be obtained from $G_2(r,0)$ by replacing $d$ by
$d+2\lambda $. 
$$\eqalign{M(\lambda )/\Gamma (\lambda )&={2^{2\lambda -1}\over
\pi^{(d+1)/2}}\Gamma\bigr(\lambda +(d-1)/2\bigr)\zeta (d-1+2\lambda
)L^{2-d-2\lambda } \cr  
&\quad +{2L\Lambda^{d-1+2\lambda }\over(2\lambda
+d-1)(4\pi)^{(d+1)/2}}-{2\Lambda^{d-2+2\lambda }\over (2\lambda
+d-2)(4\pi)^{d/2}}. \cr} $$ 
The residues of the pole in $\lambda $ of the Mellin transform yield the terms
of the small $r$ expansion 
$$Ar^\beta \mapsto A/(\lambda +\beta ).$$
Also
$$Ar^\beta \ln r\mapsto -A/(\lambda +\beta )^2.$$
\medskip
{\it Fermions.} In the case of the GN model we need two integrals. With the
help of the relations \eqns{\ezetaref,\eFTintoc} one finds 
$$\eqalignno{{\cal G}_4&={L^{4-d}\over (4\pi)^2}\int_{s_0}\d s\,
s^{1-d/2}\vartheta_1(s) \cr 
&=L^{4-d}{\Gamma(2-d/2)\over
8\pi^{4-d/2}}\left(2^{4-d}-1\right)\zeta(4-d)+{1\over d-3}{2L
\Lambda^{d-3}\over (4\pi)^{(d+1)/2} } . &\eqnd \eFTGivf \cr}
$$
The second integral is
$$\eqalignno{{\cal G}_2(0,L)&={L^{2-d}\over 4\pi}\int_{s_0}\d s\,
s^{-d/2}\vartheta_1(s) = {4L^{2-d}\over
(4\pi)^{(d+1)/2}}(1-2^{d-2})\Gamma\bigl((d-1)/2\bigr)\zeta(d-1) \cr &\quad +{1\over
d-1}{2L\Lambda^{d-1}\over (4\pi)^{(d+1)/2}}    .&\eqnd\eFTGiif \cr}
$$  
\section $\Gamma$, $\psi$, $\zeta$, $\theta$-functions: a few useful
identities 

{\it $\Gamma$, $\psi$, $\zeta$-functions.} Two useful identities on the
$\Gamma$ function are \sslbl\appGPZid 
$$\sqrt{\pi}\,\Gamma(2z)=2^{2z-1}\Gamma(z)\Gamma(z+1/2)\,,\quad
\Gamma(z)\Gamma(1-z)\sin(\pi z)=\pi\,.\eqnd\eGammadup $$
They translate into a relations for the $\psi(z)$ function,
$\psi(z)=\Gamma'(z)/\Gamma(z)$
$$2\psi(2z)=2\ln 2+\psi(z)+\psi(z+1/2),\quad \psi(z)-\psi(1-z)+\pi
/\tan(\pi z)=0\, \eqnn $$
We also need Riemann's $\zeta$ function
$$\zeta(s)=\sum_{n \ge 1}{1\over n^s} \,. \eqnd\ezetadef $$
It is useful, for what follows, to remember the reflection formula
$$\zeta(s)\Gamma(s/2)=\pi^{s-1/2}\Gamma\bigl((1-s)/2\bigr)\zeta(1-s),
\eqnd\ezetaref  $$
which can be written in different forms using $\Gamma$-function
relations. Moreover
$$\sum_{n=1}{(-1)^n\over n^s}=(2^{1-s}-1)\zeta(s),\eqnd\ezetamin  $$
Finally ($\gamma =-\psi(1)$)
$$\eqalignno{\zeta(1+\varepsilon)&={1\over\varepsilon}+\gamma,
&\eqnd\ezetaone \cr
\zeta(\varepsilon)&=-\ud(2\pi)^{\varepsilon}+O(\varepsilon^2). &
\eqnd\ezetanul \cr}$$
\medskip
{\it Jacobi's $\theta$-functions.} We define the function $\vartheta_0 (s)$
(related to Jacobi's elliptic functions, see below) 
$$\vartheta_0(s)= \sum^{+ \infty}_{n=- \infty} \e^{-\pi sn^{2}}. \eqnd\eJacobi
$$ 
Poisson's summation formula is useful in this context. Let $f(x)$ be a
function which has a Fourier transform
$$\tilde f(k)=\int\d x\,f(x)\e^{i2\pi kx}.$$
Then from 
$$\sum_{k=-\infty}^{+\infty}\e^{i2\pi kx}=\sum_{l=-\infty}^{+\infty}
\delta(x-l), $$
follows Poisson's formula
$$\sum_{k=-\infty}^{+\infty} \tilde f(k)=\sum_{l=-\infty}^{+\infty} f(l).
\eqnd\ePoissong $$
Applying this relation to the function $\e^{-\pi sx^2}$ one finds the
identity: 
$$\vartheta_0(s) = (1/s)^{1/2} \vartheta_0 \left(1/s\right).\eqnd \ePoisson $$
The two integrals, which are related by the change of variables $s\mapsto
1/s$, are needed 
$$\eqalignno{\int_{0}^\infty\d s\, s^{\alpha/2-1}\left[\vartheta_0
(s)-1\right]&= 
2\pi^{-\alpha /2}\Gamma(\alpha/2)\zeta(\alpha),&\eqnd\eFTintoa \cr
\int_{0}^\infty\d s\, s^{\alpha/2-1}\left[\vartheta_0(s)-1/\sqrt{s}\right]&=
2\pi^{\alpha/2-1/2}\Gamma[(1-\alpha)/2]\zeta(1-\alpha), &\eqnd\eFTintob\cr}
$$
where $\zeta(s)$ is Riemann's $\zeta$-function.
\par
For fermion and gauge theories calculations  we need the more
general function 
$$\vartheta_2(s;\nu,\lambda)=\e^{-\pi s\nu^2}\theta_3(\lambda+i\nu s,\e^{-\pi
s}) =\sum_n\e^{-\pi s(n+\nu)^2+2i\pi n\lambda},\eqnd\eJacobig $$
where $\theta_3$ is an elliptic Jacobi's function, which, applying Poisson's
formula, can be shown to satisfy 
$$\vartheta_2(s;\nu,\lambda)=\vartheta_2(s;-\nu,-\lambda)
=s^{-1/2}\vartheta_2(1 /s;\lambda,-\nu).\eqnd\eJacobdual $$
For fermions the relevant function is $\vartheta_1(s)$
$$\vartheta_1 (s)\equiv \vartheta_2(s;1/2,0)=\sum_n\e^{-(n+1/2)^2\pi
s}.\eqnd\eAellfer  
 $$
From \eqns{\eJacobdual} we obtain
$$\vartheta_1 (s)={1\over \sqrt{s}} \sum_n(-1)^n\e^{-\pi n^2/s}.
\eqnd\ePoissonfer $$  
Finally $(\alpha >1$)
$$\int_{0}^\infty\d s\, s^{\alpha/2-1}\vartheta_1(s)=
2(2^{\alpha }-1)\pi^{-\alpha/2 }\Gamma(\alpha/2)\zeta(\alpha)\eqnd\eFTintoc $$
\def\xib{\xi}
\section Group measure

For the example of non-abelian gauge theories we derive the form of the group
measure in the representation of group elements as exponentials of elements of
the Lie algebra. The notation and conventions are the same as in section
\ssFTnag. \sslbl\appGrmea\par 
We set 
$${\bf g}=\e^{\xib}, $$
and we will determine the invariant measure in terms of the components
$\xi^\alpha $ 
$$\xib=i\tau^\alpha \xi^\alpha .$$   
We thus introduce a time dependent group element $\bf g$ as
$${\bf g}(t)=\e^{t\xib} \,, \quad {\bf g}(1)={\bf g}\,.    $$  
We also need the element of the Lie algebra ${\bf L}^\alpha(t) $ 
$${\bf L}^\alpha(t)={\partial {\bf g}(t)\over \partial \xi^\alpha } {\bf
g}^{-1}(t).$$ 
It satisfies the differential equation
$${\d \over \d t}{\bf L}^\alpha=i\tau^\alpha  +[\xib,{\bf L}^\alpha],\quad
{\bf L}^\alpha(0)= 0\,.\eqnn $$ 
This equation can also be written in component form, setting 
$${\bf L}^\alpha=i L^{\alpha \beta }\tau^\beta  .$$
Then
$$ {\d \over \d t}L^{\alpha \beta }=\delta_{\alpha \beta }+f_{\gamma \beta
\delta } \xi^\gamma L^{\alpha \delta }.\eqnd\eGrLeq $$ 
We call $\Lambda$ the matrix of elements $L^{\alpha \beta}$ and  introduce the
antisymmetric matrix $X$ of elements 
$$X^{\alpha \beta }=f_{\alpha \beta \gamma }\xi^\gamma .$$
The solution of equation \eGrLeq~then can be written
$$\Lambda(t) =\int_0^t \d t'\e^{X(t'-t)}=X^{-1}\left(1-\e^{-tX}\right). $$
The metric tensor corresponding to the group is $g_{\alpha \beta }$
$$g_{\alpha \beta }=-\tr{\bf L}^\alpha (1){\bf L}^\beta (1)
=L^{\alpha \gamma }(1) L^{\beta \gamma }(1)=(\Lambda \Lambda^T)_{\alpha \beta
} ,\eqnn $$ 
and the group invariant measure is
$$\d{\bf g}\equiv(\det g_{\alpha \beta })^{1/2}\prod_\alpha
\d\xi^\alpha=\left(\det \Lambda \Lambda^T\right)^{1/2} 
\prod_\alpha \d\xi^\alpha .\eqnn $$
Then
$$\eqalign{\Lambda \Lambda^T&=-X^{-2}\left(1-\e^{-X}\right)
\left(1-\e^{X}\right) =-4X^{-2}\sinh^2(X/2) \cr 
&=-\prod_{n\ne 0}(1+X^2/(2\pi n)^2) ,\cr}$$
where we recognize an expression which appears in equation \eqns{\eFTgrmeas}.
\listrefs

\bye